\documentclass[10pt]{article}
\usepackage{arxiv}
\usepackage[T1]{fontenc}
\usepackage[utf8]{inputenc}
\usepackage{amsmath,amssymb,amsthm,mathtools}
\usepackage{graphicx,booktabs,tabularx,array,longtable}
\usepackage[numbers,sort&compress]{natbib}
\usepackage{xcolor,microtype,url,xurl}
\usepackage{enumitem}
\usepackage{listings}
\usepackage{placeins}
\usepackage[colorlinks=true,linkcolor=blue!45!black,citecolor=blue!45!black,urlcolor=blue!45!black]{hyperref}
\usepackage[nameinlink,noabbrev]{cleveref}
\renewcommand{\shorttitle}{\textsc{Recursive Organization Improvement}}
\renewcommand{\headeright}{}
\setlist{nosep,leftmargin=*}
\newcolumntype{Y}{>{\raggedright\arraybackslash}X}
\newtheorem{definition}{Definition}

\newcommand{\Org}{\mathcal{O}}
\newcommand{\Trace}{\mathcal{T}}
\newcommand{\Hist}{\mathcal{H}}
\newcommand{\E}{\mathbb{E}}


\title{Recursive Organization Improvement:\\A Modeling Specification for Human--Agent Organizations}
\author{
  Zilong Wang \\
  CataX AI \\
  \texttt{wangzilong@cata-x.ai}
}
\date{}
\begin{document}
\maketitle
\begin{abstract}
Stronger AI agents do not automatically produce better organizations: teams must also learn which work arrangements to retain and when to reconsider them, while those arrangements determine what evidence the team can learn from. We propose a modeling specification for recursive organization improvement and evaluate it through an executable checker, a public-record mapping, and two simulation studies. The specification connects actor-visible histories, organizational memory, decision rights, and the evidence used to retain changes. Study 1 crosses six decision rules, three memory conditions, and three task environments under fixed resource ceilings. In a stationary environment, cumulative evidence raises balanced evaluation's net value per task from 0.452 to 0.480, while selection regret falls from 0.029 to 0.002. After a workflow-ranking reversal, retained evidence delays adaptation. In post hoc acquisition-matched controls, the remaining contribution of program replacement depended on the memory rule: small and uncertain under an eight-round window, positive under cumulative memory after a reversal, and negative under reset evidence. Study 2 models human--agent review with shared errors, copying, and decision rights. It identifies a two-actor case in which public records conceal whether agreement comes from shared errors or copying, and shows how dependence changes the preferred placement of human judgment under a fixed budget. Paid observation can lower decision error while costing more than the deployment gain. Assessments of organizational improvement should track the evidence acquired and retained, using acquisition-matched comparisons to isolate the contribution of evaluator revision.
\end{abstract}
\section{Introduction}
Consider a software team that uses AI agents to draft pull requests (PRs). Faster drafting can lengthen the review queue. The team may respond by routing routine changes to automated checks and reserving human review for difficult cases. To evaluate this change, it must decide which PRs to audit, how to combine new observations with earlier evidence, and who may revise the routing or audit rule. An apparent gain can reflect a better workflow, a different sample of reviewed cases, or more effective reuse of evidence. How can the team distinguish these explanations and retain changes that improve delivery?

We propose a modeling specification for this problem and evaluate it with an executable checker, a mapping of public PR records, and two simulation studies. The running PR example connects the specification's fields to concrete organizational decisions; the studies isolate evidence acquisition and retention, then the review arrangements that generate observable judgments.

Task-level productivity studies and studies of human--AI collaboration address different parts of this problem. Experiments on professional writing and evidence from customer support document productivity effects in particular settings \citep{noy,brynjolfsson}; a meta-analysis finds substantial variation in the performance of human--AI combinations relative to their constituent members \citep{vaccaro}. Organization design supplies the intervening objects: the division and allocation of work, dependencies, and integration of effort \citep{puranam,malone}. Complementary organizational investments can also affect the returns to a new technology \citep{milgrom,jcurve}. Consequently, comparing agent models while holding the organization fixed answers a different question from comparing organizations that can revise their workflows.

Workflow revision creates a measurement problem of its own. A new review policy changes both which cases receive attention and which failures become visible. If the same selected cases determine whether that policy is retained, organizational learning can favor a locally successful but globally harmful arrangement. An improvement procedure therefore needs an explicit account of the evidence it can acquire, the changes it can authorize, and the costs it incurs. The procedure may itself become a target of revision. Study 2 examines how review visibility and decision rights shape the evidence available for judging an arrangement.

We call this problem \emph{recursive organization improvement} (ROI; distinct from return on investment): evaluating and retaining changes to organizational arrangements, including changes to the procedures through which those arrangements are improved. The research object is the human--agent organization. Agent capability, human judgment, coordination, evidence access, and authority can change at different rates and under different constraints.

This paper contributes a modeling specification that links organizational state, actor-visible histories, and the records used to accept a change. A revision names both the arrangement being changed and the procedure used to evaluate it. This makes it possible to reconstruct the evidence available at a decision, check editing rights and resource limits, and compare the outcomes of retained changes.

Study 1, \emph{evidence acquisition and retention}, holds label quality fixed and crosses reset, cumulative, and finite-window memory with fixed evaluation programs, one-time discovery, and repeated program assessment. Retention removes much of repeated assessment's disadvantage in a stable environment but delays adaptation after a workflow-ranking reversal. The acquisition-matched control is central to interpreting the apparent gain from evaluator revision: a fixed program receives the same trials and reusable labels; the remaining difference attributable to program replacement depends on the memory rule.

Study 2, \emph{review arrangements and evidence generation}, derives organizational outcomes and visible judgments from individual errors, shared error sources, upstream exposure, and decision rights. Its contribution is to distinguish mechanisms that, in a two-actor case, can look identical in public records but respond differently to blinding, separate gold-labeled measurement from human override authority, and show how error dependence changes the preferred placement of humans under a fixed budget. Paid observation policies test whether identifying a better arrangement also recovers the cost of learning about it.

Study 1 follows evidence through discovery, retention, and reassessment. Study 2 examines how that evidence is generated: blind review preserves private judgments, whereas exposed review can replace them with copied agreement. A team choosing a review arrangement therefore also chooses which judgments its improvement procedure can observe. Study 2 compares fixed observation policies for learning about these arrangements.

\section{Conceptual foundations and modeling requirements}
\label{sec:foundations}
\subsection{Organizational learning and explicit organizations}
Organizational learning connects experience to retained routines \citep{levitt,march}. The distinction between the abstract understanding of a routine and its situated performance helps explain how routines can generate both stability and change \citep{feldman}. Deliberate learning adds articulation and codification of experience \citep{zollo}, while double-loop learning addresses governing assumptions and values \citep{argyris}. Engelbart's improvement-of-improvement distinction supplies a close conceptual antecedent for examining the procedures that generate organizational change \citep{engelbart}. ROI translates this question into versions, events, change targets, and outcome comparisons.

Human--agent organizations also require an account of interdependence. Mixed-initiative interaction and levels of automation distinguish who initiates or performs a function \citep{mixed,levels}; Coactive Design places the support for interdependent activity at the center of system design \citep{coactive}. Appropriate reliance concerns when a person should use automated advice \citep{lee}. These choices alter information exposure, decision rights, and resource use, so they belong in the organizational model rather than being inferred from an actor's capability score.

Existing formalisms already supply much of the representational machinery. MOISE+ models structural, functional, and deontic organization; OperA represents organizational interaction and agent autonomy \citep{moise,opera}. Process mining uses event data for process discovery, conformance checking, and enhancement \citep{processmining}. The proposed specification connects these objects to the evaluation and retention of changes through the relations in \cref{tab:crosswalk}. A sufficiently annotated event log or state-transition model can encode the same relations. The contribution is their explicit organization into a comparison contract, whose completeness and execution can be checked.

\begin{table}[t]
\caption{Connections to established approaches. The right column identifies the records used in an ROI comparison.}
\label{tab:crosswalk}\small
\begin{tabularx}{\linewidth}{@{}p{.23\linewidth}YY@{}}\toprule
Approach & Established modeling object & Required connection for a change comparison \\\midrule
MOISE+; OperA & Roles, missions, norms, interactions & Target-specific authority, prior version, evidence and comparator for a revision. \\
Process mining & Cases, activities, event order, conformance & Organization/procedure versions, actual observation access, and sampling coverage. \\
Organizational learning & Experience, routines, retention & Link a retained routine to the experiment and criterion used to accept it. \\
Agent/workflow search & Executable candidates and evaluations & Human resources, rights, generated-output costs, and revision of the evaluator. \\
Generic event log & Arbitrary recorded attributes & The same relations, when recorded; missing fields remain unknown. \\\bottomrule
\end{tabularx}
\end{table}

Automated agent-system design and workflow optimization provide candidate-generation mechanisms \citep{adas,aflow}; linguistic feedback can support behavioral revision \citep{reflexion}. In ROI, these mechanisms are embedded in an organization with human roles, restricted evidence, and retention decisions. The central comparison concerns organizational outcomes under specified changes, whether the candidate was written by a person, generated by an agent, or selected from a finite catalog.

\subsection{From development practices to measurable arrangements}
Waterfall-like staging, agile development, and Extreme Programming suggest different settings for dependencies, batch size, feedback delay, test placement, and ownership. Royce's original account includes iteration and the risks of late system testing \citep{royce}; agile principles emphasize frequent delivery and reflection \citep{agile}; XP provides more concrete testing, pairing, and integration practices \citep{xp}. These labels identify families of configurations rather than an ordering of organizational quality. A team can retain release gates while shortening internal feedback cycles, or introduce agents into an unchanged approval pipeline.

For example, let drafting, review, and integration capacities be 4, 5, and 9 accepted-task equivalents per day. A deterministic serial capacity bound is their minimum. Doubling drafting alone raises the bound from 4 to 5; raising review capacity to 8 at the same time raises it to 8. The factorial interaction is $8-5-4+4=3$. This standard bottleneck calculation \citep{amdahl,little} isolates why technical and workflow changes can be complementary. Rework, variable quality, and demand must be modeled separately in an empirical application. Our specification records the fields needed to distinguish such joint changes from an increase in task-level speed.

\section{A modeling specification for recursive organization improvement}
\label{sec:spec}
\subsection{State, improvement procedure, and change boundary}
An instance declares a task population, environment process, outcome criterion, and assessment horizon. The environment generates tasks and outcomes; actor policies receive only the observations assigned to them. At event index $t$, define
\begin{equation}
\Org_t=(X_t,D_t,L_t,\Phi_t,\Pi_t,M_t),\qquad U_t=(g_t,a_t,e_t,s_t,r_t).
\label{eq:state}
\end{equation}
The organization state $\Org$ describes who does the work and what they can know or change. The procedure $U$ describes how the team proposes, tests, selects, and retains changes. Table~\ref{tab:notation} maps their components to the PR example. Actors also have task-specific capability profiles; approval rights are recorded separately.

\begin{table}[htbp]
\caption{Core notation, illustrated by PR review routing.}
\label{tab:notation}\small
\begin{tabularx}{\linewidth}{@{}p{.15\linewidth}YY@{}}\toprule
Symbol & Meaning & PR example \\\midrule
$X$ & Actors, roles, tasks, versioned artifacts & Reviewer, agent, PR, test report. \\
$D$ & Task and resource dependencies & Approval depends on tests and reviewer availability. \\
$L$ & Decision rights & Maintainer may change routing; review board may change auditing. \\
$\Phi$ & Observation availability & Which test results or private assessments each actor can see. \\
$\Pi$ & Interaction and allocation protocols & Assignment order, review sequence, and workload allocation. \\
$M$ & Retained evidence and versions & Audit labels, trial scores, and the adopted routing rule. \\
$U=(g,a,e,s,r)$ & Generate, acquire, evaluate, select, retain & Propose a rule; audit PRs; score and select it; monitor continued use. \\
$\mathcal B$ & Declared change boundary & Routing and auditing may change; outcome criterion stays fixed. \\
$\Hist_{i,t}$ & Evidence visible to actor $i$ before event $t$ & Reports available before approval. \\
$p,\Trace,H$ & Change contract, event trace, evaluation horizon & Recorded routing change and its follow-up period. \\\bottomrule
\end{tabularx}\end{table}

\begin{definition}[Declared boundary]
A boundary $\mathcal B$ partitions fields into operational arrangements, improvement-procedure fields, and externally fixed fields for a stated comparison. It identifies authorized editors, the evaluation horizon, and the outcome criterion.
\end{definition}
An execution event changes work artifacts. An operational adaptation changes an arrangement under the declared improvement procedure. A procedure revision changes a field of $U$ and evaluates its consequences for subsequent organizational changes or outcomes. For instance, assigning a PR to a reviewer is execution; replacing its routing rule is operational adaptation; replacing the sampling program used to evaluate routing changes is procedure revision. A fixed outer selector may govern all three. The boundary $\mathcal B$ identifies which changes count as procedure revisions.

This boundary also resolves overlapping implementations. $\Phi$ applies the current visibility rules, while $a$ requests evidence for a change evaluation. If one patch affects both, the implementation records the coupled fields in the state and cost ledger. If the outcome criterion changes, its version changes too; comparisons under the old and new criteria are reported separately.

\subsection{Events and actor-visible histories}
For the PR team, an event may be a reviewer assignment, a private assessment, or a revealed test report. An event $z$ records an identifier, logical order, actor, action, input/output artifact versions, recipients, resource use, and organization version. Actions include assignment, commitment, reveal, critique, test, escalation, approval, and patch application. Logical order represents dependency; physical timestamps can additionally represent concurrency and elapsed time.

Let $\Hist_{i,t}$ contain actor $i$'s permitted initial information and artifacts revealed to that actor before event $t$. A policy satisfies
\begin{equation}
b_{i,t}\sim\pi_i(\cdot\mid\Hist_{i,t},\Org^{\mathrm{vis}}_{i,t}),
\label{eq:information}
\end{equation}
where $\Org^{\mathrm{vis}}_{i,t}$ is the actor-visible projection of the organization state. Thus a simulator can score a decision using a hidden true error rate while the decision maker receives sampled labels only. A seed supports reproducibility without granting actors access to each other's random choices.

The ordering of commitment and reveal events represents human-first, agent-first, and parallel assessment. Parallel commitments preserve the opportunity for private judgments, while shared training data, common tools, or correlated task difficulty can still induce dependent errors. Statistical dependence is a joint property requiring measurements or explicit assumptions; it cannot be inferred from separate model names. A public review timestamp records publication; establishing what another actor read requires exposure records.

\subsection{Change contracts and retention}
A proposed routing change carries a contract: a record linking the change to its evidence, comparator, cost, and follow-up rule. Formally,
\begin{equation}
p=(\tau,v,\delta,\mathcal E,\mathcal C,c_p,H,\mathcal R_p).
\label{eq:patch}
\end{equation}
$\tau$ identifies the target and proposer; $v$ is the required input version; $\delta$ is the transformation; $\mathcal E$ identifies available evidence and its provenance; $\mathcal C$ states the comparator, population, and criterion; $c_p$ specifies resource accounting; $H$ is the horizon; and $\mathcal R_p$ specifies retention, monitoring, and rollback conditions. Contract fields may refer to separately stored artifacts. Their relationships, summarized in \cref{tab:obligations}, determine whether a comparison can be reconstructed.

\begin{table}[t]
\caption{Specification obligations and observable failure conditions. Unknown information is distinguished from a violated condition.}
\label{tab:obligations}\small
\begin{tabularx}{\linewidth}{@{}p{.19\linewidth}YY@{}}\toprule
Relation & Required condition & Failure or missing record \\\midrule
Actor--evidence & Each decision input was available to its actor beforehand. & Hidden/future input; unknown exposure history. \\
Patch--state & The target version exists and matches the expected version. & Stale patch; unversioned configuration. \\
Actor--target & The actor holds the applicable editing right at that version. & Unauthorized change; unknown historical permission. \\
Evidence--population & Sampling scope and label source support the stated population. & Uncovered stratum; unknown inclusion mechanism. \\
Outcome--criterion & Both arms use the declared criterion and horizon. & Changed success definition; censored follow-up. \\
Cost--resource & Search, outputs, labels, and transitions enter one ledger. & Omitted counterfactual output; duplicate charge. \\
Decision--memory & Accepted and rejected proposals retain evidence and versions. & Retained policy without its validation domain. \\\bottomrule
\end{tabularx}
\end{table}

Admission checks authorization, versions, available evidence, prerequisites, and resources. Efficacy is a subsequent outcome comparison. These predicates have different witnesses: permission records can justify admitting a trial, while a trial's results can justify retaining its policy. Rejected and unsuccessful revisions remain in memory because they consume resources and influence future search.

For external criterion $J$ and specified comparator $0$, the effect of a patch is
\begin{equation}
\Delta_H(p)=\E[J(\Trace_{t:t+H}^{p})-J(\Trace_{t:t+H}^{0})]-C_H(p).
\label{eq:value}
\end{equation}
$C_H$ contains costs not already included in $J$. Randomized comparisons, controlled simulation, or explicit causal assumptions are needed to estimate the two potential outcomes. Outcomes without defensible common units remain a vector with declared constraints. The simulation criteria below include all modeled costs in net value, giving $C_H=0$ in \cref{eq:value}.

\begin{figure}[!htbp]
\centering\includegraphics[width=.97\linewidth]{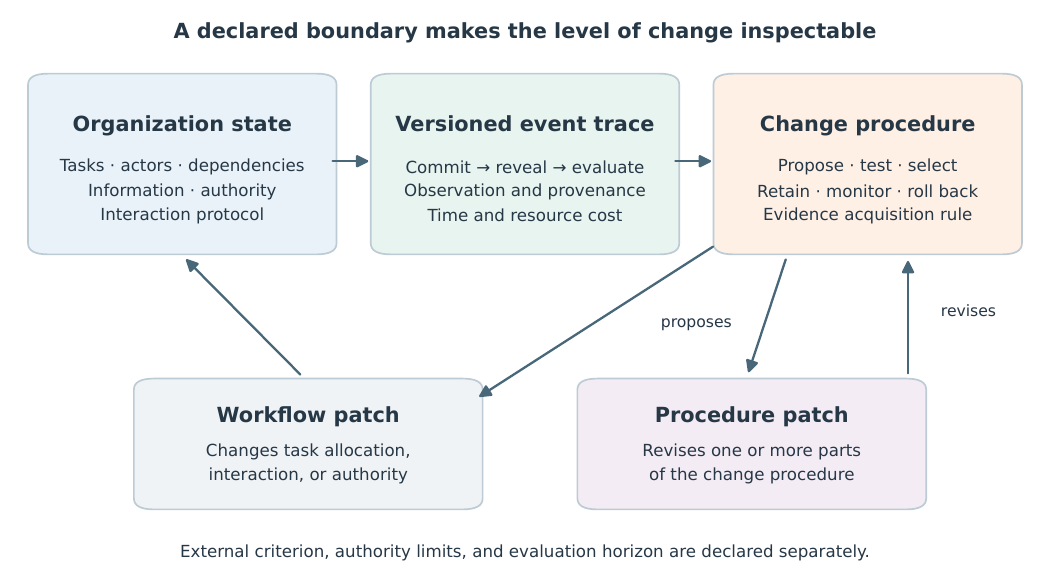}
\caption{The modeling specification connects organization state, event histories, and improvement procedures. A revision names its target, evidence, authority, and external evaluation criterion.}
\label{fig:framework}
\end{figure}

\Cref{fig:framework} connects execution to this revision loop. A software team might first change review routing, then discover that its evaluations cover only human-reviewed PRs. A second patch can alter evidence acquisition to include routine changes. That second patch is evaluated through later routing decisions and their outcomes.
\FloatBarrier

\subsection{Organizational memory as a testable design choice}
The memory field $M$ separates three retained objects: workflow evidence, evaluation-program evidence, and adopted arrangements. Keeping a program name preserves the adopted decision; accumulating knowledge requires retaining its supporting observations. Keeping observations indefinitely can preserve knowledge about a population that no longer exists. An instance therefore records evidence provenance, applicability, and expiry alongside the retained policy.

The retention component $U.r$ specifies when evidence or arrangements are reconsidered. A fixed window is one such rule; changing the rule is a procedure revision at the declared boundary. In Study 1, memory rules are experimentally assigned and held fixed within a run. Evaluation programs may then change inside each memory condition. This factorial construction separates a design intervention on retention from the execution of a program-selection procedure.

Evidence generated during organizational search can serve more than one decision. A rejected program trial may still produce valid labels about a workflow. The trace links those labels to their source and reveals them to the permitted downstream evaluator. Comparing a policy that reuses those labels with one that discards them tests an information-transfer rule. Comparing two policies with the same trials and label reuse, while allowing only one to replace its program, tests a more specific contribution of program revision.

\subsection{What the implementation checks}
The reference implementation provides a finite checker for unique event/artifact identifiers, visible inputs, version matching, protected criteria, target-specific permissions, and patch budgets. Constructed fixtures accept a valid trace and patch and reject a hidden input and five invalid patch variants. A separate target test allows a release owner to change routing but rejects the same actor's audit-rule change; granting the review board that right admits it. This checks the operational consequences of the declared boundary.

Study 1 instantiates a narrower executable subset: retained label statistics, retained trial scores, program versions, authorized patch application, label budgets, and a modeled cost ledger. It exports trajectory outcomes, program proportions, conditional block outcomes, and example version transitions. Full implementations would additionally need to enforce all relationships in \cref{tab:obligations}, including criterion and retention contracts. The checker tests these admission conditions; the simulations compare the outcomes of accepted changes.

Study 2 instantiates parts of Actor--evidence through purchased private and exposed judgments, Cost--resource through acquisition, production, and switching charges, and Decision--memory through retained observation counts and recorded arrangement choices. Study 2 selects arrangements directly from a fixed catalog, outside the versioned-patch interface of the reference checker.

Mapping public repository records supplies a complementary feasibility check. Three merged Flask PRs yield four commit artifacts and 56 mapped events: eight reviews, 42 head-commit check runs, and six opening/merge records. Artifact identifiers and recorded actions are recoverable. Historical authority, actual message exposure, effort, organization versions, and AI involvement remain unknown. Appendix~\ref{app:records} specifies the snapshot and mapping. Such a trace can expose instrumentation gaps before a team attempts a workflow-effect estimate.

\section{Study 1: Acquiring, retaining, and reconsidering evidence}
\label{sec:study}
\emph{Scope of the simulations.} Both studies examine evaluation and review arrangements in human--agent organizations using synthetic task, actor, and cost parameters. The core Study 1 design preceded execution; its acquisition-matched and change-timing controls were specified after inspection of the core results. Study 2 was designed after Study 1 and is exploratory. Task decomposition, capacity, queues, and specialist--generalist trade-offs are represented by the specification but lie outside these simulations.

\subsection{Research questions and organizational instance}
Study 1 follows a change from its discovery to its later use. It asks whether evidence retention alters the value of evaluator revision, whether a one-time discovery can substitute for repeated search, and which part of reassessment produces a measured gain. These questions instantiate $U.a$ (evidence acquisition), $M$ (retained evidence), and $U.r$ (continued use or reconsideration) in the specification.

The relations in \cref{tab:obligations} constrain specific design decisions. Decision--memory requires separate records for workflow labels, program scores, and the adopted program; we therefore intervene on these memories separately. Cost--resource requires each queried label to enter one ledger, even when both program trials and workflow selection reuse it. Actor--evidence requires current program trials to share the same pre-review information and prevents access to hidden error probabilities. Outcome--criterion keeps the deployment horizon and net-value rule common across arms. Study 1 uses these obligations to distinguish a change in the evaluator from a change in its evidence.

A proposer supplies three workflow templates, an evaluator acquires labels, and a review board may replace the evaluation program. In Study 1, the roles denote information access and decision rights and can be assigned to humans or agents; Study 2 models their individual judgments (Section~\ref{sec:actor-review}). The environment has two equally frequent observable task strata. Error probabilities are $(.20,.20)$ for the standard template, $(.16,.16)$ for broad, and initially $(.08,.46)$ for specialized. Thus broad is initially best, while specialized helps one stratum and harms the population. In \emph{stationary harm} these probabilities remain fixed. In \emph{workflow reversal}, specialized becomes $(.08,.08)$ before round 25 and becomes the best template. \emph{Uniform gain} uses that improving specialized template throughout.

Each of 48 rounds deploys one template on 4096 production tasks. Only acquired evaluation labels enter decisions; the external scorer uses exact expected production outcomes. This separates uncertainty in organizational judgment from production sampling noise. One value unit is the gross value assigned to one completed production task before errors and costs. Net value per task subtracts error penalties and modeled expenses in these normalized units. An error costs 3 units, generation costs $.02$ per production or evaluation output, a label costs $.20$, and changing evaluation programs costs $.50$. Storage and computation costs are outside this ledger.

\subsection{Memory and decision rules}
The population-weighted posterior error estimate of template $k$ is
\begin{equation}
\widehat r_{k,t}=\frac12\sum_{s=1}^{2}\frac{1+E_{ks,t}}{2+N_{ks,t}},
\label{eq:risk}
\end{equation}
where $E$ and $N$ count observed errors and labels available to the decision. All conditions begin with $\operatorname{Beta}(1,1)$. \emph{Reset} uses current-round evidence only. \emph{Cumulative} retains all earlier evidence. \emph{Window 8} retains evidence from the preceding eight rounds plus current observations. The same retention rule applies to past evaluation-program scores; a program's identity remains retained until another is selected. This separates remembering a decision from remembering the evidence supporting it.

Three evaluation programs form a fixed catalog. Biased allocates 95\% of a template's allowance to the first stratum. Balanced uses equal counts. Neyman uses a two-label-per-cell pilot and posterior variance estimates to allocate the remainder across strata \citep{neyman}. Each program chooses the template minimizing \cref{eq:risk}. Counts use population weights regardless of sampling proportions.

Workflow selection is a fixed-budget identification problem: evaluation outputs are acquired to choose a template for subsequent deployment. Successive Rejects addresses this objective by progressively eliminating candidates \citep{bai}; exploration sampling explicitly targets policy choice rather than immediate experimental outcomes \citep{kasy}. We use a between-template elimination comparator based on successive rejection, whereas Neyman allocates within each template. It observes paired stratum labels, eliminates the worst estimated candidate after a first phase, and compares the remaining two after a second phase. Historical evidence enters the same posterior estimates under the assigned memory rule. Allocation formulas and tie rules appear in Appendix~\ref{app:protocol}.

\subsection{Discovery, reuse, and fixed resource ceilings}
\Cref{tab:arms} distinguishes knowing an evaluator, discovering it once, and periodically reconsidering it. A label unit is one observed binary error outcome for one evaluated output. Every arm receives a ceiling of 2736 such observations per eight-round block. A fixed arm can allocate 342 units per round, or 114 per template for the three catalog programs. Discover once pays for program comparisons only before round 1. Repeated discovery compares programs before rounds 1, 9, 17, 25, 33, and 41. Both begin with Biased and select among the same catalog.

\begin{table}[t]
\caption{Decision rules and the comparisons they support. All arms use the same memory condition and block-level resource ceiling.}
\label{tab:arms}\small
\begin{tabularx}{\linewidth}{@{}p{.23\linewidth}YY@{}}\toprule
Arm & Program decision & Role in the study \\\midrule
Biased / Balanced / Neyman & Program specified before execution & Fixed designs with different coverage/allocation rules. \\
Successive rejection & Fixed elimination rule across templates & A between-template elimination comparator. \\
Discover once & Compare programs once; retain the winner & Discovery followed by reuse and budget reallocation. \\
Repeated discovery & Reassess every eight rounds & Incremental value of continued program assessment. \\
Trial-matched Balanced & Run the same trials; keep Balanced & Exploratory control for evidence acquisition and timing. \\\bottomrule
\end{tabularx}
\end{table}

In the base design, program trials receive at most 20\% of the block ceiling, with one trial per catalog program. Each trial screens the three workflows under the same pre-review memory snapshot, selects one, and validates it on fresh balanced labels. Its score is validation value less screening expense per production task. The highest mean retained trial score determines the program. This fixed outer selector revises the retained program within the catalog.

All acquired trial labels, including evidence from rejected programs, can enter workflow memory. Decisions about program scores precede pooling the current trials; the subsequent operational screen receives the pooled evidence. This makes failed searches potential sources of reusable information. Actual trial expenditure is deducted from the current block, and the remaining allowance is spread across its operational screens. Discover once receives the full operational allowance from block 2 onward. No arm borrows from a future block; unused units due to integer allocation are not charged.

Net value includes production generation, all acquired labels, all evaluated outputs, and program changes. Selection regret measures the gross value gap to the best available template, separately from those costs. Harmful adoption counts choices worse than standard; after reversal, this measure becomes zero for all templates, so regret and adaptation trajectories are needed to distinguish them. The primary outcome averages net value over all 48 rounds, with final-eight-round performance as a separate endpoint.

The core matrix crosses three environments, three memory rules, and six arms, using 128 independent replicate streams per cell. A separate 96-replicate grid crosses program-budget shares $.2,.4,.6$ with trial counts $1,3,6$, keeping the total ceiling fixed. It also removes program-score memory or trial-label pooling separately. The acquisition-matched controls use independent streams and include reversals before rounds 21 and 29 to assess attribution and timing. Appendix~\ref{app:protocol} specifies all conditions.

\subsection{Accumulation changes the apparent cost of improvement}
\begin{figure}[t]
\centering\includegraphics[width=\linewidth]{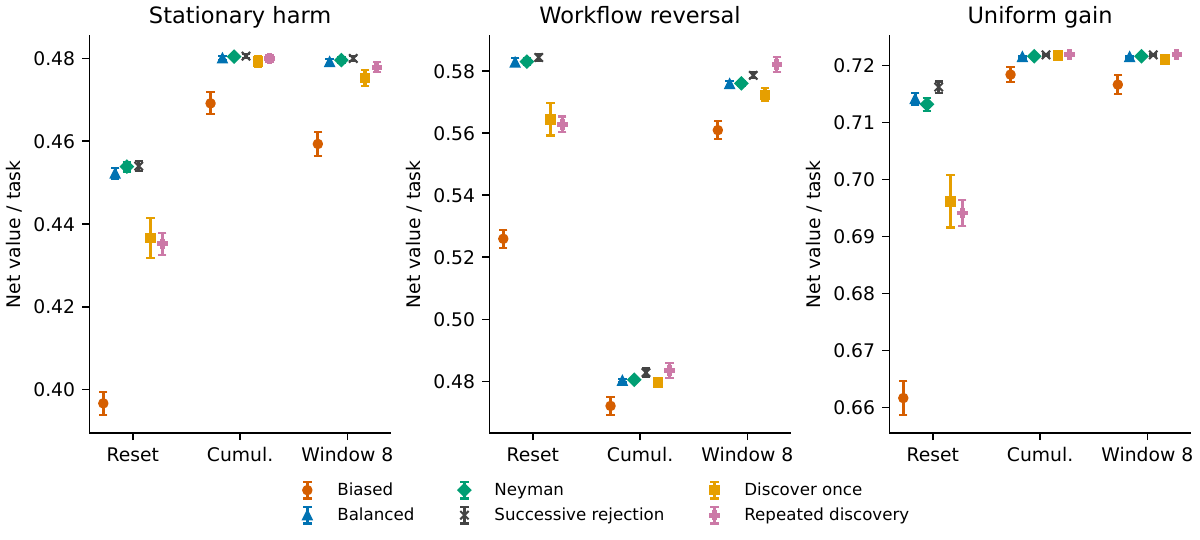}
\caption{All core comparisons: 128 replicate trajectories per cell, means and 95\% Monte Carlo intervals. Each panel uses its own value scale. ``Cumul.'' retains all acquired evidence; Window 8 retains the preceding eight rounds plus current evidence. The results concern the jointly specified memory, discovery, and acquisition rules.}
\label{fig:memory}
\end{figure}

\Cref{fig:memory} reports every core arm and memory condition. Under stationary harm, cumulative evidence raises Balanced's mean net value from $.452$ to $.480$, a paired gain of $0.0278$ (95\% Monte Carlo interval $0.0266$ to $0.0291$). Its final-eight-round value reaches $.482$, the value of consistently selecting broad under its actual evaluation expenditure. Balanced's mean selection regret falls from $.0294$ to $.00156$ per task, while its evaluation expense remains $.0184$. The net-value gain therefore equals the reduction in selection regret.

Repeated discovery is $0.0170$ below Balanced with reset evidence (interval $-0.0196$ to $-0.0144$). With cumulative evidence, both means round to $.480$ and the paired interval spans zero. Discover once reaches $.479$, while successive rejection reaches $.481$. Reusing acquired evidence thus removes most of the reset disadvantage, and fixed selection rules remain competitive with program revision. Under Uniform gain, specialized is best from the outset, so retention improves selection throughout.

The comparison to fixed programs also depends on prior knowledge. An equal-probability initial draw from the three catalog programs gives net values of $.434$ with reset evidence and $.477$ with cumulative evidence, compared with $.435$ and $.480$ for repeated discovery. Table~\ref{tab:base} reports this mixture in every core condition. Balanced provides a prespecified representative-sampling reference for distinguishing discovery from the revision of an already effective evaluation rule.

\subsection{Acquisition matching and the contribution of program replacement}
\begin{figure}[t]
\centering\includegraphics[width=\linewidth]{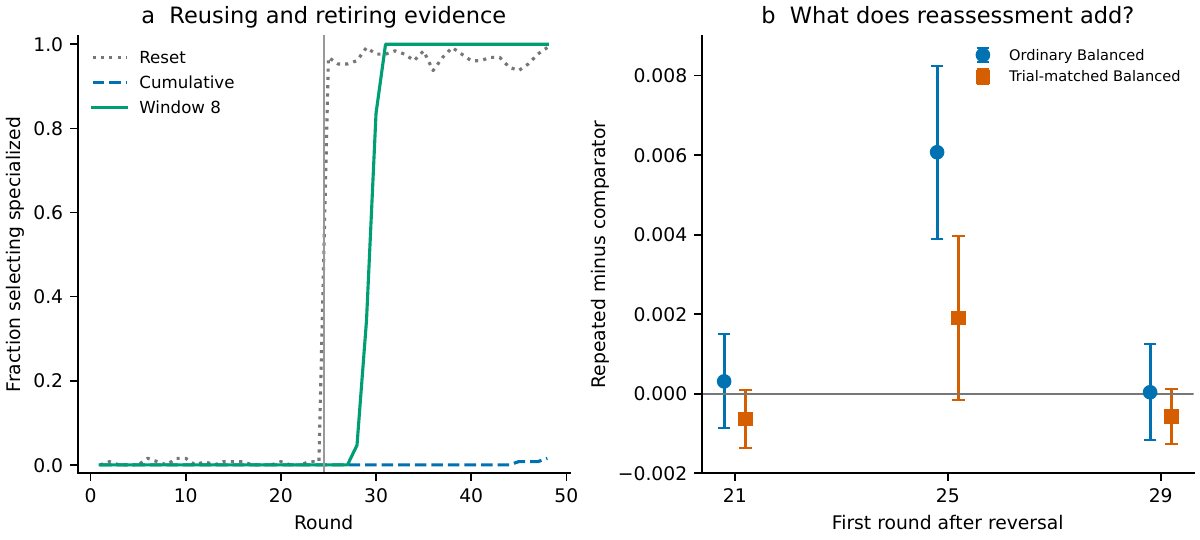}
\caption{Retention and attribution. (a) Fixed Balanced under the core reversal; the vertical line marks the change before round 25. (b) Exploratory controls with Window 8 and reversals before rounds 21, 25, or 29. Points show repeated-discovery minus comparator differences with paired 95\% Monte Carlo intervals on 128 fresh replicate streams. Matching the trials, label reuse, and timing changes the attribution of the apparent reassessment gain.}
\label{fig:attribution}
\end{figure}

Under workflow reversal, indefinite accumulation slows the replacement of broad by the newly superior specialized template. Balanced's full-horizon values are $.480$ with cumulative memory, $.576$ with Window 8, and $.583$ with reset evidence. The finite window improves considerably on unlimited accumulation, yet its transition delay makes it worse than reset over this particular horizon. In the final eight rounds it reaches $.722$, compared with $.712$ for reset and $.483$ for cumulative memory. \Cref{fig:attribution}a displays the corresponding deployment choices.

With Window 8 in the core reversal, repeated discovery exceeds discover-once by $0.0098$ (interval $0.0071$ to $0.0124$) and ordinary Balanced by $0.0062$. Program trials also concentrate acquisition near review times and supply labels that workflow selection can reuse. The trial-matched control gives Balanced these same trials, pooled labels, and acquisition timing while keeping its program fixed. On independent streams with reversal before round 25, the repeated-discovery advantage falls from $0.0061$ over ordinary Balanced (interval $0.0039$ to $0.0082$) to $0.0019$ over trial-matched Balanced (interval $-0.0002$ to $0.0040$). At reversals before rounds 21 and 29, the remaining mean differences are slightly negative and both intervals span zero. Under Window 8, acquisition matching therefore leaves a small, uncertain contribution from program replacement across these three timings. Under cumulative memory, a positive contribution remains at all three timings ($0.0017$ to $0.0113$), whereas under reset evidence it is negative; Supplementary Table S1C reports all memory conditions.

\subsection{What the allocation and retention diagnostics explain}
The full nine-cell budget-share/trial-count grid is reported in \cref{tab:grid}. Program trials can improve subsequent workflow selection by adding reusable labels to workflow memory. Removing trial-label pooling lowers repeated discovery from $.480$ to $.478$ under stationary cumulative memory, and from $.582$ to $.574$ under reversal with Window 8 on the independent sensitivity streams.

\Cref{tab:programs} relates the retained program to subsequent-block outcomes. At the last stationary review with cumulative memory, Biased, Balanced, and Neyman are retained in 38.3\%, 31.2\%, and 30.5\% of runs. The 40 runs retaining Balanced have subsequent-block net value $.482$ and zero harmful adoption. Accurate workflow selection can therefore coexist with different retained programs when their decisions share a well-informed memory. Supplementary Table S2 summarizes the final program-selection proportions across environments and memory conditions.

\section{Study 2: How review arrangements generate evidence}
\label{sec:actor-review}
\subsection{Research questions and connection to Study 1}
Study 1 prescribes workflow error rates and label quality independently of the arrangement that acquires evidence. Study 2 makes the expressed judgments and final outcomes depend on a human--agent review arrangement. The same private signals may remain available under blind review or be replaced by copied agreement after exposure. Proposers and reviewers instantiate $X$; exposure and order change $\Phi$ and $\Pi$; arbitration changes $L$; and observation retention instantiates $M$.

We vary private error dependence and copying at fixed individual accuracy, compare review arrangements under a common budget, and test observation policies for learning which arrangement to deploy.

\subsection{Private errors, copying, and bounded review arrangements}
A binary task has a proposer, agent reviewers, and, in some arrangements, a human arbiter. Humans have private error probability $q_H$; conditional on task truth, their errors are independent of one another and of the agent group. Agents have private error probability $q_A$. With probability $\rho_A$, all agents on a task, including an agent proposer, share one error draw; otherwise their errors are independent. This construction preserves individual accuracy while varying private error dependence.

Motivated by observed sycophancy in multi-agent debate \citep{consensagent}, we isolate copying (echo) without additional task information. Here $\eta_s$ is the probability that an agent reviewer copies an upstream decision from source $s\in\{H,A\}$; otherwise it retains its private judgment. The subscript denotes the upstream actor, so $\eta_H$ describes an agent copying a human. Blind protocols preserve private commitments for aggregation or escalation while retaining any intrinsic error dependence. We compare exposed and blind voting, exposed and blind disagreement-triggered human arbitration, random arbitration of unanimous cases, serial review, and single-actor baselines. Exact enumeration identifies a minimum-error arrangement within the declared candidate catalog and budget; expected cost breaks error ties. The error-first grid isolates decision quality within a hard resource ceiling. The learning comparison instead selects on deployment loss $3e+.02c$, where $e$ is error probability and $c$ is expected production call cost; its net-value score additionally deducts observation and switching costs (\cref{eq:actor-value}). Appendix~\ref{app:actor} specifies the joint error law, protocols, costs, parameter grid, and ties.

\subsection{Visible agreement and the placement of human judgment}
For an agent proposer and three agent reviewers, private accuracy of 90\% and $\rho_A=.1$, increasing $\eta_A$ from 0 to .8 raises unanimous agreement with the proposer from 69.1\% to 91.2\%, while majority-decision error rises from 3.52\% to 9.64\%. Conditional on an incorrect proposal, unanimous approval rises from 10.1\% to 59.6\%. Blind voting retains an error rate of 3.52\% under the same private-signal distribution. These curves illustrate the consequences of the copying rule. The identification problem appears when distinct mechanisms generate the same public record. With one agent proposer and one equally capable agent reviewer, $(\rho_A,\eta_A)=(.8,0)$ and $(0,.8)$ produce the same joint distribution of public votes and truth but respond differently to blinding (Appendix~\ref{app:actor}).

\begin{figure}[!htbp]
\centering\includegraphics[width=\linewidth]{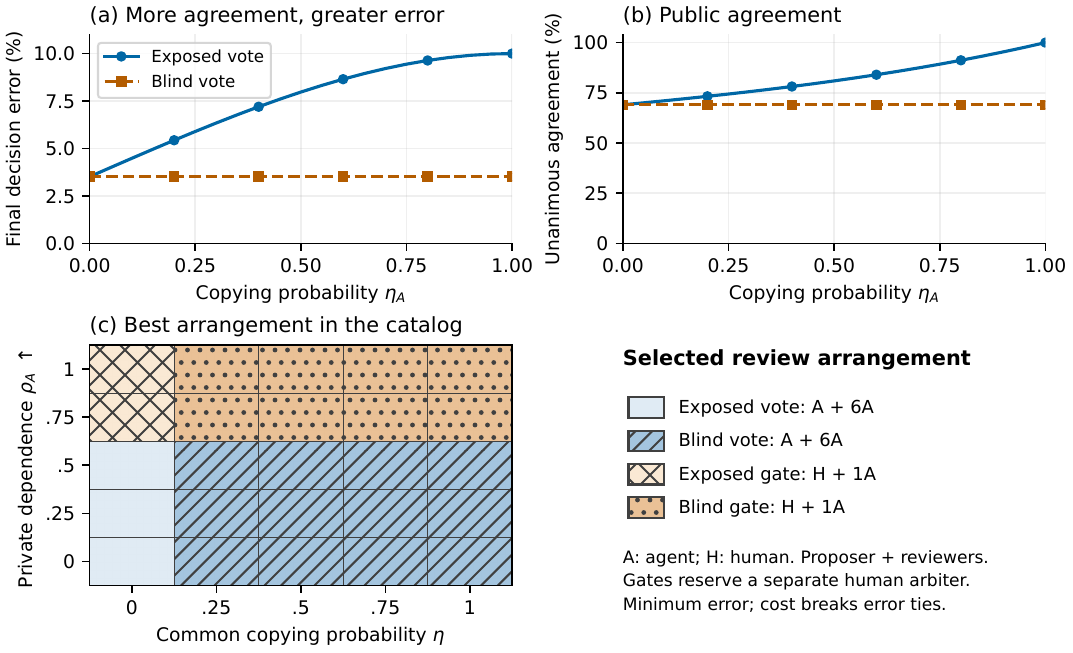}
\caption{Study 2: private judgment, upstream exposure, and review organization. (a--b) Final majority-decision error and unanimous agreement for an agent proposer and three agent reviewers, with $q_A=.1$ and $\rho_A=.1$. Exposed reviewers copy with probability $\eta_A$; blind protocols retain private votes. The curves evaluate 21 equally spaced copying probabilities. (c) The minimum-error arrangement, with cost breaking error ties, at each dependence and copying setting, for $q_H=.2$, $q_A=.1$, and a hard task budget of three human-call units. Human and agent calls cost 1 and .25; blind commitment adds .05 per reviewer. Gates reserve a separate human arbiter. Dependence increases upward. Colors and hatching identify the selected arrangements; all values are exact expectations.}
\label{fig:actor-review}
\end{figure}

\Cref{fig:actor-review} separates copying from pre-existing dependence. With $q_H=.2$, $q_A=.1$, a human-to-agent call-cost ratio of four, blind overhead of .05 per reviewer, and a hard task budget of three human-call units, the minimum-error choice at $\eta_H=\eta_A=.5$ changes from an agent proposer with six blind agent reviewers at $\rho_A=0$ to a human proposer, one blind agent reviewer, and an independent human arbiter on disagreement at $\rho_A=1$.
The preferred arrangement changes where human judgment enters the decision. In the disagreement gate, a human makes the final decision on disputed cases. Learning about the gate uses a separate gold-labeled audit. The gate changes decision authority; the audit measures performance.

\subsection{Paid observation, retention, and net value}
\label{sec:actor-learning}
We compare four fixed observation policies for selecting arrangements with unknown actor parameters. Across 48 rounds, they choose among 58 budget-feasible arrangements under reset, cumulative, or Window 8 memory. They acquire final outcomes only, paired private and post-exposure judgments, or observations from randomized blind and exposed tasks. An \emph{Omit echo} ablation receives the same paired records but sets copying probabilities to zero in its predictions. Six environments, including a copying shift before round 25 and source-dependent copying, use 128 replicate trajectories per condition. All policies share a ceiling on acquisition cost, but purchase different numbers of labels. The cost schedule is separate from Study 1's label-count budget. Observation policies are assigned treatments and do not change during learning.

In High echo with cumulative memory, paired observation achieves mean decision error 1.77\% and net value $.861$, compared with 9.82\% and $.623$ for the Omit echo ablation using the same acquired records. Randomized exposure reaches 1.78\% error and $.861$ net value. The fixed blind agent-proposer/three-reviewer baseline has higher error, 3.52\%, but higher net value, $.871$, because it purchases no continuing learning evidence. Appendix~\ref{app:actor} reports all six environments under cumulative memory; Supplementary Table S5 reports every policy--memory condition and its costs.

Across all six environments, cumulative paired observation has lower error than the fixed blind baseline, but its net value is lower in five. It spends 234 acquisition units per round, or $.0570$ per production task after amortization over 4096 tasks. In those five environments this expense exceeds the gain in deployment value, including production call cost; the switch charge is included separately. In Stronger humans, paired observation's net value is $.773$, compared with $.607$ for the fixed blind baseline.

In No echo, fixed exposed and fixed blind review have the same error rate, but net values of $.874$ and $.871$: blind review pays commitment overhead and one initial switch. Under cumulative memory, randomized exposure has a 56.7\% optimal-choice hit rate yet the same mean error as paired observation (1.77\%); its higher production call cost makes some equally accurate choices suboptimal under deployment loss. Error, regret, and net value distinguish these equally accurate but differently priced choices.

Memory effects depend on the observation policy. In Echo shift, paired observation's full-horizon net values are $.858$ with reset, $.863$ with cumulative memory, and $.863$ with Window 8; cumulative and Window 8 both reach final-16-round error 1.77\%. For randomized exposure, the corresponding net values are $.851$, $.852$, and $.857$, and final-16-round errors are 2.32\%, 1.84\%, and 1.77\%. Finite retention improves adaptation for randomized exposure, while paired observation changes little between cumulative and Window 8 memory. The value of forgetting depends on the observation procedure.

\section{Implications for human--agent organizations}
\subsection{Comparing capability and organizational change}
The specification makes an otherwise ambiguous comparison operational. A capability-only intervention changes actor profiles while keeping assignment, visibility, authority, and improvement procedures fixed. A workflow intervention changes those arrangements under a fixed procedure. A procedure intervention changes how candidate arrangements are evaluated or retained. A factorial study can compare these interventions and their interactions under the same task population and outcome criterion. The bottleneck example in \cref{sec:foundations} supplies one elementary prediction; Study 1 shows how evidence acquisition and retention alter the value of organizational search; Study 2 shows how review arrangements generate the judgments that can become evidence.

The distinction matters when AI changes both production volume and the work of supervision. Automation can redistribute monitoring demands \citep{bainbridge}; human and machine learning can become organizationally interdependent \citep{sturm}. In Study 1, the same amount of acquired evidence has different value depending on whether the organization reuses it, discards it, or retains it after the relevant task relationship changes. Evidence retention should therefore be evaluated together with the workflow that generates it.

The supplementary audit model in Appendix~\ref{app:audit} examines the earlier decision to acquire information at all. Its finite-horizon value-of-information controller can acquire little evidence under a strong prior, while Study 1 asks what happens to evidence after it has been bought. Together the models distinguish acquisition, retention, and use. Neither an audit count nor a revision count alone measures an organization's capacity to learn.

\subsection{Independence and interaction order}
Distinct agents can share models, training data, tools, and explanations, producing correlated errors \citep{clemen}. Correlated-vote extensions of Condorcet's jury theorem analyze dependent judgments \citep{ladha1992}; information-cascade theory and social-influence experiments examine how exposure changes the use and diversity of private information \citep{bikhchandani1992,lorenz}. Study 2 separates shared private errors from copying and shows that a two-actor public record can conceal this distinction. Recording initial judgments and reveal events makes the mechanisms distinguishable.

Human-first and agent-first interaction differ through the information available when a person commits an assessment. Private commitment can preserve an additional signal, while early advice can supply knowledge the person lacks. The outcome depends on capability, covariance, reliance, and the aggregation rule. The elementary calculation in Appendix~\ref{app:analytical} shows that apparent gains from exposure can equal a reweighting of private evidence. Human--AI studies of appropriate reliance and cognitive forcing motivate behavioral measurements of responses to advice \citep{lee,bansal,bucinca}; their task and participant measures can be used to estimate how exposure changes reliance.

\subsection{Preserving disagreement and separating observation from authority}

Study 2 identifies information exposure and decision rights as distinct design choices. When reviewers can copy an upstream decision, public agreement combines agreement already present in private judgments with exposure-induced conformity. An organization that escalates only visible disagreements can therefore suppress its own opportunity to correct mistakes. Retaining initial assessments, their sources, and reveal order makes these mechanisms distinguishable and supports comparisons of alternative review arrangements.

A gold-labeled audit measures the performance of a review arrangement. A human override changes its final decisions. Their costs and benefits belong to different parts of the comparison: the audit informs future arrangement choices, while the override affects the current task. This distinction helps a team decide whether it needs better measurement, a different escalation rule, or a different final decision maker.

\subsection{How the two studies connect}
\label{sec:study-connections}
Study 1's workflow reversal and Study 2's copying shift both change the relevance of earlier observations. Their different adaptation patterns show that memory must be assessed together with the process generating those observations. The acquisition-matched control separates evaluator replacement from a changed information path, and shows that its contribution depends on how long evidence is retained; the Omit echo ablation checks how a deliberately misspecified model interprets the same acquired record. Study 2 holds observation policies fixed, extending the operational comparison to review arrangements whose evidence is generated internally.

\subsection{Validity and next empirical tests}
\label{sec:validity}

The reference checker and public-data mapping demonstrate how the specification links decisions to evidence, permissions, versions, and costs. A comparative evaluation of the specification could measure model-building time, agreement about missing records, and prediction accuracy when users model the same organizations with alternative notations.

Study 1 uses a finite program catalog, known stratum weights, and retained scores from trials conducted at different knowledge states. A wider catalog and adaptive weighting of historical scores would extend the acquisition-matched comparison. Concept-drift, switching-bandit, and change-point methods provide alternatives to fixed-window retention \citep{drift,switchbandit,bocpd}; acquisition costs can be included in those comparisons \citep{costbandit}.

Study 2's next tests concern when to stop paying for observation, how deployed private judgments can supply feedback, and how production volume changes the value of that feedback. Field measurements would also need to estimate human copying of agent advice and whether recording a private commitment changes subsequent behavior. These measurements would connect the error and copying assumptions to a particular review task.

For Study 1, an organizational test would randomize review-routing changes and evaluation coverage across comparable task streams, retain failed proposals, and measure escaped errors and human attention over a declared horizon. A second stage could randomize whether teams may replace the evaluation program. Such a design would separate the value of broader evidence from the value of the procedure used to discover and retain that evidence policy.

For Study 2, an organizational test would randomize upstream visibility before private commitment across comparable review tasks, while recording actor identity, initial and final judgments, independent outcome labels, and attention costs. Separately varying escalation and override rights would distinguish the value of additional evidence from the effects of final decision authority. A later stage could randomize whether teams may replace the observation protocol, thereby testing procedure revision beyond the operational arrangement choices studied here.

\section{Conclusion}
Recursive organization improvement asks how human--agent teams change their work and learn whether those changes helped. The proposed specification links organizational arrangements to visible evidence, decision rights, retained versions, and outcome comparisons. In Study 1, evidence reuse removes much of repeated assessment's apparent disadvantage, while obsolete evidence delays adaptation. In post hoc acquisition-matched controls, the remaining contribution of program replacement depended on the memory rule: small and uncertain under an eight-round window, positive under cumulative memory after a reversal, and negative under reset evidence. Study 2 shows how review arrangements determine which judgments become observable and how shared errors and copying change the placement of human judgment under a budget. Lower decision error can still come with a net loss from continued observation. Organizational improvement should therefore be assessed through the evidence acquired and retained; acquisition-matched comparisons can isolate the contribution of revising an evaluator.

\section*{Code and data availability}
Code, protocols, trajectory results, validation records, and the public-record snapshot are available at:\par
\noindent\href{https://github.com/wizardlancet/recursive-organization-improvement}{\mbox{\texttt{https://github.com/wizardlancet/recursive-organization-improvement}}}\par
Protocols were archived at commits \texttt{75d312e} and \texttt{fdf5ec1}; neither study was externally preregistered. \hyperref[sec:supplementary-tables]{Supplementary Tables S1--S5} follow the appendices. S2 summarizes final program-selection proportions; the repository contains the full 189-group conditional results and maps each table to its source. Code is licensed under MIT; the manuscript, figures, and synthetic data under CC BY 4.0. Third-party materials retain their original terms.

\section*{Use of AI tools}
OpenAI Codex assisted with manuscript drafting and revision, simulation implementation, data analysis, and figure preparation.
\FloatBarrier
\begingroup
\small
\setlength{\bibsep}{1pt}
\bibliographystyle{plainnat}
\bibliography{bibliography}
\endgroup
\clearpage
\appendix
\section{Instantiating the specification: a three-PR feasibility mapping}
\label{app:records}
An instance begins by declaring its population, criterion, horizon, and change boundary. It then assigns actors to roles, defines dependencies and visibility, and supplies the initial arrangement and improvement procedure. Evidence references identify versioned artifacts rather than undifferentiated transcript text. Each patch records the state it expects and the evidence its proposer could inspect. A compact contract can use the following interchange form; the referenced artifacts contain the population, protocol, scores, and retention details.
\begin{lstlisting}
{
  "id": "coverage-change-1",
  "target": "evaluation.coverage_program",
  "expected_version": 0,
  "actor": "review_board",
  "transformation": {"from": "Biased", "to": "Balanced"},
  "evidence": ["program-trials-1.v1"],
  "comparison": "population-and-net-value-criterion.v1",
  "cost_ledger": "labels-outputs-transition-1.v1",
  "horizon_rounds": 8,
  "retention": "next-review-rule.v1"
}
\end{lstlisting}
The example links a proposed change to the records needed to evaluate it. The reference checker implements target, version, actor, evidence, reversibility, and admission-cost fields; the experiment protocol defines the remaining fields. A diagnostic coverage trace records sample counts, estimates, decisions, costs, and authorized program changes. Outcome equations reconstruct net value from the selected workflow and ledger.

The public mapping uses a convenience sample from \texttt{pallets/flask}, frozen on September 28, 2026. From the 100 closed PRs returned in descending creation order at retrieval, it takes the first three already merged PRs. The archive retains the candidate list, URLs, timestamps, and response hashes. Message bodies and email addresses are excluded. Actor identifiers are replaced by stable pseudonyms; the public PR links permit re-identification.

For each PR the adapter retrieves commits, reviews, and checks at its recorded head SHA, checking pagination for the subordinate lists. \Cref{tab:public} summarizes the mapped records. Commits are artifact versions; opening, review, check, and merge records are events. Endpoint definitions support these mappings \citep{githubreviews,githubchecks}. Earlier heads, test-merge commits, off-platform activity, and historical branch protection require additional sources.
\begin{table}[!htbp]
\caption{Public-record snapshot. Counts summarize the records available for mapping.}
\label{tab:public}\centering\small
\begin{tabular}{@{}lrrrr@{}}\toprule
PR & Commit artifacts & Reviews & Head checks & Events \\\midrule
\href{https://github.com/pallets/flask/pull/6133}{\#6133} & 1 & 0 & 14 & 16 \\
\href{https://github.com/pallets/flask/pull/6096}{\#6096} & 2 & 5 & 14 & 21 \\
\href{https://github.com/pallets/flask/pull/6095}{\#6095} & 1 & 3 & 14 & 19 \\\bottomrule
\end{tabular}
\end{table}
The adapter records historical permission and actual evidence exposure as unknown. The mapping identifies the links that additional instrumentation would need to recover before a workflow-effect comparison.

\section{Analytical background for interaction and evaluation}
\label{app:analytical}
\subsection{Coverage and indistinguishability}
Suppose two environments induce identical distributions over every history accessible to a procedure, but a specified workflow change improves the population in one and harms it in the other. Any binary sign decision based on those histories has the same output distribution in both environments. If its probability of reporting improvement is $q$, its two error probabilities are $1-q$ and $q$. Their sum is one, so at least one is at least $1/2$. This is the standard observation-equivalence argument.

A construction uses equally frequent flagged and unflagged cases. A change lowers flagged error from $.20$ to $.10$ in both environments; unflagged error falls to $.10$ in one and rises to $.50$ in the other. If only flagged labels are observed and no downstream signal arrives, evidence is identical although population risk changes from $.20$ to $.10$ or $.30$. With known positive inclusion probabilities $q_i$ and accurate labels $E_i$, the Horvitz--Thompson estimator $N^{-1}\sum_i Z_i E_i/q_i$ is design-unbiased for finite-population risk \citep{ht}. Small inclusion probabilities can still produce high variance. Study 1 gives every cell positive coverage and examines the resulting finite-sample selection problem.

\subsection{Dependent evidence and interaction order}
For equal-variance errors with equicorrelation $\rho$, the variance of the mean is $\sigma^2[\rho+(1-\rho)/n]$, with $-1/(n-1)\leq\rho\leq1$. Common bias contributes an additional squared-bias term to mean squared error. Information dependence therefore limits the gain from adding actors \citep{clemen}; more actors alone need not produce proportionally more information.

Let independent initial human and agent estimation errors have variances $v_H$ and $v_A$. Suppose the human sees the agent answer and replaces the private judgment by $(1-\alpha)H+\alpha A$. Equal averaging of that judgment with $A$ has risk
\begin{equation}
R(\alpha)=\{(1-\alpha)^2v_H+(1+\alpha)^2v_A\}/4.
\end{equation}
For $v_A<v_H$ and $0\leq\alpha\leq1$, the optimum is $\alpha^*=(v_H-v_A)/(v_H+v_A)$, giving $v_Hv_A/(v_H+v_A)$. This is also the risk of optimal inverse-variance weighting of the original private judgments. Thus reweighting the private judgments can reproduce the gain over equal private averaging. With initial covariance $\sigma_{HA}$, risk gains the term $(1-\alpha^2)\sigma_{HA}/2$.

\section{Study 1: complete specification}
\label{app:protocol}
\subsection{Core protocol and exploratory controls}
The core design specifies 48 rounds, a 2736-label ceiling for each eight-round block, three catalog programs, six arms, three memory conditions, three environments, and the allocation grid. The trial-matched control holds trial acquisition and label reuse fixed while varying whether the retained evaluation program can change; the timing checks move reversal to rounds 21 and 29. Code and data availability identifies the archived protocols.

The core has $3\times3\times6\times128=6912$ trajectories. The allocation grid uses 96 replicates for each share/trial-count combination and each discovery arm under stationary cumulative memory and reversal with Window 8. Four fixed references and two memory ablations in each environment bring this analysis to 4608 trajectories. The exploratory controls cross four environments (stationary and reversals before rounds 21, 25, 29), three memory conditions, four arms, and 128 replicates: 6144 trajectories. The three independent keyed root seeds are 920000, 930000, and 940000. Each random stream is indexed by root seed, round, acquisition stage, program, trial, and replicate. Conditions share underlying potential-label streams within each root; no policy receives unqueried labels.

\subsection{Evidence update and program trials}
For a template--stratum cell, cumulative memory adds each actually queried label once. Window 8 uses observations at historical rounds $t-8$ through $t-1$, plus current observations. Reset retains no earlier observations. All program trials at a review receive the same pre-review memory snapshot. Trial outputs and fresh validation labels are pooled only after program selection, then made available to the current operational screen. Validation observes only the selected template, while screening observes all templates except for the second phase of the rejection comparator. A label from a rejected program remains usable if its source and population are appropriate. The pooling ablation withholds all program-trial labels from workflow memory while retaining program scores.

Each trial score is $1-3\widehat r^{\rm val}-.22N^{\rm screen}/4096$, where validation uses equal counts across the two strata. A program's score is the mean of its trial means retained under the memory rule. Each review has the same trial count within a configuration. Ties favor Biased, then Balanced, then Neyman. Program-score memory can be disabled independently of workflow memory. Retained scores average trials performed under different historical knowledge states.

For block budget $B=2736$, program share $f$, three programs, and $q$ trials per program, a trial receives $b=\lfloor\lfloor fB\rfloor/(3q)\rfloor$ label units. Validation obtains $v=\max\{1,\lfloor\lfloor .2b\rfloor/2\rfloor\}$ labels per stratum; screening receives $b-2v$ total units. The base $(f,q)=(.2,1)$ therefore allows 182 units per trial, with 18 validation labels per stratum and 146 units for screening. Actual counts can be smaller after integer allocation. Let $M_i$ be the actual number of trial labels for replicate $i$ at a block's start. Its subsequent per-round operational allowance is $\lfloor(B-M_i)/8\rfloor$. Discover once has $M_i=0$ after block 1. This rule keeps the budget fixed when trial count or program share changes.

\subsection{Within-template and between-template allocation}
For the catalog programs, let $u$ be one template's share of the operational or trial allowance. Balanced obtains $\lfloor u/2\rfloor$ labels in each stratum. Biased obtains $\max\{1,\lfloor .95u\rfloor\}$ and $\max\{1,\lfloor .05u\rfloor\}$. All tested allocations fit the allowance. Neyman first obtains $m=\min\{2,\max\{1,\lfloor u/4\rfloor\}\}$ labels per cell. Using retained and pilot data, it estimates $p_s$, sets $v_s=p_s(1-p_s)$, and allocates
\begin{equation}
n_s=m+\left\lfloor\frac{(u-2m)\sqrt{v_s}}{\sum_\ell\sqrt{v_\ell}}\right\rfloor.
\end{equation}
Pilot labels enter the posterior once. Equal label costs make this the equal-cost stratified allocation. Catalog selection breaks posterior ties in template order standard, specialized, broad.

The successive-rejection comparator treats one pair of stratum labels as a bounded loss with mean equal to population risk. With $n$ affordable pairs and $K=3$, it uses the schedule of \citet{bai}:
\begin{equation}
\overline{\log}K=\frac12+\sum_{j=2}^{K}\frac1j,\qquad
n_k=\left\lceil\frac{n-K}{(\overline{\log}K)(K+1-k)}\right\rceil,\quad k=1,2.
\end{equation}
All three templates receive $n_1$ pairs; the worst posterior-mean template is removed, and the remaining two reach $n_2$ pairs each. Elimination ties remove the lowest indexed template; final ties use the lowest surviving index. Historical evidence enters the posterior under the memory conditions. Thus the implementation uses the published rejection schedule with declared posterior ranking and deterministic ties. At the fixed allowance of 342 labels it uses 336 labels. SR is a fixed between-template comparator, separate from the three-program discovery catalog.

\subsection{Outcome accounting, checks, and uncertainty}
For deployed template $k_t$ with true population risk $r_{k_t,t}$, total queried labels $Q_t$, and program-change indicator $S_t$, the external score is
\begin{equation}
V_t=1-3r_{k_t,t}-.02-\{.22Q_t+.50S_t\}/4096.
\label{eq:net}
\end{equation}
Each queried label corresponds to a generated evaluation output, so $.22$ includes $.20$ labeling and $.02$ generation. The separate $.02$ term prices each production output. Potential-label arrays define the common random streams; actors observe queried entries. The ledger includes rejected-program trials.

The simulator checks every block ceiling and the identity
$V_t=1-3r_t^*-.02-\mathrm{regret}_t-\mathrm{expense}_t$,
where $r_t^*$ is the best available risk. An independent scalar reconstruction checks 36 fixed-Balanced trajectories across the three environments and memory conditions to absolute tolerance $10^{-12}$. Separate fixtures check the lookback boundary, rejection allocation, and selection of a known best template. The patch checker enforces target authority and state version for program changes.

Reported half-widths are $1.96s/\sqrt N$ across independent replicate trajectories. Paired intervals use within-replicate differences under the common streams. The uniform fixed benchmark averages the conditional mean values of Biased, Balanced, and Neyman under an initial uniform program prior; its paired contrast integrates over this discrete prior analytically. Intervals are unadjusted for multiple comparisons.

\Cref{tab:base} reports all core net-value means. Supplementary Table S1 reports their Monte Carlo intervals, costs, harmful adoption, selection regret, and late outcomes; replicate-level data accompany the reproducibility materials.
\begin{table}[!htbp]
\caption{Complete core net-value means. B: Biased; Bal: Balanced; N: Neyman; Mix: uniform fixed-program mixture; SR: successive rejection; O: discover once; R: repeated discovery. Each cell uses 128 trajectories. Monte Carlo intervals and additional outcomes are reported in Supplementary Table S1.}
\label{tab:base}\centering\small
\begin{tabular}{@{}llrrrrrrr@{}}\toprule
Environment & Memory & B & Bal & N & SR & O & R & Mix \\\midrule
Stationary harm & Reset & 0.3967 & 0.4522 & 0.4539 & 0.4540 & 0.4366 & 0.4352 & 0.4343 \\
Stationary harm & Cumulative & 0.4692 & 0.4801 & 0.4805 & 0.4806 & 0.4793 & 0.4800 & 0.4766 \\
Stationary harm & Window 8 & 0.4594 & 0.4792 & 0.4796 & 0.4800 & 0.4754 & 0.4779 & 0.4727 \\
Workflow reversal & Reset & 0.5260 & 0.5828 & 0.5830 & 0.5844 & 0.5644 & 0.5628 & 0.5639 \\
Workflow reversal & Cumulative & 0.4721 & 0.4803 & 0.4805 & 0.4829 & 0.4796 & 0.4835 & 0.4776 \\
Workflow reversal & Window 8 & 0.5609 & 0.5758 & 0.5760 & 0.5786 & 0.5723 & 0.5821 & 0.5709 \\
Uniform gain & Reset & 0.6617 & 0.7141 & 0.7132 & 0.7162 & 0.6962 & 0.6941 & 0.6963 \\
Uniform gain & Cumulative & 0.7184 & 0.7215 & 0.7216 & 0.7218 & 0.7217 & 0.7219 & 0.7205 \\
Uniform gain & Window 8 & 0.7166 & 0.7215 & 0.7216 & 0.7218 & 0.7210 & 0.7219 & 0.7199 \\
 
\bottomrule\end{tabular}\end{table}

\Cref{tab:grid} reports the allocation grid. Supplementary Table S2 summarizes the retained-program proportions at each rule's final scheduled assessment; the repository contains the conditional outcomes for every assessment.
\begin{table}[!htbp]
\caption{Repeated discovery minus discover-once at fixed total budget. All program shares and trial counts are shown. Entries are paired differences $\pm$ 95\% Monte Carlo half-width; 96 fresh replicates per condition.}
\label{tab:grid}\centering\small
\begin{tabular}{@{}rrrr@{}}\toprule
Program share & Trials/program & Stationary, cumulative & Reversal, Window 8 \\\midrule
0.2 & 1 & +0.0012 $\pm$ 0.0012 & +0.0071 $\pm$ 0.0024 \\
0.2 & 3 & +0.0019 $\pm$ 0.0020 & +0.0053 $\pm$ 0.0023 \\
0.2 & 6 & +0.0006 $\pm$ 0.0008 & +0.0062 $\pm$ 0.0020 \\
0.4 & 1 & +0.0002 $\pm$ 0.0004 & +0.0105 $\pm$ 0.0025 \\
0.4 & 3 & +0.0002 $\pm$ 0.0001 & +0.0098 $\pm$ 0.0026 \\
0.4 & 6 & +0.0002 $\pm$ 0.0001 & +0.0080 $\pm$ 0.0019 \\
0.6 & 1 & +0.0007 $\pm$ 0.0011 & +0.0175 $\pm$ 0.0021 \\
0.6 & 3 & +0.0001 $\pm$ 0.0001 & +0.0158 $\pm$ 0.0022 \\
0.6 & 6 & +0.0002 $\pm$ 0.0002 & +0.0134 $\pm$ 0.0015 \\
 
\bottomrule\end{tabular}\end{table}
\begin{table}[!htbp]
\caption{Repeated discovery: retained-program percentages and subsequent eight-round outcomes conditional on retaining Balanced. Stationary and uniform gain use cumulative memory; reversal uses Window 8. B/Bal/N are Biased/Balanced/Neyman; $n_{\rm Bal}$ is the number retaining Balanced among 128 replicates. Harm is the percentage of rounds deploying a template worse than standard. }
\label{tab:programs}\centering\small
\begin{tabular}{@{}lrrrrrrr@{}}\toprule
Environment & Review & B (\%) & Bal (\%) & N (\%) & $n_{\rm Bal}$ & Net given Bal & Harm (\%) \\\midrule
Stationary & 1 & 27.3 & 37.5 & 35.2 & 48 & 0.4747 & 0.00 \\
Stationary & 9 & 26.6 & 35.9 & 37.5 & 46 & 0.4819 & 0.00 \\
Stationary & 17 & 33.6 & 34.4 & 32.0 & 44 & 0.4819 & 0.00 \\
Stationary & 25 & 33.6 & 35.9 & 30.5 & 46 & 0.4819 & 0.00 \\
Stationary & 33 & 35.9 & 32.0 & 32.0 & 41 & 0.4819 & 0.00 \\
Stationary & 41 & 38.3 & 31.2 & 30.5 & 40 & 0.4819 & 0.00 \\
Reversal & 1 & 27.3 & 37.5 & 35.2 & 48 & 0.4747 & 0.00 \\
Reversal & 9 & 26.6 & 35.9 & 37.5 & 46 & 0.4783 & 0.00 \\
Reversal & 17 & 43.8 & 27.3 & 28.9 & 35 & 0.4819 & 0.00 \\
Reversal & 25 & 39.8 & 30.5 & 29.7 & 39 & 0.6188 & 0.00 \\
Reversal & 33 & 43.8 & 28.1 & 28.1 & 36 & 0.7219 & 0.00 \\
Reversal & 41 & 47.7 & 20.3 & 32.0 & 26 & 0.7219 & 0.00 \\
Uniform gain & 1 & 35.9 & 27.3 & 36.7 & 35 & 0.7219 & 0.00 \\
Uniform gain & 9 & 26.6 & 30.5 & 43.0 & 39 & 0.7219 & 0.00 \\
Uniform gain & 17 & 25.8 & 29.7 & 44.5 & 38 & 0.7219 & 0.00 \\
Uniform gain & 25 & 28.1 & 30.5 & 41.4 & 39 & 0.7219 & 0.00 \\
Uniform gain & 33 & 28.1 & 28.1 & 43.8 & 36 & 0.7219 & 0.00 \\
Uniform gain & 41 & 30.5 & 25.8 & 43.8 & 33 & 0.7219 & 0.00 \\
 
\bottomrule\end{tabular}\end{table}
\FloatBarrier

\section{Supplementary audit-control model}
\label{app:audit}
The supplementary audit model isolates longitudinal evidence acquisition under a fixed controller. It contains 160 periods of 32 tasks. Route A has base value 1 and error probability $.04$ (stationary) or alternating $.04/.25$ blocks of 40 periods; route B has base value $.82$ and error probability $.08$. An error costs 3, an audit costs $.20$, and an audit-rate change costs $.50$ per batch. Only audits reveal A's errors, available next period; executing B can still be paired with acquiring an A audit. Observations update a sliding-window Beta estimate. The base prior is $\operatorname{Beta}(1,9)$ and window 12.

The payoff-optimal routing threshold is $.14$, since $1-3p=.82-3(.08)=.58$. The heuristic adaptive controller audits at rate $.40$ when posterior $P(p>.14)$ lies between $.10$ and $.90$, and at $.025$ otherwise. Periodic dense auditing uses $.40$ for the first three periods of each 12-period cycle and $.025$ otherwise. The base audit comparison uses 400 seeds from 20260928, common across policies. Stationary net values are $.86089$ for fixed $.025$, $.86395$ for fixed $.05$, $.85841$ for fixed $.10$, and $.87976$ for no auditing. No auditing retains A under the base prior, the optimal route in that stationary case; audits cannot repair current outputs.

With slow shifts, adaptive auditing exceeds fixed $.10$ by $.00857$ per task (paired interval $.00676$--$.01038$). A separate 200-seed grid, beginning at 20360928, crosses windows 6, 12, 24 with stationary, 40-period, and 10-period regimes. It includes no audit, fixed $.05/.10/.40$, periodic dense, and adaptive policies. Supplementary Table S3 reports all 54 combinations of environment, memory window, and audit policy.

\subsection{Finite-horizon EVSI controller}
The finite-horizon expected-value-of-sample-information (EVSI) controller evaluates acquiring $m\in\{0,1,4,16\}$ labels. Given current Beta parameters $(a,b)$, a beta-binomial predictive distribution over $k$ observed errors gives posterior mean $(a+k)/(a+b+m)$. Let $L_\theta(p)$ be $1-3p$ if $p\leq\theta$ and $.58$ otherwise. The controller maximizes
\begin{equation}
h\{\E_k[L_\theta((a+k)/(a+b+m))]-L_\theta(a/(a+b))\}
-\frac{\bigl[(.20+g_c I_B)m+.50I_{\rm change}\bigr]}{32}.
\end{equation}
$I_{\rm change}$ equals 1 when the audit rate changes and 0 otherwise. Both costs in the bracket are subtracted. $I_B$ indicates executing B; $g_c$ is the cost of generating a counterfactual A output to audit. The predictive sum is exact under the current Beta model. The controller approximates future value with horizon $h$ and a sliding window, so it is a finite-horizon EVSI heuristic under drift.

Development fixes the prior at $\operatorname{Beta}(1,9)$, the threshold at $.14$, and $g_c=0$. Seeds 710000--710063 choose $h$ from $1,4,8,16$ using stationary and slow-shift mean value; $h=16$ is selected. Testing uses 96 independent seeds 810000--810095, three priors, thresholds $.14/.16$, generation costs $g_c=0/.20$, two environments, and three policies: 6912 trajectories. Threshold $.16$ is a misspecified routing-rule sensitivity under unchanged payoffs. Nonzero $g_c$ charges selectively generated A outputs while B is executed; it is distinct from a common cost charged for producing every candidate under every policy.

\Cref{tab:audit} reports all prior/threshold means at $g_c=0$. Supplementary Table S4 reports both candidate-generation cost conditions, marginal Monte Carlo intervals, and paired policy contrasts. EVSI's performance depends strongly on the prior: with no initial feedback, information acquisition itself can stall. Four regression cases match a reference scalar implementation to $10^{-12}$.
\begin{table}[!htbp]
\caption{Supplementary audit sensitivity at zero additional candidate-generation cost. Columns give net-value means over 96 trajectories; F: fixed $.10$, A: adaptive heuristic, E: EVSI. Uncertainty estimates and results at $g_c=.20$ are reported in Supplementary Table S4.}
\label{tab:audit}\centering\small
\begin{tabular}{@{}llrrrrrr@{}}\toprule
 & & \multicolumn{3}{c}{Stationary} & \multicolumn{3}{c}{Slow shifts} \\
Prior $(a:b)$ & Threshold & F & A & E & F & A & E \\\midrule
1:1 & 0.14 & 0.8479 & 0.8530 & 0.8338 & 0.6616 & 0.6753 & 0.6758 \\
1:1 & 0.16 & 0.8523 & 0.8571 & 0.8342 & 0.6597 & 0.6729 & 0.6757 \\
1:9 & 0.14 & 0.8581 & 0.8649 & 0.8656 & 0.6620 & 0.6706 & 0.6599 \\
1:9 & 0.16 & 0.8592 & 0.8686 & 0.8660 & 0.6515 & 0.6642 & 0.6526 \\
1:19 & 0.14 & 0.8595 & 0.8720 & 0.8795 & 0.6468 & 0.6571 & 0.5673 \\
1:19 & 0.16 & 0.8595 & 0.8735 & 0.8795 & 0.6263 & 0.6396 & 0.5673 \\
 
\bottomrule\end{tabular}
\end{table}

\FloatBarrier
\section{Study 2: complete specification}
\label{app:actor}
\subsection{Private judgments, exposure, and decision rights}
The review and learning protocols below define the actor error law, observation access, candidate arrangements, costs, and retention rules used in Study 2.

A task has balanced binary truth $y\in\{0,1\}$. A private judgment equals $y$ unless its actor makes an error. Human errors are independent Bernoulli draws with probability $q_H$, conditional on $y$, and independent of the agent group. For each task, draw a shared-mode indicator with probability $\rho_A$. In shared mode, all agent errors equal one Bernoulli draw with probability $q_A$; otherwise each agent draws an independent Bernoulli error with that probability. Tasks are independent. An agent proposer belongs to the same group as its reviewers. This mixture preserves marginal agent error $q_A$ and gives pairwise error correlation $\rho_A$; the full mixture law determines collective error.

Let $d_0$ be the displayed upstream judgment and $d_i^{\rm priv}$ reviewer $i$'s private judgment. An exposed reviewer reports
\begin{equation}
d_i^{\rm pub}=(1-C_i)d_i^{\rm priv}+C_i d_0,
\qquad C_i\sim\operatorname{Bernoulli}(\eta_s),\quad s\in\{H,A\}.
\label{eq:echo}
\end{equation}
Here $\eta_s$ is the probability that an agent reviewer copies an upstream decision from source $s$; $s$ identifies a human or agent upstream actor. Copy indicators are independent of private errors and of one another. The copying rule selects either the private judgment or the displayed upstream judgment. Blind protocols use private commitments for aggregation or escalation, so changing $\eta_s$ does not change their decisions. In serial review, the immediately preceding review replaces $d_0$ at each step, and all sources after the first step are agents.

\Cref{tab:actor-protocols} defines the six protocol families and two single-actor baselines. A gate retains the proposal only if all relevant reviews agree with it; otherwise a separate human makes the final decision. The human arbiter has error probability $q_H$ and is independent of the proposer and reviewers. A human proposer and a human arbiter therefore require two distinct humans. A production audit grants the arbiter override authority, whereas a gold label supplies correct measurement for learning.
\begin{table}[!htbp]
\caption{Review protocols and final decision rights. Voting includes the proposer and all reviewers; a tied vote uses a fair random choice. All reviewers are agents. Gates call an additional human only on the indicated cases.}
\label{tab:actor-protocols}\small
\begin{tabularx}{\linewidth}{@{}p{.23\linewidth}YY@{}}\toprule
Protocol & Evidence used & Final decision \\\midrule
Exposed vote & Reviews after seeing the proposal & Majority vote. \\
Blind vote & Private commitments & Majority vote. \\
Exposed disagreement gate & Public reviews versus proposal & Human arbitration on any disagreement. \\
Blind disagreement gate & Private reviews versus proposal & Human arbitration on any disagreement. \\
Gate with random arbitration & Public reviews versus proposal & Arbitrate all disagreements and a random 25\% of unanimous cases. \\
Serial last-reviewer decision & Each reviewer sees only its immediate predecessor & Last review; no collective vote. \\
Single actor & One human or agent private judgment & That actor's judgment. \\\bottomrule
\end{tabularx}
\end{table}
These rules separate private information, public disagreement, and final authority. For example, the change in error from randomly overriding a unanimous case is $q_H-P(d_0\ne y\mid\text{unanimous})$. An independent human override helps only when this difference is negative.

The exact search crosses $q_H\in\{.1,.2,.3\}$, $q_A\in\{.05,.1,.2,.3\}$, $\rho_A\in\{0,.25,.5,.75,1\}$ and $\eta\in\{0,.25,.5,.75,1\}$ with $\eta_H=\eta_A=\eta$, human-to-agent call-cost ratios $2,4,8$, blind overheads $0,.05,.2$ per reviewer, task budgets $1,2,3$, and hard versus expected budget constraints. Human calls define one cost unit. A hard budget reserves every potentially called actor; an expected budget constrains the mean cost including escalation probabilities. The catalog has $2\times7\times6+2=86$ arrangements: two proposer types, one to seven reviewers, six protocols, and two single-actor baselines. Exact finite-state probabilities identify the minimum-error feasible arrangement in each of 16,200 cells. Errors within $10^{-12}$ count as ties; lower expected cost breaks error ties. Remaining ties use catalog order: human then agent baseline, then human/agent proposer, increasing reviewer count, and the six protocols in table order. The deployment-loss optimum is recorded separately.

The fixed-roster comparison retains all 25,200 combinations of the accuracy, dependence and exposure grid, two proposer types, one to seven reviewers, and six protocols. A separate source-sensitivity grid uses $\eta_H,\eta_A\in\{0,.2,.5,.8,1\}$, $\rho_A\in\{0,.5,.9\}$, $q_H=.2$, and $q_A=.1$, giving 6,450 arrangement evaluations. The curves in \cref{fig:actor-review}a--b evaluate 21 equally spaced exposure values.

\subsection{Paid observations and retained evidence}
An observer can learn individual ability and copying only from the information it buys. For an agent proposer and one equally capable agent reviewer,
\[
P(d_1^{\rm pub}\ne d_0)=2q_A(1-q_A)(1-\rho_A)(1-\eta_A).
\]
Consequently, at the same $q_A$, $(\rho_A,\eta_A)=(.8,0)$ and $(0,.8)$ give the same joint distribution of public votes and truth in this two-actor case, but different responses to blinding. Private observations or randomized visibility can distinguish the mechanisms.

Four fixed observation policies select among 58 arrangements with at most five reviewers and hard task cost at most three. Human, agent, gold-label, and per-reviewer blind-commitment costs are $1,.25,1,.05$, respectively. Each round permits at most 240 acquisition-cost units, hence 1920 per eight-round block, with no borrowing. These cost units differ from the label counts used in Appendix~\ref{app:protocol}. All current observations are acquired and charged before the selected arrangement is deployed; production outcomes do not provide free learning feedback.

\emph{Paired observation} records two human and six agent private judgments, ten additional exposed agent responses (five for each proposer type), a gold label, and six private-commitment overheads on each task. Its cost is $2+6(.25)+10(.25)+1+6(.05)=7.3$, allowing 32 tasks and costing 233.6 per round. The same task supplies pre- and post-exposure judgments. Among reviews that initially disagree with a source, the fraction switching to that source estimates $\eta_s$. Recording a private commitment leaves the later copying probability unchanged in this model.

\emph{Randomized exposure} acquires 50 tasks, randomized in balanced cells crossing human/agent source and blind/exposed review. Each task purchases two human responses, six agent responses, and a gold label. Blind tasks additionally pay six commitment overheads. The four cells have 12 or 13 tasks, with 25 blind tasks in total, costing $25(4.8)+25(4.5)=232.5$ per round. Private ability and dependence use only blind tasks. For each source, the estimate of copying is one minus exposed disagreement divided by blind disagreement, clipped to $[0,1]$. This design uses disjoint blind and exposed tasks and assumes exchangeability across their randomized assignments.

Both model-based policies estimate $q_H$ and $q_A$ from gold-labeled private errors with one error and one success pseudocount, then clip estimates to $[.001,.499]$. Agent dependence is the private pairwise error moment minus squared raw error frequency, divided by the raw Bernoulli variance and clipped to $[0,1]$; zero estimated variance gives zero dependence. Paired copying estimates use half an event in each outcome as pseudocounts. Randomized disagreement rates likewise use half-event pseudocounts before taking their ratio. Each policy pools the requisite counts over its assigned memory interval and predicts every feasible arrangement's error and call cost under the known model family. The true parameters are used only by the external scorer and catalog benchmark.

\emph{Omit echo ablation} purchases exactly the paired observations but fixes $\eta_H=\eta_A=0$ when predicting arrangements. \emph{Final-outcome sampling} instead buys an arrangement's executed final outcome and a gold label, cycling through candidates. At zero-based round $r$, trial $j$ uses catalog index $(83r+j)\bmod58$. A trial reserves its worst possible call cost plus one gold label; actual calls and that label are charged. It stops before exceeding the ceiling. For a candidate with $n$ retained observations, $e$ errors, total call cost $c$, and worst call cost $c_{\max}$, its score is $3(1+e)/(2+n)+.02(c+c_{\max})/(1+n)$. Thus unobserved candidates also have defined scores.

Reset retains current-round counts; cumulative memory retains all acquired counts; Window 8 retains the current round and eight preceding rounds. Observation policies and memory rules are assigned research treatments and remain fixed. Each selected arrangement changes operational visibility, aggregation, or authority under the assigned procedure.

\subsection{Evaluation, reporting, and validation}
The six learning environments in \cref{tab:actor-worlds} isolate copying, private dependence, source asymmetry, and relative actor ability. The copying shift occurs before acquisition in round 25. Each environment crosses four observation policies and three memories, plus two fixed baselines, with 128 replicate trajectories of 48 rounds: $6(4\times3+2)128=10{,}752$ trajectories. Fixed baselines use an agent proposer and three agent reviewers, either exposed or blind, and buy no learning evidence.
\begin{table}[!htbp]
\caption{Study 2 environments. In the environment and policy names, echo means copying. All other costs and resource ceilings are held fixed as specified in Appendix~\ref{app:actor}. }
\label{tab:actor-worlds}\centering\small
\begin{tabular}{@{}lrrrrrl@{}}\toprule
Environment & $q_H$ & $q_A$ & $\rho_A$ & $\eta_H$ & $\eta_A$ & Change \\\midrule
No echo & .2 & .1 & .1 & 0 & 0 & None \\
High echo & .2 & .1 & .1 & .8 & .8 & None \\
Both dependences & .2 & .1 & .8 & .8 & .8 & None \\
Echo shift & .2 & .1 & .1 & 0 & 0 & Both $\eta_s$ become .8 at round 25 \\
Source asymmetry & .2 & .1 & .2 & .9 & .1 & None \\
Stronger humans & .1 & .2 & .2 & .8 & .8 & None \\\bottomrule
\end{tabular}
\end{table}
The model-based learner selects the arrangement minimizing estimated deployment loss $3e+.02c$, where $e$ is final error probability and $c$ is expected production call cost. Ties use the filtered catalog order. Selection regret compares this loss with the feasible catalog minimum under the true environment parameters; an optimal-choice hit accepts every loss within $10^{-10}$ of that minimum. This objective differs from the error-first grid search and excludes acquisition and switching costs. Net value additionally charges those costs:
\begin{equation}
V_t^{\rm actor}=1-3e_t-.02c_t-\frac{A_t+.5S_t}{4096},
\label{eq:actor-value}
\end{equation}
where $A_t$ is purchased-observation cost and $S_t$ indicates a change from the previous arrangement. All conditions start from the exposed agent-proposer/three-reviewer arrangement; the fixed blind baseline therefore pays one initial switch. The denominator 4096 is the production-volume scale over which observation and switching costs are amortized. The scorer uses expected production outcomes.

Learning uses root seed 2026100204, keyed by environment and round, with independent replicate rows and common potential-task panels across policies. Only purchased entries are revealed. Full-horizon and final-16-round summaries average within a trajectory before computing means and $1.96s/\sqrt{128}$ Monte Carlo half-widths across trajectories. Paired comparisons use within-replicate differences. The fixed baselines have deterministic expected-outcome scores.

\Cref{tab:actor-results} reports all six environments under cumulative memory; Supplementary Table S5 gives all 84 policy--memory--environment conditions, late outcomes, costs, regret, and paired comparisons. Error and net value are reported together because better selection need not repay continuing acquisition costs.
\begingroup
\small
\setlength{\tabcolsep}{4pt}
\setlength{\LTcapwidth}{\linewidth}
\begin{longtable}{@{}llrrr@{}}
\caption{Study 2: selection error and net value under paid observation. All six environments are shown; learning policies use cumulative memory. Entries are means $\pm$ 95\% Monte Carlo half-widths over 128 trajectories of 48 rounds. Fixed baselines have exact expected-outcome scores under the model. Error is a percentage; hit is the percentage of rounds attaining a deployment-loss minimum within $10^{-10}$.}\label{tab:actor-results}\\
\toprule Environment & Observation / fixed rule & Error (\%) & Net value & Hit (\%) \\\midrule
\endfirsthead
\multicolumn{5}{l}{\tablename\ \thetable\ (continued)}\\
\toprule Environment & Observation / fixed rule & Error (\%) & Net value & Hit (\%) \\\midrule
\endhead
\midrule\multicolumn{5}{r}{Continued on next page}\\\endfoot
\bottomrule\endlastfoot
No echo & Paired & $1.772 \pm 0.002$ & $0.86475 \pm 0.00008$ & 98.70 \\*
 & Randomized & $1.772 \pm 0.003$ & $0.86333 \pm 0.00026$ & 56.67 \\*
 & Omit echo & $1.772 \pm 0.002$ & $0.86480 \pm 0.00008$ & 99.97 \\*
 & Final only & $4.919 \pm 0.197$ & $0.78045 \pm 0.00563$ & 14.47 \\*
 & Fixed exposed & $3.520$ & $0.87440$ & 0.00 \\*
 & Fixed blind & $3.520$ & $0.87140$ & 0.00 \\
\addlinespace[3pt]
High echo & Paired & $1.773 \pm 0.003$ & $0.86077 \pm 0.00009$ & 99.95 \\*
 & Randomized & $1.782 \pm 0.008$ & $0.86078 \pm 0.00024$ & 99.82 \\*
 & Omit echo & $9.825 \pm 0.004$ & $0.62323 \pm 0.00013$ & 0.00 \\*
 & Final only & $5.983 \pm 0.217$ & $0.74760 \pm 0.00640$ & 10.89 \\*
 & Fixed exposed & $9.637$ & $0.69089$ & 0.00 \\*
 & Fixed blind & $3.520$ & $0.87140$ & 0.00 \\
\addlinespace[3pt]
Both dependences & Paired & $7.423 \pm 0.049$ & $0.69099 \pm 0.00105$ & 82.29 \\*
 & Randomized & $7.498 \pm 0.063$ & $0.68945 \pm 0.00135$ & 77.10 \\*
 & Omit echo & $15.939 \pm 0.317$ & $0.43965 \pm 0.00969$ & 0.00 \\*
 & Final only & $9.728 \pm 0.136$ & $0.63630 \pm 0.00404$ & 5.49 \\*
 & Fixed exposed & $9.919$ & $0.68242$ & 0.00 \\*
 & Fixed blind & $8.560$ & $0.72020$ & 0.00 \\
\addlinespace[3pt]
Echo shift & Paired & $1.779 \pm 0.006$ & $0.86255 \pm 0.00018$ & 98.54 \\*
 & Randomized & $2.113 \pm 0.083$ & $0.85206 \pm 0.00238$ & 72.40 \\*
 & Omit echo & $5.797 \pm 0.002$ & $0.74404 \pm 0.00008$ & 49.97 \\*
 & Final only & $5.648 \pm 0.217$ & $0.75770 \pm 0.00621$ & 11.20 \\*
 & Fixed exposed & $6.579$ & $0.78264$ & 0.00 \\*
 & Fixed blind & $3.520$ & $0.87140$ & 0.00 \\
\addlinespace[3pt]
Source asymmetry & Paired & $2.693 \pm 0.004$ & $0.83318 \pm 0.00013$ & 99.76 \\*
 & Randomized & $2.936 \pm 0.046$ & $0.82728 \pm 0.00116$ & 70.91 \\*
 & Omit echo & $3.546 \pm 0.015$ & $0.81160 \pm 0.00044$ & 0.00 \\*
 & Final only & $6.352 \pm 0.191$ & $0.73785 \pm 0.00535$ & 5.34 \\*
 & Fixed exposed & $5.101$ & $0.82697$ & 0.00 \\*
 & Fixed blind & $4.240$ & $0.84980$ & 0.00 \\
\addlinespace[3pt]
Stronger humans & Paired & $4.623 \pm 0.009$ & $0.77306 \pm 0.00028$ & 99.37 \\*
 & Randomized & $4.639 \pm 0.014$ & $0.77286 \pm 0.00041$ & 98.96 \\*
 & Omit echo & $8.969 \pm 0.024$ & $0.64781 \pm 0.00072$ & 0.00 \\*
 & Final only & $9.522 \pm 0.263$ & $0.63271 \pm 0.00748$ & 10.04 \\*
 & Fixed exposed & $19.570$ & $0.39290$ & 0.00 \\*
 & Fixed blind & $12.320$ & $0.60740$ & 0.00 \\
\end{longtable}
\endgroup

The error and cost comparisons are discussed in Section~\ref{sec:actor-learning}.

An independent bit-level enumeration checks 144 arrangement--environment cases to maximum absolute difference $1.11\times10^{-15}$. A second implementation simulates 150,000 tasks in each of four independent environments and checks 344 arrangements on the shared panels; each metric must agree within seven estimated standard errors plus $2\times10^{-5}$. The separate validation seed is 2026100205. Further checks cover the memory lookback boundary, copying endpoints, both resource ceilings, and the net-value identity. Exact enumeration results have no sampling interval; the Monte Carlo comparison checks the enumeration against a separately implemented sampler.

\FloatBarrier
\clearpage
\begingroup
\renewcommand{\arraystretch}{1.08}
\renewcommand{\footnotesize}{\fontsize{8}{10}\selectfont}
\section*{Supplementary Tables}
\label{sec:supplementary-tables}
\addcontentsline{toc}{section}{Supplementary Tables}
Tables S1--S5 report the simulation results and supplementary audit controls. Uncertainty entries are 95\% Monte Carlo half-widths across independent trajectories. Cost units and reporting horizons are defined separately for the two studies.

\subsection*{Table S1: Study 1: evidence acquisition and retention}
\label{supp:S1}
\addcontentsline{toc}{subsection}{Table S1: Study 1: evidence acquisition and retention}
{\footnotesize Source: \texttt{results/learning\_summary.csv}; \texttt{results/learning\_contrasts.csv}.}\par Core design: 54 conditions, each with 128 independent replicate trajectories of 48 rounds. Window 8 includes eight previous rounds plus current observations. Net value, regret and expenses are per production task. Harm is the percentage of rounds selecting a template worse than standard. Late net value covers the final eight rounds. Panel C separates core and exploratory control contrasts; intervals use paired replicate differences.

\subsubsection*{A. Net value, harmful adoption, and selection regret}
\begingroup\footnotesize\setlength{\tabcolsep}{4pt}

\begin{longtable}{@{}llllll@{}}
\toprule Environment & Memory & Arm (SR = Successive rejection) & Net value & Harm (\%) & Regret \\\midrule\endfirsthead
\toprule Environment & Memory & Arm (SR = Successive rejection) & Net value & Harm (\%) & Regret \\\midrule\endhead
\midrule\multicolumn{6}{r}{Continued on next page}\\\endfoot
\bottomrule\endlastfoot
Stationary & Reset & Biased & $0.39666 \pm 0.00278$ & $15.755 \pm 0.800$ & $0.08514 \pm 0.00278$ \\
Stationary & Reset & Balanced & $0.45224 \pm 0.00123$ & $0.537 \pm 0.177$ & $0.02939 \pm 0.00123$ \\
Stationary & Reset & Neyman & $0.45389 \pm 0.00118$ & $0.521 \pm 0.175$ & $0.02781 \pm 0.00118$ \\
Stationary & Reset & SR & $0.45400 \pm 0.00128$ & $0.505 \pm 0.174$ & $0.02795 \pm 0.00128$ \\
Stationary & Reset & Once & $0.43664 \pm 0.00476$ & $4.704 \pm 1.256$ & $0.04509 \pm 0.00476$ \\
Stationary & Reset & Repeated & $0.43522 \pm 0.00261$ & $4.980 \pm 0.718$ & $0.04667 \pm 0.00261$ \\
Stationary & Cumulative & Biased & $0.46917 \pm 0.00267$ & $1.009 \pm 0.376$ & $0.01263 \pm 0.00267$ \\
Stationary & Cumulative & Balanced & $0.48007 \pm 0.00050$ & $0.000 \pm 0.000$ & $0.00156 \pm 0.00050$ \\
Stationary & Cumulative & Neyman & $0.48046 \pm 0.00048$ & $0.000 \pm 0.000$ & $0.00133 \pm 0.00048$ \\
Stationary & Cumulative & SR & $0.48059 \pm 0.00054$ & $0.000 \pm 0.000$ & $0.00137 \pm 0.00054$ \\
Stationary & Cumulative & Once & $0.47928 \pm 0.00124$ & $0.016 \pm 0.032$ & $0.00248 \pm 0.00124$ \\
Stationary & Cumulative & Repeated & $0.48000 \pm 0.00080$ & $0.016 \pm 0.032$ & $0.00191 \pm 0.00080$ \\
Stationary & Window 8 & Biased & $0.45936 \pm 0.00284$ & $1.514 \pm 0.496$ & $0.02244 \pm 0.00284$ \\
Stationary & Window 8 & Balanced & $0.47921 \pm 0.00073$ & $0.000 \pm 0.000$ & $0.00242 \pm 0.00073$ \\
Stationary & Window 8 & Neyman & $0.47962 \pm 0.00071$ & $0.000 \pm 0.000$ & $0.00217 \pm 0.00071$ \\
Stationary & Window 8 & SR & $0.48002 \pm 0.00069$ & $0.000 \pm 0.000$ & $0.00193 \pm 0.00069$ \\
Stationary & Window 8 & Once & $0.47537 \pm 0.00194$ & $0.309 \pm 0.280$ & $0.00639 \pm 0.00194$ \\
Stationary & Window 8 & Repeated & $0.47790 \pm 0.00118$ & $0.016 \pm 0.032$ & $0.00400 \pm 0.00118$ \\
Reversal & Reset & Biased & $0.52595 \pm 0.00282$ & $8.203 \pm 0.594$ & $0.07584 \pm 0.00282$ \\
Reversal & Reset & Balanced & $0.58280 \pm 0.00128$ & $0.244 \pm 0.117$ & $0.01883 \pm 0.00128$ \\
Reversal & Reset & Neyman & $0.58300 \pm 0.00121$ & $0.244 \pm 0.117$ & $0.01870 \pm 0.00121$ \\
Reversal & Reset & SR & $0.58435 \pm 0.00115$ & $0.244 \pm 0.117$ & $0.01760 \pm 0.00115$ \\
Reversal & Reset & Once & $0.56442 \pm 0.00519$ & $2.702 \pm 0.754$ & $0.03731 \pm 0.00520$ \\
Reversal & Reset & Repeated & $0.56282 \pm 0.00253$ & $2.523 \pm 0.522$ & $0.03907 \pm 0.00253$ \\
Reversal & Cumulative & Biased & $0.47211 \pm 0.00288$ & $1.009 \pm 0.376$ & $0.12968 \pm 0.00288$ \\
Reversal & Cumulative & Balanced & $0.48026 \pm 0.00060$ & $0.000 \pm 0.000$ & $0.12137 \pm 0.00060$ \\
Reversal & Cumulative & Neyman & $0.48054 \pm 0.00051$ & $0.000 \pm 0.000$ & $0.12125 \pm 0.00051$ \\
Reversal & Cumulative & SR & $0.48285 \pm 0.00131$ & $0.000 \pm 0.000$ & $0.11910 \pm 0.00131$ \\
Reversal & Cumulative & Once & $0.47963 \pm 0.00134$ & $0.016 \pm 0.032$ & $0.12212 \pm 0.00134$ \\
Reversal & Cumulative & Repeated & $0.48347 \pm 0.00235$ & $0.016 \pm 0.032$ & $0.11843 \pm 0.00235$ \\
Reversal & Window 8 & Biased & $0.56094 \pm 0.00291$ & $1.270 \pm 0.460$ & $0.04085 \pm 0.00291$ \\
Reversal & Window 8 & Balanced & $0.57583 \pm 0.00089$ & $0.000 \pm 0.000$ & $0.02580 \pm 0.00089$ \\
Reversal & Window 8 & Neyman & $0.57601 \pm 0.00081$ & $0.000 \pm 0.000$ & $0.02578 \pm 0.00081$ \\
Reversal & Window 8 & SR & $0.57859 \pm 0.00096$ & $0.000 \pm 0.000$ & $0.02336 \pm 0.00096$ \\
Reversal & Window 8 & Once & $0.57231 \pm 0.00206$ & $0.212 \pm 0.247$ & $0.02945 \pm 0.00207$ \\
Reversal & Window 8 & Repeated & $0.58206 \pm 0.00231$ & $0.016 \pm 0.032$ & $0.01984 \pm 0.00231$ \\
Uniform gain & Reset & Biased & $0.66166 \pm 0.00295$ & $0.000 \pm 0.000$ & $0.06014 \pm 0.00295$ \\
Uniform gain & Reset & Balanced & $0.71411 \pm 0.00111$ & $0.000 \pm 0.000$ & $0.00752 \pm 0.00111$ \\
Uniform gain & Reset & Neyman & $0.71317 \pm 0.00114$ & $0.000 \pm 0.000$ & $0.00852 \pm 0.00114$ \\
Uniform gain & Reset & SR & $0.71619 \pm 0.00097$ & $0.000 \pm 0.000$ & $0.00576 \pm 0.00097$ \\
Uniform gain & Reset & Once & $0.69615 \pm 0.00458$ & $0.000 \pm 0.000$ & $0.02559 \pm 0.00459$ \\
Uniform gain & Reset & Repeated & $0.69412 \pm 0.00234$ & $0.000 \pm 0.000$ & $0.02777 \pm 0.00234$ \\
Uniform gain & Cumulative & Biased & $0.71837 \pm 0.00126$ & $0.000 \pm 0.000$ & $0.00342 \pm 0.00126$ \\
Uniform gain & Cumulative & Balanced & $0.72147 \pm 0.00019$ & $0.000 \pm 0.000$ & $0.00016 \pm 0.00019$ \\
Uniform gain & Cumulative & Neyman & $0.72159 \pm 0.00025$ & $0.000 \pm 0.000$ & $0.00020 \pm 0.00025$ \\
Uniform gain & Cumulative & SR & $0.72184 \pm 0.00017$ & $0.000 \pm 0.000$ & $0.00012 \pm 0.00017$ \\
Uniform gain & Cumulative & Once & $0.72173 \pm 0.00008$ & $0.000 \pm 0.000$ & $0.00004 \pm 0.00008$ \\
Uniform gain & Cumulative & Repeated & $0.72187 \pm 0.00008$ & $0.000 \pm 0.000$ & $0.00004 \pm 0.00008$ \\
Uniform gain & Window 8 & Biased & $0.71662 \pm 0.00165$ & $0.000 \pm 0.000$ & $0.00518 \pm 0.00165$ \\
Uniform gain & Window 8 & Balanced & $0.72147 \pm 0.00019$ & $0.000 \pm 0.000$ & $0.00016 \pm 0.00019$ \\
Uniform gain & Window 8 & Neyman & $0.72159 \pm 0.00025$ & $0.000 \pm 0.000$ & $0.00020 \pm 0.00025$ \\
Uniform gain & Window 8 & SR & $0.72184 \pm 0.00017$ & $0.000 \pm 0.000$ & $0.00012 \pm 0.00017$ \\
Uniform gain & Window 8 & Once & $0.72101 \pm 0.00080$ & $0.000 \pm 0.000$ & $0.00076 \pm 0.00080$ \\
Uniform gain & Window 8 & Repeated & $0.72186 \pm 0.00008$ & $0.000 \pm 0.000$ & $0.00004 \pm 0.00008$ \\
\end{longtable}\endgroup
\subsubsection*{B. Evaluation expense, program-trial expense, and late outcome}
\begingroup\footnotesize\setlength{\tabcolsep}{4pt}

\begin{longtable}{@{}llllll@{}}
\toprule Environment & Memory & Arm (SR = Successive rejection) & Expense & Trial expense & Late net \\\midrule\endfirsthead
\toprule Environment & Memory & Arm (SR = Successive rejection) & Expense & Trial expense & Late net \\\midrule\endhead
\midrule\multicolumn{6}{r}{Continued on next page}\\\endfoot
\bottomrule\endlastfoot
Stationary & Reset & Biased & $0.01821 \pm 0.00000$ & $0.00000 \pm 0.00000$ & $0.39935 \pm 0.00766$ \\
Stationary & Reset & Balanced & $0.01837 \pm 0.00000$ & $0.00000 \pm 0.00000$ & $0.45327 \pm 0.00317$ \\
Stationary & Reset & Neyman & $0.01830 \pm 0.00000$ & $0.00000 \pm 0.00000$ & $0.45577 \pm 0.00305$ \\
Stationary & Reset & SR & $0.01805 \pm 0.00000$ & $0.00000 \pm 0.00000$ & $0.45670 \pm 0.00320$ \\
Stationary & Reset & Once & $0.01827 \pm 0.00001$ & $0.00060 \pm 0.00000$ & $0.44309 \pm 0.00568$ \\
Stationary & Reset & Repeated & $0.01811 \pm 0.00000$ & $0.00361 \pm 0.00000$ & $0.43967 \pm 0.00566$ \\
Stationary & Cumulative & Biased & $0.01821 \pm 0.00000$ & $0.00000 \pm 0.00000$ & $0.48062 \pm 0.00186$ \\
Stationary & Cumulative & Balanced & $0.01837 \pm 0.00000$ & $0.00000 \pm 0.00000$ & $0.48163 \pm 0.00000$ \\
Stationary & Cumulative & Neyman & $0.01821 \pm 0.00000$ & $0.00000 \pm 0.00000$ & $0.48179 \pm 0.00000$ \\
Stationary & Cumulative & SR & $0.01805 \pm 0.00000$ & $0.00000 \pm 0.00000$ & $0.48195 \pm 0.00000$ \\
Stationary & Cumulative & Once & $0.01824 \pm 0.00001$ & $0.00060 \pm 0.00000$ & $0.48173 \pm 0.00001$ \\
Stationary & Cumulative & Repeated & $0.01809 \pm 0.00000$ & $0.00359 \pm 0.00000$ & $0.48191 \pm 0.00000$ \\
Stationary & Window 8 & Biased & $0.01821 \pm 0.00000$ & $0.00000 \pm 0.00000$ & $0.46571 \pm 0.00562$ \\
Stationary & Window 8 & Balanced & $0.01837 \pm 0.00000$ & $0.00000 \pm 0.00000$ & $0.47976 \pm 0.00216$ \\
Stationary & Window 8 & Neyman & $0.01821 \pm 0.00000$ & $0.00000 \pm 0.00000$ & $0.47992 \pm 0.00211$ \\
Stationary & Window 8 & SR & $0.01805 \pm 0.00000$ & $0.00000 \pm 0.00000$ & $0.48066 \pm 0.00185$ \\
Stationary & Window 8 & Once & $0.01824 \pm 0.00001$ & $0.00060 \pm 0.00000$ & $0.47719 \pm 0.00265$ \\
Stationary & Window 8 & Repeated & $0.01810 \pm 0.00000$ & $0.00359 \pm 0.00000$ & $0.47968 \pm 0.00185$ \\
Reversal & Reset & Biased & $0.01821 \pm 0.00000$ & $0.00000 \pm 0.00000$ & $0.65722 \pm 0.00820$ \\
Reversal & Reset & Balanced & $0.01837 \pm 0.00000$ & $0.00000 \pm 0.00000$ & $0.71237 \pm 0.00297$ \\
Reversal & Reset & Neyman & $0.01830 \pm 0.00000$ & $0.00000 \pm 0.00000$ & $0.71102 \pm 0.00318$ \\
Reversal & Reset & SR & $0.01805 \pm 0.00000$ & $0.00000 \pm 0.00000$ & $0.71445 \pm 0.00250$ \\
Reversal & Reset & Once & $0.01827 \pm 0.00001$ & $0.00060 \pm 0.00000$ & $0.69498 \pm 0.00676$ \\
Reversal & Reset & Repeated & $0.01811 \pm 0.00000$ & $0.00361 \pm 0.00000$ & $0.69142 \pm 0.00633$ \\
Reversal & Cumulative & Biased & $0.01821 \pm 0.00000$ & $0.00000 \pm 0.00000$ & $0.49808 \pm 0.00972$ \\
Reversal & Cumulative & Balanced & $0.01837 \pm 0.00000$ & $0.00000 \pm 0.00000$ & $0.48280 \pm 0.00189$ \\
Reversal & Cumulative & Neyman & $0.01821 \pm 0.00000$ & $0.00000 \pm 0.00000$ & $0.48226 \pm 0.00092$ \\
Reversal & Cumulative & SR & $0.01805 \pm 0.00000$ & $0.00000 \pm 0.00000$ & $0.49484 \pm 0.00631$ \\
Reversal & Cumulative & Once & $0.01824 \pm 0.00001$ & $0.00060 \pm 0.00000$ & $0.48384 \pm 0.00307$ \\
Reversal & Cumulative & Repeated & $0.01809 \pm 0.00000$ & $0.00359 \pm 0.00000$ & $0.49785 \pm 0.00945$ \\
Reversal & Window 8 & Biased & $0.01821 \pm 0.00000$ & $0.00000 \pm 0.00000$ & $0.72062 \pm 0.00152$ \\
Reversal & Window 8 & Balanced & $0.01837 \pm 0.00000$ & $0.00000 \pm 0.00000$ & $0.72163 \pm 0.00000$ \\
Reversal & Window 8 & Neyman & $0.01821 \pm 0.00000$ & $0.00000 \pm 0.00000$ & $0.72179 \pm 0.00000$ \\
Reversal & Window 8 & SR & $0.01805 \pm 0.00000$ & $0.00000 \pm 0.00000$ & $0.72195 \pm 0.00000$ \\
Reversal & Window 8 & Once & $0.01824 \pm 0.00001$ & $0.00060 \pm 0.00000$ & $0.72056 \pm 0.00151$ \\
Reversal & Window 8 & Repeated & $0.01810 \pm 0.00000$ & $0.00359 \pm 0.00000$ & $0.72190 \pm 0.00000$ \\
Uniform gain & Reset & Biased & $0.01821 \pm 0.00000$ & $0.00000 \pm 0.00000$ & $0.65722 \pm 0.00820$ \\
Uniform gain & Reset & Balanced & $0.01837 \pm 0.00000$ & $0.00000 \pm 0.00000$ & $0.71237 \pm 0.00297$ \\
Uniform gain & Reset & Neyman & $0.01831 \pm 0.00000$ & $0.00000 \pm 0.00000$ & $0.71102 \pm 0.00318$ \\
Uniform gain & Reset & SR & $0.01805 \pm 0.00000$ & $0.00000 \pm 0.00000$ & $0.71445 \pm 0.00250$ \\
Uniform gain & Reset & Once & $0.01826 \pm 0.00001$ & $0.00060 \pm 0.00000$ & $0.69077 \pm 0.00763$ \\
Uniform gain & Reset & Repeated & $0.01811 \pm 0.00000$ & $0.00361 \pm 0.00000$ & $0.69142 \pm 0.00633$ \\
Uniform gain & Cumulative & Biased & $0.01821 \pm 0.00000$ & $0.00000 \pm 0.00000$ & $0.72179 \pm 0.00000$ \\
Uniform gain & Cumulative & Balanced & $0.01837 \pm 0.00000$ & $0.00000 \pm 0.00000$ & $0.72163 \pm 0.00000$ \\
Uniform gain & Cumulative & Neyman & $0.01821 \pm 0.00000$ & $0.00000 \pm 0.00000$ & $0.72179 \pm 0.00000$ \\
Uniform gain & Cumulative & SR & $0.01805 \pm 0.00000$ & $0.00000 \pm 0.00000$ & $0.72195 \pm 0.00000$ \\
Uniform gain & Cumulative & Once & $0.01823 \pm 0.00001$ & $0.00060 \pm 0.00000$ & $0.72175 \pm 0.00001$ \\
Uniform gain & Cumulative & Repeated & $0.01809 \pm 0.00000$ & $0.00359 \pm 0.00000$ & $0.72191 \pm 0.00000$ \\
Uniform gain & Window 8 & Biased & $0.01821 \pm 0.00000$ & $0.00000 \pm 0.00000$ & $0.72062 \pm 0.00152$ \\
Uniform gain & Window 8 & Balanced & $0.01837 \pm 0.00000$ & $0.00000 \pm 0.00000$ & $0.72163 \pm 0.00000$ \\
Uniform gain & Window 8 & Neyman & $0.01821 \pm 0.00000$ & $0.00000 \pm 0.00000$ & $0.72179 \pm 0.00000$ \\
Uniform gain & Window 8 & SR & $0.01805 \pm 0.00000$ & $0.00000 \pm 0.00000$ & $0.72195 \pm 0.00000$ \\
Uniform gain & Window 8 & Once & $0.01823 \pm 0.00001$ & $0.00060 \pm 0.00000$ & $0.72175 \pm 0.00001$ \\
Uniform gain & Window 8 & Repeated & $0.01810 \pm 0.00000$ & $0.00359 \pm 0.00000$ & $0.72190 \pm 0.00000$ \\
\end{longtable}\endgroup
Reversal labels in Panel C give the one-based round before which the change occurs.

\subsubsection*{C. Paired net-value contrasts}
\begingroup\footnotesize\setlength{\tabcolsep}{4pt}

\renewcommand{\arraystretch}{1.0}
\begin{longtable}{@{}lp{.19\linewidth}lp{.29\linewidth}rr@{}}
\toprule Study & Environment & Memory & Comparison & $N$ & Difference \\\midrule\endfirsthead
\toprule Study & Environment & Memory & Comparison & $N$ & Difference \\\midrule\endhead
\midrule\multicolumn{6}{r}{Continued on next page}\\\endfoot
\bottomrule\endlastfoot
core & Stationary & Reset & Repeated - Balanced & 128 & $-0.01702 \pm 0.00262$ \\
core & Stationary & Reset & Repeated - Successive rejection & 128 & $-0.01878 \pm 0.00266$ \\
core & Stationary & Reset & Repeated - Discover once & 128 & $-0.00142 \pm 0.00441$ \\
core & Stationary & Reset & Repeated - uniform fixed mixture & 128 & $0.00096 \pm 0.00252$ \\
core & Stationary & Cumulative & Repeated - Balanced & 128 & $-0.00007 \pm 0.00092$ \\
core & Stationary & Cumulative & Repeated - Successive rejection & 128 & $-0.00059 \pm 0.00097$ \\
core & Stationary & Cumulative & Repeated - Discover once & 128 & $0.00071 \pm 0.00070$ \\
core & Stationary & Cumulative & Repeated - uniform fixed mixture & 128 & $0.00343 \pm 0.00117$ \\
core & Stationary & Window 8 & Repeated - Balanced & 128 & $-0.00130 \pm 0.00114$ \\
core & Stationary & Window 8 & Repeated - Successive rejection & 128 & $-0.00212 \pm 0.00115$ \\
core & Stationary & Window 8 & Repeated - Discover once & 128 & $0.00254 \pm 0.00166$ \\
core & Stationary & Window 8 & Repeated - uniform fixed mixture & 128 & $0.00518 \pm 0.00137$ \\
core & Stationary & Cumulative & Balanced - Reset & 128 & $0.02783 \pm 0.00124$ \\
core & Stationary & Window 8 & Balanced - Reset & 128 & $0.02697 \pm 0.00126$ \\
core & Stationary & Cumulative & Repeated discovery - Reset & 128 & $0.04478 \pm 0.00283$ \\
core & Stationary & Window 8 & Repeated discovery - Reset & 128 & $0.04268 \pm 0.00291$ \\
core & Stationary & Cumulative & Discover once - Reset & 128 & $0.04264 \pm 0.00456$ \\
core & Stationary & Window 8 & Discover once - Reset & 128 & $0.03873 \pm 0.00364$ \\
core & Reversal at 25 & Reset & Repeated - Balanced & 128 & $-0.01997 \pm 0.00235$ \\
core & Reversal at 25 & Reset & Repeated - Successive rejection & 128 & $-0.02153 \pm 0.00241$ \\
core & Reversal at 25 & Reset & Repeated - Discover once & 128 & $-0.00159 \pm 0.00500$ \\
core & Reversal at 25 & Reset & Repeated - uniform fixed mixture & 128 & $-0.00109 \pm 0.00224$ \\
core & Reversal at 25 & Cumulative & Repeated - Balanced & 128 & $0.00321 \pm 0.00245$ \\
core & Reversal at 25 & Cumulative & Repeated - Successive rejection & 128 & $0.00062 \pm 0.00273$ \\
core & Reversal at 25 & Cumulative & Repeated - Discover once & 128 & $0.00384 \pm 0.00250$ \\
core & Reversal at 25 & Cumulative & Repeated - uniform fixed mixture & 128 & $0.00583 \pm 0.00259$ \\
core & Reversal at 25 & Window 8 & Repeated - Balanced & 128 & $0.00623 \pm 0.00240$ \\
core & Reversal at 25 & Window 8 & Repeated - Successive rejection & 128 & $0.00347 \pm 0.00252$ \\
core & Reversal at 25 & Window 8 & Repeated - Discover once & 128 & $0.00975 \pm 0.00267$ \\
core & Reversal at 25 & Window 8 & Repeated - uniform fixed mixture & 128 & $0.01114 \pm 0.00255$ \\
core & Reversal at 25 & Cumulative & Balanced - Reset & 128 & $-0.10253 \pm 0.00132$ \\
core & Reversal at 25 & Window 8 & Balanced - Reset & 128 & $-0.00697 \pm 0.00136$ \\
core & Reversal at 25 & Cumulative & Repeated discovery - Reset & 128 & $-0.07935 \pm 0.00345$ \\
core & Reversal at 25 & Window 8 & Repeated discovery - Reset & 128 & $0.01924 \pm 0.00351$ \\
core & Reversal at 25 & Cumulative & Discover once - Reset & 128 & $-0.08478 \pm 0.00493$ \\
core & Reversal at 25 & Window 8 & Discover once - Reset & 128 & $0.00789 \pm 0.00423$ \\
core & Uniform gain & Reset & Repeated - Balanced & 128 & $-0.02000 \pm 0.00240$ \\
core & Uniform gain & Reset & Repeated - Successive rejection & 128 & $-0.02208 \pm 0.00239$ \\
core & Uniform gain & Reset & Repeated - Discover once & 128 & $-0.00204 \pm 0.00464$ \\
core & Uniform gain & Reset & Repeated - uniform fixed mixture & 128 & $-0.00220 \pm 0.00223$ \\
core & Uniform gain & Cumulative & Repeated - Balanced & 128 & $0.00039 \pm 0.00013$ \\
core & Uniform gain & Cumulative & Repeated - Successive rejection & 128 & $0.00003 \pm 0.00011$ \\
core & Uniform gain & Cumulative & Repeated - Discover once & 128 & $0.00013 \pm 0.00001$ \\
core & Uniform gain & Cumulative & Repeated - uniform fixed mixture & 128 & $0.00139 \pm 0.00044$ \\
core & Uniform gain & Window 8 & Repeated - Balanced & 128 & $0.00039 \pm 0.00013$ \\
core & Uniform gain & Window 8 & Repeated - Successive rejection & 128 & $0.00003 \pm 0.00011$ \\
core & Uniform gain & Window 8 & Repeated - Discover once & 128 & $0.00085 \pm 0.00080$ \\
core & Uniform gain & Window 8 & Repeated - uniform fixed mixture & 128 & $0.00197 \pm 0.00056$ \\
core & Uniform gain & Cumulative & Balanced - Reset & 128 & $0.00736 \pm 0.00110$ \\
core & Uniform gain & Window 8 & Balanced - Reset & 128 & $0.00736 \pm 0.00110$ \\
core & Uniform gain & Cumulative & Repeated discovery - Reset & 128 & $0.02775 \pm 0.00235$ \\
core & Uniform gain & Window 8 & Repeated discovery - Reset & 128 & $0.02775 \pm 0.00235$ \\
core & Uniform gain & Cumulative & Discover once - Reset & 128 & $0.02558 \pm 0.00459$ \\
core & Uniform gain & Window 8 & Discover once - Reset & 128 & $0.02486 \pm 0.00437$ \\
exploratory control & Stationary & Reset & Repeated - Balanced & 128 & $-0.01665 \pm 0.00255$ \\
exploratory control & Stationary & Reset & Repeated - Trial-matched Balanced & 128 & $-0.01382 \pm 0.00238$ \\
exploratory control & Stationary & Reset & Repeated - Discover once & 128 & $0.00035 \pm 0.00448$ \\
exploratory control & Stationary & Cumulative & Repeated - Balanced & 128 & $0.00078 \pm 0.00091$ \\
exploratory control & Stationary & Cumulative & Repeated - Trial-matched Balanced & 128 & $-0.00016 \pm 0.00046$ \\
exploratory control & Stationary & Cumulative & Repeated - Discover once & 128 & $0.00110 \pm 0.00135$ \\
exploratory control & Stationary & Window 8 & Repeated - Balanced & 128 & $0.00023 \pm 0.00104$ \\
exploratory control & Stationary & Window 8 & Repeated - Trial-matched Balanced & 128 & $-0.00116 \pm 0.00058$ \\
exploratory control & Stationary & Window 8 & Repeated - Discover once & 128 & $0.00538 \pm 0.00210$ \\
exploratory control & Reversal at 25 & Reset & Repeated - Balanced & 128 & $-0.01808 \pm 0.00239$ \\
exploratory control & Reversal at 25 & Reset & Repeated - Trial-matched Balanced & 128 & $-0.01525 \pm 0.00224$ \\
exploratory control & Reversal at 25 & Reset & Repeated - Discover once & 128 & $-0.00223 \pm 0.00438$ \\
exploratory control & Reversal at 25 & Cumulative & Repeated - Balanced & 128 & $0.00922 \pm 0.00353$ \\
exploratory control & Reversal at 25 & Cumulative & Repeated - Trial-matched Balanced & 128 & $0.00832 \pm 0.00363$ \\
exploratory control & Reversal at 25 & Cumulative & Repeated - Discover once & 128 & $0.00926 \pm 0.00397$ \\
exploratory control & Reversal at 25 & Window 8 & Repeated - Balanced & 128 & $0.00607 \pm 0.00217$ \\
exploratory control & Reversal at 25 & Window 8 & Repeated - Trial-matched Balanced & 128 & $0.00191 \pm 0.00207$ \\
exploratory control & Reversal at 25 & Window 8 & Repeated - Discover once & 128 & $0.00889 \pm 0.00220$ \\
exploratory control & Reversal at 21 & Reset & Repeated - Balanced & 128 & $-0.01817 \pm 0.00254$ \\
exploratory control & Reversal at 21 & Reset & Repeated - Trial-matched Balanced & 128 & $-0.01586 \pm 0.00228$ \\
exploratory control & Reversal at 21 & Reset & Repeated - Discover once & 128 & $-0.00242 \pm 0.00439$ \\
exploratory control & Reversal at 21 & Cumulative & Repeated - Balanced & 128 & $0.00840 \pm 0.00513$ \\
exploratory control & Reversal at 21 & Cumulative & Repeated - Trial-matched Balanced & 128 & $0.01132 \pm 0.00526$ \\
exploratory control & Reversal at 21 & Cumulative & Repeated - Discover once & 128 & $0.00891 \pm 0.00550$ \\
exploratory control & Reversal at 21 & Window 8 & Repeated - Balanced & 128 & $0.00031 \pm 0.00118$ \\
exploratory control & Reversal at 21 & Window 8 & Repeated - Trial-matched Balanced & 128 & $-0.00063 \pm 0.00072$ \\
exploratory control & Reversal at 21 & Window 8 & Repeated - Discover once & 128 & $0.00257 \pm 0.00137$ \\
exploratory control & Reversal at 29 & Reset & Repeated - Balanced & 128 & $-0.01700 \pm 0.00252$ \\
exploratory control & Reversal at 29 & Reset & Repeated - Trial-matched Balanced & 128 & $-0.01399 \pm 0.00235$ \\
exploratory control & Reversal at 29 & Reset & Repeated - Discover once & 128 & $-0.00116 \pm 0.00433$ \\
exploratory control & Reversal at 29 & Cumulative & Repeated - Balanced & 128 & $0.00262 \pm 0.00148$ \\
exploratory control & Reversal at 29 & Cumulative & Repeated - Trial-matched Balanced & 128 & $0.00168 \pm 0.00136$ \\
exploratory control & Reversal at 29 & Cumulative & Repeated - Discover once & 128 & $0.00293 \pm 0.00225$ \\
exploratory control & Reversal at 29 & Window 8 & Repeated - Balanced & 128 & $0.00004 \pm 0.00120$ \\
exploratory control & Reversal at 29 & Window 8 & Repeated - Trial-matched Balanced & 128 & $-0.00057 \pm 0.00069$ \\
exploratory control & Reversal at 29 & Window 8 & Repeated - Discover once & 128 & $0.00342 \pm 0.00177$ \\
\end{longtable}\endgroup
\clearpage
\subsection*{Table S2: Program selection at the final scheduled assessment}
\label{supp:S2}
\addcontentsline{toc}{subsection}{Table S2: Program selection at the final scheduled assessment}
{\footnotesize Source: \texttt{results/learning\_conditional.csv}.}\par Each row summarizes 128 runs at the final program assessment: round 1 for discover-once and round 41 for repeated discovery. Program columns give the percentage retaining Biased, Balanced, or Neyman. The source table also records the conditional block outcomes and earlier assessments.

\subsubsection*{Final program-selection proportions}
\begingroup\footnotesize\setlength{\tabcolsep}{4pt}

\begin{longtable}{@{}lllllll@{}}
\toprule Environment & Memory & Arm & Review & Biased (\%) & Balanced (\%) & Neyman (\%) \\\midrule\endfirsthead
\toprule Environment & Memory & Arm & Review & Biased (\%) & Balanced (\%) & Neyman (\%) \\\midrule\endhead
\midrule\multicolumn{7}{r}{Continued on next page}\\\endfoot
\bottomrule\endlastfoot
Stationary & Reset & Once & 1 & 27.3 & 37.5 & 35.2 \\
Stationary & Reset & Repeated & 41 & 27.3 & 43.0 & 29.7 \\
Stationary & Cumulative & Once & 1 & 27.3 & 37.5 & 35.2 \\
Stationary & Cumulative & Repeated & 41 & 38.3 & 31.2 & 30.5 \\
Stationary & Window 8 & Once & 1 & 27.3 & 37.5 & 35.2 \\
Stationary & Window 8 & Repeated & 41 & 32.0 & 29.7 & 38.3 \\
Reversal & Reset & Once & 1 & 27.3 & 37.5 & 35.2 \\
Reversal & Reset & Repeated & 41 & 47.7 & 22.7 & 29.7 \\
Reversal & Cumulative & Once & 1 & 27.3 & 37.5 & 35.2 \\
Reversal & Cumulative & Repeated & 41 & 38.3 & 31.2 & 30.5 \\
Reversal & Window 8 & Once & 1 & 27.3 & 37.5 & 35.2 \\
Reversal & Window 8 & Repeated & 41 & 47.7 & 20.3 & 32.0 \\
Uniform gain & Reset & Once & 1 & 35.9 & 27.3 & 36.7 \\
Uniform gain & Reset & Repeated & 41 & 47.7 & 22.7 & 29.7 \\
Uniform gain & Cumulative & Once & 1 & 35.9 & 27.3 & 36.7 \\
Uniform gain & Cumulative & Repeated & 41 & 30.5 & 25.8 & 43.8 \\
Uniform gain & Window 8 & Once & 1 & 35.9 & 27.3 & 36.7 \\
Uniform gain & Window 8 & Repeated & 41 & 47.7 & 20.3 & 32.0 \\
\end{longtable}\endgroup
\clearpage
\subsection*{Table S3: Audit policy and memory-window grid}
\label{supp:S3}
\addcontentsline{toc}{subsection}{Table S3: Audit policy and memory-window grid}
{\footnotesize Source: \texttt{results/robustness\_summary.csv}.}\par All 54 combinations of three environments, three memory windows, and six audit policies. Each cell uses 200 independent seeds. Paired differences compare with fixed audit rate .10 within the same environment and window. The outcome is net value per task under the supplementary audit model.

\subsubsection*{Complete audit grid}
\begingroup\footnotesize\setlength{\tabcolsep}{4pt}

\renewcommand{\arraystretch}{.93}
\begin{longtable}{@{}lllll@{}}
\toprule Environment & Window & Policy & Net value & Difference vs. fixed .10 \\\midrule\endfirsthead
\toprule Environment & Window & Policy & Net value & Difference vs. fixed .10 \\\midrule\endhead
\midrule\multicolumn{5}{r}{Continued on next page}\\\endfoot
\bottomrule\endlastfoot
Stationary & 6 & No audit & $0.88054 \pm 0.00117$ & $0.02452 \pm 0.00080$ \\
Stationary & 6 & Fixed 0.05 & $0.85918 \pm 0.00181$ & $0.00316 \pm 0.00119$ \\
Stationary & 6 & Fixed 0.10 & $0.85602 \pm 0.00152$ & $0.00000 \pm 0.00000$ \\
Stationary & 6 & Fixed 0.40 & $0.80061 \pm 0.00120$ & $-0.05541 \pm 0.00078$ \\
Stationary & 6 & Periodic dense & $0.84527 \pm 0.00148$ & $-0.01075 \pm 0.00108$ \\
Stationary & 6 & Adaptive audit & $0.85356 \pm 0.00135$ & $-0.00246 \pm 0.00082$ \\
Stationary & 12 & No audit & $0.88054 \pm 0.00117$ & $0.02118 \pm 0.00047$ \\
Stationary & 12 & Fixed 0.05 & $0.86517 \pm 0.00177$ & $0.00580 \pm 0.00109$ \\
Stationary & 12 & Fixed 0.10 & $0.85937 \pm 0.00130$ & $0.00000 \pm 0.00000$ \\
Stationary & 12 & Fixed 0.40 & $0.80066 \pm 0.00119$ & $-0.05871 \pm 0.00049$ \\
Stationary & 12 & Periodic dense & $0.85290 \pm 0.00120$ & $-0.00647 \pm 0.00051$ \\
Stationary & 12 & Adaptive audit & $0.86681 \pm 0.00127$ & $0.00744 \pm 0.00060$ \\
Stationary & 24 & No audit & $0.88054 \pm 0.00117$ & $0.02037 \pm 0.00027$ \\
Stationary & 24 & Fixed 0.05 & $0.86869 \pm 0.00147$ & $0.00851 \pm 0.00067$ \\
Stationary & 24 & Fixed 0.10 & $0.86018 \pm 0.00120$ & $0.00000 \pm 0.00000$ \\
Stationary & 24 & Fixed 0.40 & $0.80066 \pm 0.00119$ & $-0.05952 \pm 0.00031$ \\
Stationary & 24 & Periodic dense & $0.85319 \pm 0.00118$ & $-0.00698 \pm 0.00028$ \\
Stationary & 24 & Adaptive audit & $0.87162 \pm 0.00125$ & $0.01144 \pm 0.00037$ \\
Slow shifts & 6 & No audit & $0.56603 \pm 0.00195$ & $-0.09302 \pm 0.00284$ \\
Slow shifts & 6 & Fixed 0.05 & $0.64384 \pm 0.00319$ & $-0.01521 \pm 0.00231$ \\
Slow shifts & 6 & Fixed 0.10 & $0.65905 \pm 0.00249$ & $0.00000 \pm 0.00000$ \\
Slow shifts & 6 & Fixed 0.40 & $0.62772 \pm 0.00153$ & $-0.03133 \pm 0.00222$ \\
Slow shifts & 6 & Periodic dense & $0.64667 \pm 0.00239$ & $-0.01238 \pm 0.00245$ \\
Slow shifts & 6 & Adaptive audit & $0.66438 \pm 0.00198$ & $0.00533 \pm 0.00234$ \\
Slow shifts & 12 & No audit & $0.56603 \pm 0.00195$ & $-0.09364 \pm 0.00277$ \\
Slow shifts & 12 & Fixed 0.05 & $0.65240 \pm 0.00322$ & $-0.00727 \pm 0.00266$ \\
Slow shifts & 12 & Fixed 0.10 & $0.65967 \pm 0.00241$ & $0.00000 \pm 0.00000$ \\
Slow shifts & 12 & Fixed 0.40 & $0.61216 \pm 0.00148$ & $-0.04751 \pm 0.00202$ \\
Slow shifts & 12 & Periodic dense & $0.66029 \pm 0.00204$ & $0.00062 \pm 0.00236$ \\
Slow shifts & 12 & Adaptive audit & $0.66785 \pm 0.00239$ & $0.00818 \pm 0.00263$ \\
Slow shifts & 24 & No audit & $0.56603 \pm 0.00195$ & $-0.06813 \pm 0.00271$ \\
Slow shifts & 24 & Fixed 0.05 & $0.63665 \pm 0.00326$ & $0.00249 \pm 0.00266$ \\
Slow shifts & 24 & Fixed 0.10 & $0.63416 \pm 0.00242$ & $0.00000 \pm 0.00000$ \\
Slow shifts & 24 & Fixed 0.40 & $0.57706 \pm 0.00157$ & $-0.05710 \pm 0.00205$ \\
Slow shifts & 24 & Periodic dense & $0.63149 \pm 0.00278$ & $-0.00267 \pm 0.00300$ \\
Slow shifts & 24 & Adaptive audit & $0.65398 \pm 0.00290$ & $0.01982 \pm 0.00316$ \\
Fast shifts & 6 & No audit & $0.56512 \pm 0.00210$ & $-0.02974 \pm 0.00259$ \\
Fast shifts & 6 & Fixed 0.05 & $0.59498 \pm 0.00279$ & $0.00012 \pm 0.00226$ \\
Fast shifts & 6 & Fixed 0.10 & $0.59486 \pm 0.00251$ & $0.00000 \pm 0.00000$ \\
Fast shifts & 6 & Fixed 0.40 & $0.54718 \pm 0.00180$ & $-0.04768 \pm 0.00209$ \\
Fast shifts & 6 & Periodic dense & $0.58839 \pm 0.00256$ & $-0.00647 \pm 0.00247$ \\
Fast shifts & 6 & Adaptive audit & $0.59544 \pm 0.00234$ & $0.00058 \pm 0.00247$ \\
Fast shifts & 12 & No audit & $0.56512 \pm 0.00210$ & $0.03871 \pm 0.00253$ \\
Fast shifts & 12 & Fixed 0.05 & $0.54272 \pm 0.00307$ & $0.01632 \pm 0.00243$ \\
Fast shifts & 12 & Fixed 0.10 & $0.52640 \pm 0.00295$ & $0.00000 \pm 0.00000$ \\
Fast shifts & 12 & Fixed 0.40 & $0.45976 \pm 0.00214$ & $-0.06665 \pm 0.00221$ \\
Fast shifts & 12 & Periodic dense & $0.52383 \pm 0.00234$ & $-0.00257 \pm 0.00265$ \\
Fast shifts & 12 & Adaptive audit & $0.52467 \pm 0.00273$ & $-0.00173 \pm 0.00289$ \\
Fast shifts & 24 & No audit & $0.56512 \pm 0.00210$ & $-0.01550 \pm 0.00239$ \\
Fast shifts & 24 & Fixed 0.05 & $0.58156 \pm 0.00266$ & $0.00093 \pm 0.00242$ \\
Fast shifts & 24 & Fixed 0.10 & $0.58062 \pm 0.00237$ & $0.00000 \pm 0.00000$ \\
Fast shifts & 24 & Fixed 0.40 & $0.54112 \pm 0.00239$ & $-0.03950 \pm 0.00270$ \\
Fast shifts & 24 & Periodic dense & $0.57131 \pm 0.00240$ & $-0.00931 \pm 0.00244$ \\
Fast shifts & 24 & Adaptive audit & $0.56229 \pm 0.00220$ & $-0.01834 \pm 0.00251$ \\
\end{longtable}\endgroup
\clearpage
\subsection*{Table S4: Expected-value-of-information sensitivity}
\label{supp:S4}
\addcontentsline{toc}{subsection}{Table S4: Expected-value-of-information sensitivity}
{\footnotesize Source: \texttt{results/audit\_extensions\_summary.csv}.}\par All 72 combinations of three priors, two routing thresholds, two candidate-generation costs, two environments, and three policies. Each cell uses 96 independent test seeds; development seeds selected the horizon separately. Nonzero generation cost charges counterfactual A outputs when route B is executed. Paired differences use fixed audit rate .10 within the same condition.

\subsubsection*{Complete prior, threshold, and generation-cost grid}
\begingroup\footnotesize\setlength{\tabcolsep}{4pt}

\begin{longtable}{@{}lllllll@{}}
\toprule Prior & Threshold & $g_c$ & Environment & Policy & Net value & Difference vs. fixed .10 \\\midrule\endfirsthead
\toprule Prior & Threshold & $g_c$ & Environment & Policy & Net value & Difference vs. fixed .10 \\\midrule\endhead
\midrule\multicolumn{7}{r}{Continued on next page}\\\endfoot
\bottomrule\endlastfoot
1:1 & 0.14 & 0 & Stationary & Fixed 0.10 & $0.84794 \pm 0.00236$ & $0.00000 \pm 0.00000$ \\
1:1 & 0.14 & 0 & Stationary & Adaptive audit & $0.85300 \pm 0.00213$ & $0.00506 \pm 0.00173$ \\
1:1 & 0.14 & 0 & Stationary & EVSI & $0.83379 \pm 0.00359$ & $-0.01415 \pm 0.00326$ \\
1:1 & 0.14 & 0 & Slow shifts & Fixed 0.10 & $0.66159 \pm 0.00308$ & $0.00000 \pm 0.00000$ \\
1:1 & 0.14 & 0 & Slow shifts & Adaptive audit & $0.67534 \pm 0.00293$ & $0.01375 \pm 0.00317$ \\
1:1 & 0.14 & 0 & Slow shifts & EVSI & $0.67584 \pm 0.00569$ & $0.01426 \pm 0.00540$ \\
1:1 & 0.14 & 0.2 & Stationary & Fixed 0.10 & $0.84718 \pm 0.00244$ & $0.00000 \pm 0.00000$ \\
1:1 & 0.14 & 0.2 & Stationary & Adaptive audit & $0.85173 \pm 0.00220$ & $0.00455 \pm 0.00184$ \\
1:1 & 0.14 & 0.2 & Stationary & EVSI & $0.82623 \pm 0.00362$ & $-0.02094 \pm 0.00338$ \\
1:1 & 0.14 & 0.2 & Slow shifts & Fixed 0.10 & $0.65182 \pm 0.00305$ & $0.00000 \pm 0.00000$ \\
1:1 & 0.14 & 0.2 & Slow shifts & Adaptive audit & $0.66587 \pm 0.00299$ & $0.01405 \pm 0.00323$ \\
1:1 & 0.14 & 0.2 & Slow shifts & EVSI & $0.66778 \pm 0.00568$ & $0.01596 \pm 0.00542$ \\
1:1 & 0.16 & 0 & Stationary & Fixed 0.10 & $0.85228 \pm 0.00191$ & $0.00000 \pm 0.00000$ \\
1:1 & 0.16 & 0 & Stationary & Adaptive audit & $0.85706 \pm 0.00194$ & $0.00479 \pm 0.00134$ \\
1:1 & 0.16 & 0 & Stationary & EVSI & $0.83421 \pm 0.00355$ & $-0.01807 \pm 0.00313$ \\
1:1 & 0.16 & 0 & Slow shifts & Fixed 0.10 & $0.65968 \pm 0.00334$ & $0.00000 \pm 0.00000$ \\
1:1 & 0.16 & 0 & Slow shifts & Adaptive audit & $0.67286 \pm 0.00277$ & $0.01318 \pm 0.00371$ \\
1:1 & 0.16 & 0 & Slow shifts & EVSI & $0.67570 \pm 0.00566$ & $0.01602 \pm 0.00547$ \\
1:1 & 0.16 & 0.2 & Stationary & Fixed 0.10 & $0.85180 \pm 0.00195$ & $0.00000 \pm 0.00000$ \\
1:1 & 0.16 & 0.2 & Stationary & Adaptive audit & $0.85607 \pm 0.00199$ & $0.00427 \pm 0.00145$ \\
1:1 & 0.16 & 0.2 & Stationary & EVSI & $0.82674 \pm 0.00357$ & $-0.02506 \pm 0.00321$ \\
1:1 & 0.16 & 0.2 & Slow shifts & Fixed 0.10 & $0.65064 \pm 0.00327$ & $0.00000 \pm 0.00000$ \\
1:1 & 0.16 & 0.2 & Slow shifts & Adaptive audit & $0.66208 \pm 0.00286$ & $0.01143 \pm 0.00367$ \\
1:1 & 0.16 & 0.2 & Slow shifts & EVSI & $0.66745 \pm 0.00563$ & $0.01681 \pm 0.00541$ \\
1:9 & 0.14 & 0 & Stationary & Fixed 0.10 & $0.85814 \pm 0.00172$ & $0.00000 \pm 0.00000$ \\
1:9 & 0.14 & 0 & Stationary & Adaptive audit & $0.86494 \pm 0.00192$ & $0.00680 \pm 0.00087$ \\
1:9 & 0.14 & 0 & Stationary & EVSI & $0.86557 \pm 0.00207$ & $0.00743 \pm 0.00147$ \\
1:9 & 0.14 & 0 & Slow shifts & Fixed 0.10 & $0.66200 \pm 0.00333$ & $0.00000 \pm 0.00000$ \\
1:9 & 0.14 & 0 & Slow shifts & Adaptive audit & $0.67056 \pm 0.00285$ & $0.00856 \pm 0.00352$ \\
1:9 & 0.14 & 0 & Slow shifts & EVSI & $0.65987 \pm 0.00606$ & $-0.00213 \pm 0.00667$ \\
1:9 & 0.14 & 0.2 & Stationary & Fixed 0.10 & $0.85805 \pm 0.00173$ & $0.00000 \pm 0.00000$ \\
1:9 & 0.14 & 0.2 & Stationary & Adaptive audit & $0.86480 \pm 0.00195$ & $0.00675 \pm 0.00094$ \\
1:9 & 0.14 & 0.2 & Stationary & EVSI & $0.86507 \pm 0.00222$ & $0.00702 \pm 0.00164$ \\
1:9 & 0.14 & 0.2 & Slow shifts & Fixed 0.10 & $0.65364 \pm 0.00323$ & $0.00000 \pm 0.00000$ \\
1:9 & 0.14 & 0.2 & Slow shifts & Adaptive audit & $0.65984 \pm 0.00290$ & $0.00619 \pm 0.00345$ \\
1:9 & 0.14 & 0.2 & Slow shifts & EVSI & $0.65465 \pm 0.00638$ & $0.00100 \pm 0.00689$ \\
1:9 & 0.16 & 0 & Stationary & Fixed 0.10 & $0.85918 \pm 0.00162$ & $0.00000 \pm 0.00000$ \\
1:9 & 0.16 & 0 & Stationary & Adaptive audit & $0.86861 \pm 0.00183$ & $0.00943 \pm 0.00053$ \\
1:9 & 0.16 & 0 & Stationary & EVSI & $0.86603 \pm 0.00210$ & $0.00685 \pm 0.00134$ \\
1:9 & 0.16 & 0 & Slow shifts & Fixed 0.10 & $0.65150 \pm 0.00390$ & $0.00000 \pm 0.00000$ \\
1:9 & 0.16 & 0 & Slow shifts & Adaptive audit & $0.66416 \pm 0.00362$ & $0.01266 \pm 0.00468$ \\
1:9 & 0.16 & 0 & Slow shifts & EVSI & $0.65263 \pm 0.00621$ & $0.00114 \pm 0.00705$ \\
1:9 & 0.16 & 0.2 & Stationary & Fixed 0.10 & $0.85915 \pm 0.00162$ & $0.00000 \pm 0.00000$ \\
1:9 & 0.16 & 0.2 & Stationary & Adaptive audit & $0.86858 \pm 0.00184$ & $0.00943 \pm 0.00055$ \\
1:9 & 0.16 & 0.2 & Stationary & EVSI & $0.86603 \pm 0.00210$ & $0.00687 \pm 0.00135$ \\
1:9 & 0.16 & 0.2 & Slow shifts & Fixed 0.10 & $0.64420 \pm 0.00377$ & $0.00000 \pm 0.00000$ \\
1:9 & 0.16 & 0.2 & Slow shifts & Adaptive audit & $0.65145 \pm 0.00364$ & $0.00725 \pm 0.00449$ \\
1:9 & 0.16 & 0.2 & Slow shifts & EVSI & $0.65263 \pm 0.00621$ & $0.00844 \pm 0.00697$ \\
1:19 & 0.14 & 0 & Stationary & Fixed 0.10 & $0.85951 \pm 0.00162$ & $0.00000 \pm 0.00000$ \\
1:19 & 0.14 & 0 & Stationary & Adaptive audit & $0.87199 \pm 0.00177$ & $0.01248 \pm 0.00036$ \\
1:19 & 0.14 & 0 & Stationary & EVSI & $0.87950 \pm 0.00163$ & $0.02000 \pm 0.00020$ \\
1:19 & 0.14 & 0 & Slow shifts & Fixed 0.10 & $0.64678 \pm 0.00420$ & $0.00000 \pm 0.00000$ \\
1:19 & 0.14 & 0 & Slow shifts & Adaptive audit & $0.65715 \pm 0.00397$ & $0.01037 \pm 0.00451$ \\
1:19 & 0.14 & 0 & Slow shifts & EVSI & $0.56729 \pm 0.00321$ & $-0.07949 \pm 0.00502$ \\
1:19 & 0.14 & 0.2 & Stationary & Fixed 0.10 & $0.85951 \pm 0.00162$ & $0.00000 \pm 0.00000$ \\
1:19 & 0.14 & 0.2 & Stationary & Adaptive audit & $0.87198 \pm 0.00177$ & $0.01247 \pm 0.00037$ \\
1:19 & 0.14 & 0.2 & Stationary & EVSI & $0.87950 \pm 0.00163$ & $0.02000 \pm 0.00020$ \\
1:19 & 0.14 & 0.2 & Slow shifts & Fixed 0.10 & $0.63993 \pm 0.00406$ & $0.00000 \pm 0.00000$ \\
1:19 & 0.14 & 0.2 & Slow shifts & Adaptive audit & $0.64531 \pm 0.00381$ & $0.00539 \pm 0.00416$ \\
1:19 & 0.14 & 0.2 & Slow shifts & EVSI & $0.56729 \pm 0.00321$ & $-0.07263 \pm 0.00487$ \\
1:19 & 0.16 & 0 & Stationary & Fixed 0.10 & $0.85954 \pm 0.00163$ & $0.00000 \pm 0.00000$ \\
1:19 & 0.16 & 0 & Stationary & Adaptive audit & $0.87348 \pm 0.00167$ & $0.01394 \pm 0.00026$ \\
1:19 & 0.16 & 0 & Stationary & EVSI & $0.87950 \pm 0.00163$ & $0.01997 \pm 0.00019$ \\
1:19 & 0.16 & 0 & Slow shifts & Fixed 0.10 & $0.62630 \pm 0.00474$ & $0.00000 \pm 0.00000$ \\
1:19 & 0.16 & 0 & Slow shifts & Adaptive audit & $0.63964 \pm 0.00445$ & $0.01334 \pm 0.00580$ \\
1:19 & 0.16 & 0 & Slow shifts & EVSI & $0.56729 \pm 0.00321$ & $-0.05901 \pm 0.00535$ \\
1:19 & 0.16 & 0.2 & Stationary & Fixed 0.10 & $0.85954 \pm 0.00163$ & $0.00000 \pm 0.00000$ \\
1:19 & 0.16 & 0.2 & Stationary & Adaptive audit & $0.87348 \pm 0.00167$ & $0.01394 \pm 0.00026$ \\
1:19 & 0.16 & 0.2 & Stationary & EVSI & $0.87950 \pm 0.00163$ & $0.01997 \pm 0.00019$ \\
1:19 & 0.16 & 0.2 & Slow shifts & Fixed 0.10 & $0.62102 \pm 0.00454$ & $0.00000 \pm 0.00000$ \\
1:19 & 0.16 & 0.2 & Slow shifts & Adaptive audit & $0.62680 \pm 0.00414$ & $0.00578 \pm 0.00529$ \\
1:19 & 0.16 & 0.2 & Slow shifts & EVSI & $0.56729 \pm 0.00321$ & $-0.05373 \pm 0.00515$ \\
\end{longtable}\endgroup
\clearpage
\subsection*{Table S5: Study 2: review arrangements and evidence generation}
\label{supp:S5}
\addcontentsline{toc}{subsection}{Table S5: Study 2: review arrangements and evidence generation}
{\footnotesize Source: \texttt{study2/results/learning\_summary.csv}; \texttt{study2/results/paired\_contrasts.csv}.}\par All 84 environment--policy--memory conditions: six environments, four paid policies crossed with three memories, and two fixed baselines per environment. Each learning condition uses 128 trajectories of 48 rounds; late outcomes cover the final 16 rounds. Error and optimal-choice hit are percentages. Regret is deployment loss above the finite-catalog minimum, excluding acquisition and switching. Acquisition cost is per round; production cost is per task in human-call units. Fixed baselines use exact expected-outcome scores. Paired, Randomized, Omit echo and Final only denote the four observation policies defined in Appendix E.

\subsubsection*{A. Full-horizon selection and net value}
\begingroup\footnotesize\setlength{\tabcolsep}{4pt}

\begin{longtable}{@{}llllll@{}}
\toprule Environment & Policy & Memory & Error (\%) & Net value & Hit (\%) \\\midrule\endfirsthead
\toprule Environment & Policy & Memory & Error (\%) & Net value & Hit (\%) \\\midrule\endhead
\midrule\multicolumn{6}{r}{Continued on next page}\\\endfoot
\bottomrule\endlastfoot
No echo & Final only & Reset & $10.472 \pm 0.056$ & $0.62353 \pm 0.00174$ & $0.03 \pm 0.04$ \\
No echo & Final only & Cumulative & $4.919 \pm 0.197$ & $0.78045 \pm 0.00563$ & $14.47 \pm 3.55$ \\
No echo & Final only & Window 8 & $7.116 \pm 0.137$ & $0.71668 \pm 0.00390$ & $6.38 \pm 0.76$ \\
No echo & Paired & Reset & $1.864 \pm 0.018$ & $0.85928 \pm 0.00056$ & $33.06 \pm 1.16$ \\
No echo & Paired & Cumulative & $1.772 \pm 0.002$ & $0.86475 \pm 0.00008$ & $98.70 \pm 0.18$ \\
No echo & Paired & Window 8 & $1.772 \pm 0.002$ & $0.86475 \pm 0.00008$ & $98.70 \pm 0.18$ \\
No echo & Randomized & Reset & $1.944 \pm 0.022$ & $0.85786 \pm 0.00068$ & $50.85 \pm 1.21$ \\
No echo & Randomized & Cumulative & $1.772 \pm 0.003$ & $0.86333 \pm 0.00026$ & $56.67 \pm 6.18$ \\
No echo & Randomized & Window 8 & $1.772 \pm 0.003$ & $0.86321 \pm 0.00016$ & $53.99 \pm 3.14$ \\
No echo & Omit echo & Reset & $1.857 \pm 0.018$ & $0.86217 \pm 0.00056$ & $98.36 \pm 0.33$ \\
No echo & Omit echo & Cumulative & $1.772 \pm 0.002$ & $0.86480 \pm 0.00008$ & $99.97 \pm 0.04$ \\
No echo & Omit echo & Window 8 & $1.772 \pm 0.002$ & $0.86480 \pm 0.00008$ & $99.97 \pm 0.04$ \\
No echo & Fixed exposed & Fixed & $3.520 \pm 0.000$ & $0.87440 \pm 0.00000$ & $0.00 \pm 0.00$ \\
No echo & Fixed blind & Fixed & $3.520 \pm 0.000$ & $0.87140 \pm 0.00000$ & $0.00 \pm 0.00$ \\
High echo & Final only & Reset & $11.109 \pm 0.058$ & $0.60564 \pm 0.00184$ & $0.00 \pm 0.00$ \\
High echo & Final only & Cumulative & $5.983 \pm 0.217$ & $0.74760 \pm 0.00640$ & $10.89 \pm 3.09$ \\
High echo & Final only & Window 8 & $8.788 \pm 0.124$ & $0.66901 \pm 0.00365$ & $2.70 \pm 0.50$ \\
High echo & Paired & Reset & $1.880 \pm 0.017$ & $0.85753 \pm 0.00052$ & $97.92 \pm 0.31$ \\
High echo & Paired & Cumulative & $1.773 \pm 0.003$ & $0.86077 \pm 0.00009$ & $99.95 \pm 0.05$ \\
High echo & Paired & Window 8 & $1.773 \pm 0.003$ & $0.86077 \pm 0.00009$ & $99.95 \pm 0.05$ \\
High echo & Randomized & Reset & $2.267 \pm 0.045$ & $0.84636 \pm 0.00134$ & $92.53 \pm 0.64$ \\
High echo & Randomized & Cumulative & $1.782 \pm 0.008$ & $0.86078 \pm 0.00024$ & $99.82 \pm 0.12$ \\
High echo & Randomized & Window 8 & $1.782 \pm 0.008$ & $0.86078 \pm 0.00024$ & $99.82 \pm 0.12$ \\
High echo & Omit echo & Reset & $9.961 \pm 0.023$ & $0.61912 \pm 0.00070$ & $0.00 \pm 0.00$ \\
High echo & Omit echo & Cumulative & $9.825 \pm 0.004$ & $0.62323 \pm 0.00013$ & $0.00 \pm 0.00$ \\
High echo & Omit echo & Window 8 & $9.825 \pm 0.004$ & $0.62323 \pm 0.00013$ & $0.00 \pm 0.00$ \\
High echo & Fixed exposed & Fixed & $9.637 \pm 0.000$ & $0.69089 \pm 0.00000$ & $0.00 \pm 0.00$ \\
High echo & Fixed blind & Fixed & $3.520 \pm 0.000$ & $0.87140 \pm 0.00000$ & $0.00 \pm 0.00$ \\
Both dependences & Final only & Reset & $11.045 \pm 0.054$ & $0.60834 \pm 0.00170$ & $0.00 \pm 0.00$ \\
Both dependences & Final only & Cumulative & $9.728 \pm 0.136$ & $0.63630 \pm 0.00404$ & $5.49 \pm 2.55$ \\
Both dependences & Final only & Window 8 & $10.295 \pm 0.079$ & $0.62548 \pm 0.00251$ & $1.01 \pm 0.27$ \\
Both dependences & Paired & Reset & $7.822 \pm 0.017$ & $0.68223 \pm 0.00036$ & $51.84 \pm 1.28$ \\
Both dependences & Paired & Cumulative & $7.423 \pm 0.049$ & $0.69099 \pm 0.00105$ & $82.29 \pm 3.92$ \\
Both dependences & Paired & Window 8 & $7.563 \pm 0.034$ & $0.68789 \pm 0.00073$ & $70.72 \pm 2.72$ \\
Both dependences & Randomized & Reset & $8.265 \pm 0.041$ & $0.66972 \pm 0.00115$ & $46.96 \pm 1.21$ \\
Both dependences & Randomized & Cumulative & $7.498 \pm 0.063$ & $0.68945 \pm 0.00135$ & $77.10 \pm 4.85$ \\
Both dependences & Randomized & Window 8 & $7.631 \pm 0.040$ & $0.68650 \pm 0.00087$ & $66.28 \pm 3.03$ \\
Both dependences & Omit echo & Reset & $13.749 \pm 0.098$ & $0.50789 \pm 0.00302$ & $0.00 \pm 0.00$ \\
Both dependences & Omit echo & Cumulative & $15.939 \pm 0.317$ & $0.43965 \pm 0.00969$ & $0.00 \pm 0.00$ \\
Both dependences & Omit echo & Window 8 & $15.125 \pm 0.212$ & $0.46465 \pm 0.00649$ & $0.00 \pm 0.00$ \\
Both dependences & Fixed exposed & Fixed & $9.919 \pm 0.000$ & $0.68242 \pm 0.00000$ & $0.00 \pm 0.00$ \\
Both dependences & Fixed blind & Fixed & $8.560 \pm 0.000$ & $0.72020 \pm 0.00000$ & $0.00 \pm 0.00$ \\
Echo shift & Final only & Reset & $10.722 \pm 0.054$ & $0.61665 \pm 0.00170$ & $0.00 \pm 0.00$ \\
Echo shift & Final only & Cumulative & $5.648 \pm 0.217$ & $0.75770 \pm 0.00621$ & $11.20 \pm 2.78$ \\
Echo shift & Final only & Window 8 & $7.826 \pm 0.118$ & $0.69639 \pm 0.00341$ & $4.82 \pm 0.57$ \\
Echo shift & Paired & Reset & $1.886 \pm 0.020$ & $0.85798 \pm 0.00060$ & $65.32 \pm 0.88$ \\
Echo shift & Paired & Cumulative & $1.779 \pm 0.006$ & $0.86255 \pm 0.00018$ & $98.54 \pm 0.18$ \\
Echo shift & Paired & Window 8 & $1.772 \pm 0.002$ & $0.86274 \pm 0.00008$ & $98.62 \pm 0.17$ \\
Echo shift & Randomized & Reset & $2.148 \pm 0.036$ & $0.85081 \pm 0.00108$ & $71.22 \pm 0.95$ \\
Echo shift & Randomized & Cumulative & $2.113 \pm 0.083$ & $0.85206 \pm 0.00238$ & $72.40 \pm 2.66$ \\
Echo shift & Randomized & Window 8 & $1.960 \pm 0.040$ & $0.85659 \pm 0.00117$ & $74.54 \pm 2.11$ \\
Echo shift & Omit echo & Reset & $5.926 \pm 0.023$ & $0.74011 \pm 0.00071$ & $48.96 \pm 0.26$ \\
Echo shift & Omit echo & Cumulative & $5.797 \pm 0.002$ & $0.74404 \pm 0.00008$ & $49.97 \pm 0.04$ \\
Echo shift & Omit echo & Window 8 & $5.797 \pm 0.002$ & $0.74404 \pm 0.00008$ & $49.97 \pm 0.04$ \\
Echo shift & Fixed exposed & Fixed & $6.579 \pm 0.000$ & $0.78264 \pm 0.00000$ & $0.00 \pm 0.00$ \\
Echo shift & Fixed blind & Fixed & $3.520 \pm 0.000$ & $0.87140 \pm 0.00000$ & $0.00 \pm 0.00$ \\
Source asymmetry & Final only & Reset & $11.009 \pm 0.058$ & $0.60849 \pm 0.00181$ & $0.00 \pm 0.00$ \\
Source asymmetry & Final only & Cumulative & $6.352 \pm 0.191$ & $0.73785 \pm 0.00535$ & $5.34 \pm 2.06$ \\
Source asymmetry & Final only & Window 8 & $8.247 \pm 0.134$ & $0.68513 \pm 0.00385$ & $1.84 \pm 0.41$ \\
Source asymmetry & Paired & Reset & $2.983 \pm 0.028$ & $0.82450 \pm 0.00084$ & $90.67 \pm 0.77$ \\
Source asymmetry & Paired & Cumulative & $2.693 \pm 0.004$ & $0.83318 \pm 0.00013$ & $99.76 \pm 0.13$ \\
Source asymmetry & Paired & Window 8 & $2.693 \pm 0.004$ & $0.83318 \pm 0.00013$ & $99.76 \pm 0.13$ \\
Source asymmetry & Randomized & Reset & $3.414 \pm 0.035$ & $0.81334 \pm 0.00107$ & $48.91 \pm 1.38$ \\
Source asymmetry & Randomized & Cumulative & $2.936 \pm 0.046$ & $0.82728 \pm 0.00116$ & $70.91 \pm 5.42$ \\
Source asymmetry & Randomized & Window 8 & $3.009 \pm 0.027$ & $0.82545 \pm 0.00069$ & $62.27 \pm 3.18$ \\
Source asymmetry & Omit echo & Reset & $4.386 \pm 0.094$ & $0.78638 \pm 0.00281$ & $0.00 \pm 0.00$ \\
Source asymmetry & Omit echo & Cumulative & $3.546 \pm 0.015$ & $0.81160 \pm 0.00044$ & $0.00 \pm 0.00$ \\
Source asymmetry & Omit echo & Window 8 & $3.546 \pm 0.015$ & $0.81160 \pm 0.00044$ & $0.00 \pm 0.00$ \\
Source asymmetry & Fixed exposed & Fixed & $5.101 \pm 0.000$ & $0.82697 \pm 0.00000$ & $0.00 \pm 0.00$ \\
Source asymmetry & Fixed blind & Fixed & $4.240 \pm 0.000$ & $0.84980 \pm 0.00000$ & $0.00 \pm 0.00$ \\
Stronger humans & Final only & Reset & $18.205 \pm 0.073$ & $0.39097 \pm 0.00207$ & $0.00 \pm 0.00$ \\
Stronger humans & Final only & Cumulative & $9.522 \pm 0.263$ & $0.63271 \pm 0.00748$ & $10.04 \pm 3.28$ \\
Stronger humans & Final only & Window 8 & $12.956 \pm 0.201$ & $0.53738 \pm 0.00577$ & $3.79 \pm 0.79$ \\
Stronger humans & Paired & Reset & $5.413 \pm 0.040$ & $0.74875 \pm 0.00121$ & $77.65 \pm 1.07$ \\
Stronger humans & Paired & Cumulative & $4.623 \pm 0.009$ & $0.77306 \pm 0.00028$ & $99.37 \pm 0.25$ \\
Stronger humans & Paired & Window 8 & $4.628 \pm 0.011$ & $0.77289 \pm 0.00032$ & $99.20 \pm 0.29$ \\
Stronger humans & Randomized & Reset & $5.629 \pm 0.044$ & $0.74241 \pm 0.00133$ & $72.31 \pm 1.12$ \\
Stronger humans & Randomized & Cumulative & $4.639 \pm 0.014$ & $0.77286 \pm 0.00041$ & $98.96 \pm 0.35$ \\
Stronger humans & Randomized & Window 8 & $4.646 \pm 0.014$ & $0.77262 \pm 0.00043$ & $98.73 \pm 0.38$ \\
Stronger humans & Omit echo & Reset & $10.693 \pm 0.101$ & $0.59515 \pm 0.00304$ & $0.00 \pm 0.00$ \\
Stronger humans & Omit echo & Cumulative & $8.969 \pm 0.024$ & $0.64781 \pm 0.00072$ & $0.00 \pm 0.00$ \\
Stronger humans & Omit echo & Window 8 & $8.982 \pm 0.027$ & $0.64741 \pm 0.00082$ & $0.00 \pm 0.00$ \\
Stronger humans & Fixed exposed & Fixed & $19.570 \pm 0.000$ & $0.39290 \pm 0.00000$ & $0.00 \pm 0.00$ \\
Stronger humans & Fixed blind & Fixed & $12.320 \pm 0.000$ & $0.60740 \pm 0.00000$ & $0.00 \pm 0.00$ \\
\end{longtable}\endgroup
\subsubsection*{B. Full-horizon costs and regret}
\begingroup\footnotesize\setlength{\tabcolsep}{4pt}

\begin{longtable}{@{}llllll@{}}
\toprule Environment & Policy & Memory & Acquisition cost & Production cost & Regret \\\midrule\endfirsthead
\toprule Environment & Policy & Memory & Acquisition cost & Production cost & Regret \\\midrule\endhead
\midrule\multicolumn{6}{r}{Continued on next page}\\\endfoot
\bottomrule\endlastfoot
No echo & Final only & Reset & $214.788 \pm 0.076$ & $0.4879 \pm 0.0049$ & $0.24580 \pm 0.00174$ \\
No echo & Final only & Cumulative & $214.788 \pm 0.076$ & $0.9764 \pm 0.0341$ & $0.08897 \pm 0.00563$ \\
No echo & Final only & Window 8 & $214.788 \pm 0.076$ & $0.8658 \pm 0.0193$ & $0.15270 \pm 0.00390$ \\
No echo & Paired & Reset & $233.600 \pm 0.000$ & $1.3853 \pm 0.0024$ & $0.00552 \pm 0.00056$ \\
No echo & Paired & Cumulative & $233.600 \pm 0.000$ & $1.2526 \pm 0.0004$ & $0.00011 \pm 0.00008$ \\
No echo & Paired & Window 8 & $233.600 \pm 0.000$ & $1.2526 \pm 0.0004$ & $0.00011 \pm 0.00008$ \\
No echo & Randomized & Reset & $232.500 \pm 0.000$ & $1.3501 \pm 0.0028$ & $0.00720 \pm 0.00068$ \\
No echo & Randomized & Cumulative & $232.500 \pm 0.000$ & $1.3367 \pm 0.0124$ & $0.00179 \pm 0.00026$ \\
No echo & Randomized & Window 8 & $232.500 \pm 0.000$ & $1.3420 \pm 0.0063$ & $0.00189 \pm 0.00016$ \\
No echo & Omit echo & Reset & $233.600 \pm 0.000$ & $1.2538 \pm 0.0010$ & $0.00268 \pm 0.00056$ \\
No echo & Omit echo & Cumulative & $233.600 \pm 0.000$ & $1.2501 \pm 0.0001$ & $0.00005 \pm 0.00008$ \\
No echo & Omit echo & Window 8 & $233.600 \pm 0.000$ & $1.2501 \pm 0.0001$ & $0.00005 \pm 0.00008$ \\
No echo & Fixed exposed & Fixed & $0.000 \pm 0.000$ & $1.0000 \pm 0.0000$ & $0.04749 \pm 0.00000$ \\
No echo & Fixed blind & Fixed & $0.000 \pm 0.000$ & $1.1500 \pm 0.0000$ & $0.05049 \pm 0.00000$ \\
High echo & Final only & Reset & $209.381 \pm 0.065$ & $0.4931 \pm 0.0055$ & $0.26102 \pm 0.00184$ \\
High echo & Final only & Cumulative & $209.381 \pm 0.065$ & $1.0879 \pm 0.0388$ & $0.11915 \pm 0.00640$ \\
High echo & Final only & Window 8 & $209.381 \pm 0.065$ & $0.8080 \pm 0.0206$ & $0.19770 \pm 0.00365$ \\
High echo & Paired & Reset & $233.600 \pm 0.000$ & $1.4516 \pm 0.0006$ & $0.00332 \pm 0.00052$ \\
High echo & Paired & Cumulative & $233.600 \pm 0.000$ & $1.4501 \pm 0.0001$ & $0.00008 \pm 0.00009$ \\
High echo & Paired & Window 8 & $233.600 \pm 0.000$ & $1.4501 \pm 0.0001$ & $0.00008 \pm 0.00009$ \\
High echo & Randomized & Reset & $232.500 \pm 0.000$ & $1.4428 \pm 0.0016$ & $0.01475 \pm 0.00134$ \\
High echo & Randomized & Cumulative & $232.500 \pm 0.000$ & $1.4498 \pm 0.0003$ & $0.00035 \pm 0.00024$ \\
High echo & Randomized & Window 8 & $232.500 \pm 0.000$ & $1.4498 \pm 0.0003$ & $0.00035 \pm 0.00024$ \\
High echo & Omit echo & Reset & $233.600 \pm 0.000$ & $1.2507 \pm 0.0003$ & $0.24173 \pm 0.00070$ \\
High echo & Omit echo & Cumulative & $233.600 \pm 0.000$ & $1.2500 \pm 0.0000$ & $0.23763 \pm 0.00013$ \\
High echo & Omit echo & Window 8 & $233.600 \pm 0.000$ & $1.2500 \pm 0.0000$ & $0.23763 \pm 0.00013$ \\
High echo & Fixed exposed & Fixed & $0.000 \pm 0.000$ & $1.0000 \pm 0.0000$ & $0.22700 \pm 0.00000$ \\
High echo & Fixed blind & Fixed & $0.000 \pm 0.000$ & $1.1500 \pm 0.0000$ & $0.04649 \pm 0.00000$ \\
Both dependences & Final only & Reset & $206.524 \pm 0.051$ & $0.4882 \pm 0.0051$ & $0.09392 \pm 0.00170$ \\
Both dependences & Final only & Cumulative & $206.524 \pm 0.051$ & $1.0708 \pm 0.0502$ & $0.06605 \pm 0.00404$ \\
Both dependences & Final only & Window 8 & $206.524 \pm 0.051$ & $0.7599 \pm 0.0200$ & $0.07684 \pm 0.00251$ \\
Both dependences & Paired & Reset & $233.600 \pm 0.000$ & $1.3006 \pm 0.0085$ & $0.01346 \pm 0.00036$ \\
Both dependences & Paired & Cumulative & $233.600 \pm 0.000$ & $1.4639 \pm 0.0232$ & $0.00477 \pm 0.00105$ \\
Both dependences & Paired & Window 8 & $233.600 \pm 0.000$ & $1.4090 \pm 0.0160$ & $0.00786 \pm 0.00073$ \\
Both dependences & Randomized & Reset & $232.500 \pm 0.000$ & $1.2746 \pm 0.0091$ & $0.02623 \pm 0.00115$ \\
Both dependences & Randomized & Cumulative & $232.500 \pm 0.000$ & $1.4414 \pm 0.0294$ & $0.00658 \pm 0.00135$ \\
Both dependences & Randomized & Window 8 & $232.500 \pm 0.000$ & $1.3894 \pm 0.0180$ & $0.00951 \pm 0.00087$ \\
Both dependences & Omit echo & Reset & $233.600 \pm 0.000$ & $1.1266 \pm 0.0064$ & $0.18780 \pm 0.00302$ \\
Both dependences & Omit echo & Cumulative & $233.600 \pm 0.000$ & $1.2569 \pm 0.0123$ & $0.25611 \pm 0.00969$ \\
Both dependences & Omit echo & Window 8 & $233.600 \pm 0.000$ & $1.2274 \pm 0.0097$ & $0.23109 \pm 0.00649$ \\
Both dependences & Fixed exposed & Fixed & $0.000 \pm 0.000$ & $1.0000 \pm 0.0000$ & $0.07038 \pm 0.00000$ \\
Both dependences & Fixed blind & Fixed & $0.000 \pm 0.000$ & $1.1500 \pm 0.0000$ & $0.03260 \pm 0.00000$ \\
Echo shift & Final only & Reset & $212.034 \pm 0.072$ & $0.4900 \pm 0.0051$ & $0.25136 \pm 0.00170$ \\
Echo shift & Final only & Cumulative & $212.034 \pm 0.072$ & $1.0534 \pm 0.0368$ & $0.11040 \pm 0.00621$ \\
Echo shift & Final only & Window 8 & $212.034 \pm 0.072$ & $0.8499 \pm 0.0191$ & $0.17166 \pm 0.00341$ \\
Echo shift & Paired & Reset & $233.600 \pm 0.000$ & $1.4192 \pm 0.0018$ & $0.00484 \pm 0.00060$ \\
Echo shift & Paired & Cumulative & $233.600 \pm 0.000$ & $1.3526 \pm 0.0004$ & $0.00030 \pm 0.00018$ \\
Echo shift & Paired & Window 8 & $233.600 \pm 0.000$ & $1.3528 \pm 0.0004$ & $0.00011 \pm 0.00008$ \\
Echo shift & Randomized & Reset & $232.500 \pm 0.000$ & $1.3970 \pm 0.0023$ & $0.01227 \pm 0.00108$ \\
Echo shift & Randomized & Cumulative & $232.500 \pm 0.000$ & $1.3888 \pm 0.0077$ & $0.01105 \pm 0.00238$ \\
Echo shift & Randomized & Window 8 & $232.500 \pm 0.000$ & $1.3922 \pm 0.0045$ & $0.00652 \pm 0.00117$ \\
Echo shift & Omit echo & Reset & $233.600 \pm 0.000$ & $1.2530 \pm 0.0008$ & $0.12274 \pm 0.00071$ \\
Echo shift & Omit echo & Cumulative & $233.600 \pm 0.000$ & $1.2501 \pm 0.0001$ & $0.11881 \pm 0.00008$ \\
Echo shift & Omit echo & Window 8 & $233.600 \pm 0.000$ & $1.2501 \pm 0.0001$ & $0.11881 \pm 0.00008$ \\
Echo shift & Fixed exposed & Fixed & $0.000 \pm 0.000$ & $1.0000 \pm 0.0000$ & $0.13724 \pm 0.00000$ \\
Echo shift & Fixed blind & Fixed & $0.000 \pm 0.000$ & $1.1500 \pm 0.0000$ & $0.04849 \pm 0.00000$ \\
Source asymmetry & Final only & Reset & $210.669 \pm 0.070$ & $0.4852 \pm 0.0054$ & $0.23042 \pm 0.00181$ \\
Source asymmetry & Final only & Cumulative & $210.669 \pm 0.070$ & $1.0058 \pm 0.0389$ & $0.10114 \pm 0.00535$ \\
Source asymmetry & Final only & Window 8 & $210.669 \pm 0.070$ & $0.7979 \pm 0.0204$ & $0.15384 \pm 0.00385$ \\
Source asymmetry & Paired & Reset & $233.600 \pm 0.000$ & $1.4483 \pm 0.0014$ & $0.00890 \pm 0.00084$ \\
Source asymmetry & Paired & Cumulative & $233.600 \pm 0.000$ & $1.4500 \pm 0.0002$ & $0.00024 \pm 0.00013$ \\
Source asymmetry & Paired & Window 8 & $233.600 \pm 0.000$ & $1.4500 \pm 0.0002$ & $0.00024 \pm 0.00013$ \\
Source asymmetry & Randomized & Reset & $232.500 \pm 0.000$ & $1.3701 \pm 0.0030$ & $0.02028 \pm 0.00107$ \\
Source asymmetry & Randomized & Cumulative & $232.500 \pm 0.000$ & $1.3925 \pm 0.0108$ & $0.00640 \pm 0.00116$ \\
Source asymmetry & Randomized & Window 8 & $232.500 \pm 0.000$ & $1.3752 \pm 0.0064$ & $0.00822 \pm 0.00069$ \\
Source asymmetry & Omit echo & Reset & $233.600 \pm 0.000$ & $1.2504 \pm 0.0006$ & $0.04703 \pm 0.00281$ \\
Source asymmetry & Omit echo & Cumulative & $233.600 \pm 0.000$ & $1.2500 \pm 0.0000$ & $0.02183 \pm 0.00044$ \\
Source asymmetry & Omit echo & Window 8 & $233.600 \pm 0.000$ & $1.2500 \pm 0.0000$ & $0.02183 \pm 0.00044$ \\
Source asymmetry & Fixed exposed & Fixed & $0.000 \pm 0.000$ & $1.0000 \pm 0.0000$ & $0.06349 \pm 0.00000$ \\
Source asymmetry & Fixed blind & Fixed & $0.000 \pm 0.000$ & $1.1500 \pm 0.0000$ & $0.04066 \pm 0.00000$ \\
Stronger humans & Final only & Reset & $211.370 \pm 0.066$ & $0.5579 \pm 0.0063$ & $0.38811 \pm 0.00207$ \\
Stronger humans & Final only & Cumulative & $211.370 \pm 0.066$ & $1.4993 \pm 0.0418$ & $0.14645 \pm 0.00748$ \\
Stronger humans & Final only & Window 8 & $211.370 \pm 0.066$ & $1.1133 \pm 0.0188$ & $0.24175 \pm 0.00576$ \\
Stronger humans & Paired & Reset & $233.600 \pm 0.000$ & $1.5894 \pm 0.0041$ & $0.02498 \pm 0.00121$ \\
Stronger humans & Paired & Cumulative & $233.600 \pm 0.000$ & $1.5611 \pm 0.0008$ & $0.00070 \pm 0.00028$ \\
Stronger humans & Paired & Window 8 & $233.600 \pm 0.000$ & $1.5614 \pm 0.0010$ & $0.00088 \pm 0.00032$ \\
Stronger humans & Randomized & Reset & $232.500 \pm 0.000$ & $1.5957 \pm 0.0047$ & $0.03157 \pm 0.00133$ \\
Stronger humans & Randomized & Cumulative & $232.500 \pm 0.000$ & $1.5609 \pm 0.0009$ & $0.00118 \pm 0.00041$ \\
Stronger humans & Randomized & Window 8 & $232.500 \pm 0.000$ & $1.5618 \pm 0.0013$ & $0.00142 \pm 0.00043$ \\
Stronger humans & Omit echo & Reset & $233.600 \pm 0.000$ & $1.3492 \pm 0.0046$ & $0.17857 \pm 0.00304$ \\
Stronger humans & Omit echo & Cumulative & $233.600 \pm 0.000$ & $1.3039 \pm 0.0012$ & $0.12595 \pm 0.00072$ \\
Stronger humans & Omit echo & Window 8 & $233.600 \pm 0.000$ & $1.3053 \pm 0.0015$ & $0.12636 \pm 0.00082$ \\
Stronger humans & Fixed exposed & Fixed & $0.000 \pm 0.000$ & $1.0000 \pm 0.0000$ & $0.43790 \pm 0.00000$ \\
Stronger humans & Fixed blind & Fixed & $0.000 \pm 0.000$ & $1.1500 \pm 0.0000$ & $0.22340 \pm 0.00000$ \\
\end{longtable}\endgroup
\subsubsection*{C. Final-16-round outcomes}
\begingroup\footnotesize\setlength{\tabcolsep}{4pt}

\begin{longtable}{@{}llllll@{}}
\toprule Environment & Policy & Memory & Late error (\%) & Late net & Late hit (\%) \\\midrule\endfirsthead
\toprule Environment & Policy & Memory & Late error (\%) & Late net & Late hit (\%) \\\midrule\endhead
\midrule\multicolumn{6}{r}{Continued on next page}\\\endfoot
\bottomrule\endlastfoot
No echo & Final only & Reset & $10.445 \pm 0.097$ & $0.62531 \pm 0.00303$ & $0.05 \pm 0.10$ \\
No echo & Final only & Cumulative & $3.130 \pm 0.228$ & $0.83006 \pm 0.00638$ & $27.78 \pm 6.96$ \\
No echo & Final only & Window 8 & $7.290 \pm 0.265$ & $0.71205 \pm 0.00757$ & $5.71 \pm 1.04$ \\
No echo & Paired & Reset & $1.863 \pm 0.031$ & $0.85928 \pm 0.00096$ & $32.86 \pm 2.15$ \\
No echo & Paired & Cumulative & $1.770 \pm 0.000$ & $0.86486 \pm 0.00000$ & $100.00 \pm 0.00$ \\
No echo & Paired & Window 8 & $1.770 \pm 0.000$ & $0.86486 \pm 0.00000$ & $100.00 \pm 0.00$ \\
No echo & Randomized & Reset & $1.947 \pm 0.042$ & $0.85773 \pm 0.00129$ & $50.15 \pm 2.11$ \\
No echo & Randomized & Cumulative & $1.770 \pm 0.000$ & $0.86349 \pm 0.00031$ & $59.33 \pm 7.63$ \\
No echo & Randomized & Window 8 & $1.770 \pm 0.000$ & $0.86324 \pm 0.00021$ & $53.47 \pm 5.24$ \\
No echo & Omit echo & Reset & $1.861 \pm 0.031$ & $0.86206 \pm 0.00096$ & $98.34 \pm 0.57$ \\
No echo & Omit echo & Cumulative & $1.770 \pm 0.000$ & $0.86486 \pm 0.00000$ & $100.00 \pm 0.00$ \\
No echo & Omit echo & Window 8 & $1.770 \pm 0.000$ & $0.86486 \pm 0.00000$ & $100.00 \pm 0.00$ \\
No echo & Fixed exposed & Fixed & $3.520 \pm 0.000$ & $0.87440 \pm 0.00000$ & $0.00 \pm 0.00$ \\
No echo & Fixed blind & Fixed & $3.520 \pm 0.000$ & $0.87140 \pm 0.00000$ & $0.00 \pm 0.00$ \\
High echo & Final only & Reset & $10.631 \pm 0.100$ & $0.62109 \pm 0.00317$ & $0.00 \pm 0.00$ \\
High echo & Final only & Cumulative & $3.308 \pm 0.294$ & $0.82076 \pm 0.00821$ & $21.53 \pm 6.14$ \\
High echo & Final only & Window 8 & $9.179 \pm 0.191$ & $0.65758 \pm 0.00552$ & $5.08 \pm 1.09$ \\
High echo & Paired & Reset & $1.895 \pm 0.037$ & $0.85708 \pm 0.00111$ & $97.61 \pm 0.68$ \\
High echo & Paired & Cumulative & $1.770 \pm 0.000$ & $0.86086 \pm 0.00000$ & $100.00 \pm 0.00$ \\
High echo & Paired & Window 8 & $1.770 \pm 0.000$ & $0.86086 \pm 0.00000$ & $100.00 \pm 0.00$ \\
High echo & Randomized & Reset & $2.263 \pm 0.071$ & $0.84645 \pm 0.00212$ & $92.68 \pm 1.01$ \\
High echo & Randomized & Cumulative & $1.770 \pm 0.000$ & $0.86113 \pm 0.00000$ & $100.00 \pm 0.00$ \\
High echo & Randomized & Window 8 & $1.770 \pm 0.000$ & $0.86113 \pm 0.00000$ & $100.00 \pm 0.00$ \\
High echo & Omit echo & Reset & $9.977 \pm 0.049$ & $0.61864 \pm 0.00147$ & $0.00 \pm 0.00$ \\
High echo & Omit echo & Cumulative & $9.821 \pm 0.000$ & $0.62334 \pm 0.00000$ & $0.00 \pm 0.00$ \\
High echo & Omit echo & Window 8 & $9.821 \pm 0.000$ & $0.62334 \pm 0.00000$ & $0.00 \pm 0.00$ \\
High echo & Fixed exposed & Fixed & $9.637 \pm 0.000$ & $0.69089 \pm 0.00000$ & $0.00 \pm 0.00$ \\
High echo & Fixed blind & Fixed & $3.520 \pm 0.000$ & $0.87140 \pm 0.00000$ & $0.00 \pm 0.00$ \\
Both dependences & Final only & Reset & $10.605 \pm 0.090$ & $0.62261 \pm 0.00284$ & $0.00 \pm 0.00$ \\
Both dependences & Final only & Cumulative & $9.145 \pm 0.183$ & $0.65054 \pm 0.00502$ & $8.01 \pm 4.00$ \\
Both dependences & Final only & Window 8 & $10.305 \pm 0.142$ & $0.62560 \pm 0.00439$ & $0.98 \pm 0.44$ \\
Both dependences & Paired & Reset & $7.830 \pm 0.030$ & $0.68209 \pm 0.00064$ & $51.46 \pm 2.24$ \\
Both dependences & Paired & Cumulative & $7.297 \pm 0.047$ & $0.69368 \pm 0.00102$ & $92.19 \pm 3.82$ \\
Both dependences & Paired & Window 8 & $7.580 \pm 0.056$ & $0.68748 \pm 0.00119$ & $69.04 \pm 4.45$ \\
Both dependences & Randomized & Reset & $8.269 \pm 0.078$ & $0.66957 \pm 0.00222$ & $47.17 \pm 2.10$ \\
Both dependences & Randomized & Cumulative & $7.372 \pm 0.065$ & $0.69228 \pm 0.00141$ & $85.94 \pm 5.29$ \\
Both dependences & Randomized & Window 8 & $7.605 \pm 0.056$ & $0.68720 \pm 0.00119$ & $67.09 \pm 4.45$ \\
Both dependences & Omit echo & Reset & $13.735 \pm 0.170$ & $0.50836 \pm 0.00526$ & $0.00 \pm 0.00$ \\
Both dependences & Omit echo & Cumulative & $16.695 \pm 0.321$ & $0.41632 \pm 0.00971$ & $0.00 \pm 0.00$ \\
Both dependences & Omit echo & Window 8 & $15.002 \pm 0.339$ & $0.46839 \pm 0.01040$ & $0.00 \pm 0.00$ \\
Both dependences & Fixed exposed & Fixed & $9.919 \pm 0.000$ & $0.68242 \pm 0.00000$ & $0.00 \pm 0.00$ \\
Both dependences & Fixed blind & Fixed & $8.560 \pm 0.000$ & $0.72020 \pm 0.00000$ & $0.00 \pm 0.00$ \\
Echo shift & Final only & Reset & $10.644 \pm 0.094$ & $0.62065 \pm 0.00299$ & $0.00 \pm 0.00$ \\
Echo shift & Final only & Cumulative & $4.285 \pm 0.411$ & $0.79352 \pm 0.01167$ & $24.41 \pm 6.42$ \\
Echo shift & Final only & Window 8 & $9.043 \pm 0.201$ & $0.66161 \pm 0.00589$ & $6.40 \pm 1.25$ \\
Echo shift & Paired & Reset & $1.890 \pm 0.034$ & $0.85722 \pm 0.00104$ & $97.80 \pm 0.63$ \\
Echo shift & Paired & Cumulative & $1.770 \pm 0.000$ & $0.86086 \pm 0.00000$ & $100.00 \pm 0.00$ \\
Echo shift & Paired & Window 8 & $1.770 \pm 0.000$ & $0.86086 \pm 0.00000$ & $100.00 \pm 0.00$ \\
Echo shift & Randomized & Reset & $2.321 \pm 0.080$ & $0.84473 \pm 0.00238$ & $91.89 \pm 1.12$ \\
Echo shift & Randomized & Cumulative & $1.837 \pm 0.057$ & $0.85915 \pm 0.00169$ & $99.17 \pm 0.71$ \\
Echo shift & Randomized & Window 8 & $1.770 \pm 0.000$ & $0.86113 \pm 0.00000$ & $100.00 \pm 0.00$ \\
Echo shift & Omit echo & Reset & $9.981 \pm 0.048$ & $0.61851 \pm 0.00144$ & $0.00 \pm 0.00$ \\
Echo shift & Omit echo & Cumulative & $9.821 \pm 0.000$ & $0.62334 \pm 0.00000$ & $0.00 \pm 0.00$ \\
Echo shift & Omit echo & Window 8 & $9.821 \pm 0.000$ & $0.62334 \pm 0.00000$ & $0.00 \pm 0.00$ \\
Echo shift & Fixed exposed & Fixed & $9.637 \pm 0.000$ & $0.69089 \pm 0.00000$ & $0.00 \pm 0.00$ \\
Echo shift & Fixed blind & Fixed & $3.520 \pm 0.000$ & $0.87140 \pm 0.00000$ & $0.00 \pm 0.00$ \\
Source asymmetry & Final only & Reset & $10.490 \pm 0.088$ & $0.62523 \pm 0.00279$ & $0.00 \pm 0.00$ \\
Source asymmetry & Final only & Cumulative & $4.391 \pm 0.239$ & $0.79002 \pm 0.00652$ & $12.65 \pm 4.99$ \\
Source asymmetry & Final only & Window 8 & $8.316 \pm 0.244$ & $0.68341 \pm 0.00695$ & $4.30 \pm 1.01$ \\
Source asymmetry & Paired & Reset & $2.964 \pm 0.052$ & $0.82508 \pm 0.00158$ & $90.92 \pm 1.38$ \\
Source asymmetry & Paired & Cumulative & $2.685 \pm 0.000$ & $0.83342 \pm 0.00000$ & $100.00 \pm 0.00$ \\
Source asymmetry & Paired & Window 8 & $2.685 \pm 0.000$ & $0.83342 \pm 0.00000$ & $100.00 \pm 0.00$ \\
Source asymmetry & Randomized & Reset & $3.393 \pm 0.054$ & $0.81400 \pm 0.00163$ & $49.02 \pm 2.20$ \\
Source asymmetry & Randomized & Cumulative & $2.887 \pm 0.055$ & $0.82859 \pm 0.00139$ & $75.83 \pm 6.61$ \\
Source asymmetry & Randomized & Window 8 & $3.004 \pm 0.043$ & $0.82562 \pm 0.00107$ & $61.77 \pm 5.09$ \\
Source asymmetry & Omit echo & Reset & $4.310 \pm 0.173$ & $0.78864 \pm 0.00519$ & $0.00 \pm 0.00$ \\
Source asymmetry & Omit echo & Cumulative & $3.521 \pm 0.000$ & $0.81234 \pm 0.00000$ & $0.00 \pm 0.00$ \\
Source asymmetry & Omit echo & Window 8 & $3.521 \pm 0.000$ & $0.81234 \pm 0.00000$ & $0.00 \pm 0.00$ \\
Source asymmetry & Fixed exposed & Fixed & $5.101 \pm 0.000$ & $0.82697 \pm 0.00000$ & $0.00 \pm 0.00$ \\
Source asymmetry & Fixed blind & Fixed & $4.240 \pm 0.000$ & $0.84980 \pm 0.00000$ & $0.00 \pm 0.00$ \\
Stronger humans & Final only & Reset & $18.684 \pm 0.136$ & $0.37746 \pm 0.00385$ & $0.00 \pm 0.00$ \\
Stronger humans & Final only & Cumulative & $7.391 \pm 0.370$ & $0.69146 \pm 0.01067$ & $15.82 \pm 5.67$ \\
Stronger humans & Final only & Window 8 & $12.709 \pm 0.342$ & $0.54463 \pm 0.00987$ & $3.61 \pm 1.19$ \\
Stronger humans & Paired & Reset & $5.481 \pm 0.079$ & $0.74670 \pm 0.00243$ & $75.88 \pm 2.18$ \\
Stronger humans & Paired & Cumulative & $4.600 \pm 0.000$ & $0.77377 \pm 0.00000$ & $100.00 \pm 0.00$ \\
Stronger humans & Paired & Window 8 & $4.615 \pm 0.017$ & $0.77329 \pm 0.00052$ & $99.56 \pm 0.45$ \\
Stronger humans & Randomized & Reset & $5.591 \pm 0.079$ & $0.74361 \pm 0.00240$ & $73.10 \pm 2.05$ \\
Stronger humans & Randomized & Cumulative & $4.600 \pm 0.000$ & $0.77404 \pm 0.00000$ & $100.00 \pm 0.00$ \\
Stronger humans & Randomized & Window 8 & $4.613 \pm 0.013$ & $0.77362 \pm 0.00041$ & $99.61 \pm 0.38$ \\
Stronger humans & Omit echo & Reset & $10.863 \pm 0.191$ & $0.59001 \pm 0.00574$ & $0.00 \pm 0.00$ \\
Stronger humans & Omit echo & Cumulative & $8.920 \pm 0.000$ & $0.64933 \pm 0.00000$ & $0.00 \pm 0.00$ \\
Stronger humans & Omit echo & Window 8 & $8.952 \pm 0.043$ & $0.64831 \pm 0.00129$ & $0.00 \pm 0.00$ \\
Stronger humans & Fixed exposed & Fixed & $19.570 \pm 0.000$ & $0.39290 \pm 0.00000$ & $0.00 \pm 0.00$ \\
Stronger humans & Fixed blind & Fixed & $12.320 \pm 0.000$ & $0.60740 \pm 0.00000$ & $0.00 \pm 0.00$ \\
\end{longtable}\endgroup
\subsubsection*{D. Final-16-round costs and regret}
\begingroup\footnotesize\setlength{\tabcolsep}{4pt}

\begin{longtable}{@{}llllll@{}}
\toprule Environment & Policy & Memory & Late acquisition & Late production & Late regret \\\midrule\endfirsthead
\toprule Environment & Policy & Memory & Late acquisition & Late production & Late regret \\\midrule\endhead
\midrule\multicolumn{6}{r}{Continued on next page}\\\endfoot
\bottomrule\endlastfoot
No echo & Final only & Reset & $215.187 \pm 0.137$ & $0.4351 \pm 0.0094$ & $0.24393 \pm 0.00303$ \\
No echo & Final only & Cumulative & $215.187 \pm 0.137$ & $1.1752 \pm 0.0603$ & $0.03929 \pm 0.00639$ \\
No echo & Final only & Window 8 & $215.187 \pm 0.137$ & $0.8326 \pm 0.0325$ & $0.15724 \pm 0.00757$ \\
No echo & Paired & Reset & $233.600 \pm 0.000$ & $1.3864 \pm 0.0044$ & $0.00552 \pm 0.00096$ \\
No echo & Paired & Cumulative & $233.600 \pm 0.000$ & $1.2500 \pm 0.0000$ & $0.00000 \pm 0.00000$ \\
No echo & Paired & Window 8 & $233.600 \pm 0.000$ & $1.2500 \pm 0.0000$ & $0.00000 \pm 0.00000$ \\
No echo & Randomized & Reset & $232.500 \pm 0.000$ & $1.3511 \pm 0.0048$ & $0.00733 \pm 0.00129$ \\
No echo & Randomized & Cumulative & $232.500 \pm 0.000$ & $1.3313 \pm 0.0153$ & $0.00163 \pm 0.00031$ \\
No echo & Randomized & Window 8 & $232.500 \pm 0.000$ & $1.3431 \pm 0.0105$ & $0.00186 \pm 0.00021$ \\
No echo & Omit echo & Reset & $233.600 \pm 0.000$ & $1.2543 \pm 0.0015$ & $0.00279 \pm 0.00096$ \\
No echo & Omit echo & Cumulative & $233.600 \pm 0.000$ & $1.2500 \pm 0.0000$ & $0.00000 \pm 0.00000$ \\
No echo & Omit echo & Window 8 & $233.600 \pm 0.000$ & $1.2500 \pm 0.0000$ & $0.00000 \pm 0.00000$ \\
No echo & Fixed exposed & Fixed & $0.000 \pm 0.000$ & $1.0000 \pm 0.0000$ & $0.04749 \pm 0.00000$ \\
No echo & Fixed blind & Fixed & $0.000 \pm 0.000$ & $1.1500 \pm 0.0000$ & $0.05049 \pm 0.00000$ \\
High echo & Final only & Reset & $209.793 \pm 0.110$ & $0.4320 \pm 0.0089$ & $0.24546 \pm 0.00317$ \\
High echo & Final only & Cumulative & $209.793 \pm 0.110$ & $1.4389 \pm 0.0728$ & $0.04590 \pm 0.00821$ \\
High echo & Final only & Window 8 & $209.793 \pm 0.110$ & $0.7887 \pm 0.0335$ & $0.20904 \pm 0.00553$ \\
High echo & Paired & Reset & $233.600 \pm 0.000$ & $1.4516 \pm 0.0013$ & $0.00377 \pm 0.00111$ \\
High echo & Paired & Cumulative & $233.600 \pm 0.000$ & $1.4500 \pm 0.0000$ & $0.00000 \pm 0.00000$ \\
High echo & Paired & Window 8 & $233.600 \pm 0.000$ & $1.4500 \pm 0.0000$ & $0.00000 \pm 0.00000$ \\
High echo & Randomized & Reset & $232.500 \pm 0.000$ & $1.4446 \pm 0.0025$ & $0.01466 \pm 0.00212$ \\
High echo & Randomized & Cumulative & $232.500 \pm 0.000$ & $1.4500 \pm 0.0000$ & $0.00000 \pm 0.00000$ \\
High echo & Randomized & Window 8 & $232.500 \pm 0.000$ & $1.4500 \pm 0.0000$ & $0.00000 \pm 0.00000$ \\
High echo & Omit echo & Reset & $233.600 \pm 0.000$ & $1.2506 \pm 0.0008$ & $0.24221 \pm 0.00147$ \\
High echo & Omit echo & Cumulative & $233.600 \pm 0.000$ & $1.2500 \pm 0.0000$ & $0.23752 \pm 0.00000$ \\
High echo & Omit echo & Window 8 & $233.600 \pm 0.000$ & $1.2500 \pm 0.0000$ & $0.23752 \pm 0.00000$ \\
High echo & Fixed exposed & Fixed & $0.000 \pm 0.000$ & $1.0000 \pm 0.0000$ & $0.22700 \pm 0.00000$ \\
High echo & Fixed blind & Fixed & $0.000 \pm 0.000$ & $1.1500 \pm 0.0000$ & $0.04649 \pm 0.00000$ \\
Both dependences & Final only & Reset & $206.910 \pm 0.106$ & $0.4305 \pm 0.0082$ & $0.07956 \pm 0.00284$ \\
Both dependences & Final only & Cumulative & $206.910 \pm 0.106$ & $1.2295 \pm 0.0790$ & $0.05173 \pm 0.00502$ \\
Both dependences & Final only & Window 8 & $206.910 \pm 0.106$ & $0.7337 \pm 0.0334$ & $0.07663 \pm 0.00439$ \\
Both dependences & Paired & Reset & $233.600 \pm 0.000$ & $1.2952 \pm 0.0152$ & $0.01360 \pm 0.00063$ \\
Both dependences & Paired & Cumulative & $233.600 \pm 0.000$ & $1.5183 \pm 0.0217$ & $0.00209 \pm 0.00102$ \\
Both dependences & Paired & Window 8 & $233.600 \pm 0.000$ & $1.4029 \pm 0.0269$ & $0.00827 \pm 0.00119$ \\
Both dependences & Randomized & Reset & $232.500 \pm 0.000$ & $1.2762 \pm 0.0154$ & $0.02638 \pm 0.00222$ \\
Both dependences & Randomized & Cumulative & $232.500 \pm 0.000$ & $1.4903 \pm 0.0303$ & $0.00376 \pm 0.00141$ \\
Both dependences & Randomized & Window 8 & $232.500 \pm 0.000$ & $1.3928 \pm 0.0275$ & $0.00881 \pm 0.00119$ \\
Both dependences & Omit echo & Reset & $233.600 \pm 0.000$ & $1.1245 \pm 0.0118$ & $0.18733 \pm 0.00527$ \\
Both dependences & Omit echo & Cumulative & $233.600 \pm 0.000$ & $1.2902 \pm 0.0084$ & $0.27945 \pm 0.00972$ \\
Both dependences & Omit echo & Window 8 & $233.600 \pm 0.000$ & $1.2252 \pm 0.0164$ & $0.22736 \pm 0.01040$ \\
Both dependences & Fixed exposed & Fixed & $0.000 \pm 0.000$ & $1.0000 \pm 0.0000$ & $0.07038 \pm 0.00000$ \\
Both dependences & Fixed blind & Fixed & $0.000 \pm 0.000$ & $1.1500 \pm 0.0000$ & $0.03260 \pm 0.00000$ \\
Echo shift & Final only & Reset & $209.670 \pm 0.124$ & $0.4359 \pm 0.0092$ & $0.24594 \pm 0.00299$ \\
Echo shift & Final only & Cumulative & $209.670 \pm 0.124$ & $1.3365 \pm 0.0699$ & $0.07317 \pm 0.01168$ \\
Echo shift & Final only & Window 8 & $209.670 \pm 0.124$ & $0.7926 \pm 0.0303$ & $0.20503 \pm 0.00589$ \\
Echo shift & Paired & Reset & $233.600 \pm 0.000$ & $1.4524 \pm 0.0007$ & $0.00363 \pm 0.00104$ \\
Echo shift & Paired & Cumulative & $233.600 \pm 0.000$ & $1.4500 \pm 0.0000$ & $0.00000 \pm 0.00000$ \\
Echo shift & Paired & Window 8 & $233.600 \pm 0.000$ & $1.4500 \pm 0.0000$ & $0.00000 \pm 0.00000$ \\
Echo shift & Randomized & Reset & $232.500 \pm 0.000$ & $1.4427 \pm 0.0025$ & $0.01638 \pm 0.00238$ \\
Echo shift & Randomized & Cumulative & $232.500 \pm 0.000$ & $1.4483 \pm 0.0014$ & $0.00197 \pm 0.00169$ \\
Echo shift & Randomized & Window 8 & $232.500 \pm 0.000$ & $1.4500 \pm 0.0000$ & $0.00000 \pm 0.00000$ \\
Echo shift & Omit echo & Reset & $233.600 \pm 0.000$ & $1.2511 \pm 0.0003$ & $0.24234 \pm 0.00144$ \\
Echo shift & Omit echo & Cumulative & $233.600 \pm 0.000$ & $1.2500 \pm 0.0000$ & $0.23752 \pm 0.00000$ \\
Echo shift & Omit echo & Window 8 & $233.600 \pm 0.000$ & $1.2500 \pm 0.0000$ & $0.23752 \pm 0.00000$ \\
Echo shift & Fixed exposed & Fixed & $0.000 \pm 0.000$ & $1.0000 \pm 0.0000$ & $0.22700 \pm 0.00000$ \\
Echo shift & Fixed blind & Fixed & $0.000 \pm 0.000$ & $1.1500 \pm 0.0000$ & $0.04649 \pm 0.00000$ \\
Source asymmetry & Final only & Reset & $211.038 \pm 0.117$ & $0.4223 \pm 0.0082$ & $0.21360 \pm 0.00278$ \\
Source asymmetry & Final only & Cumulative & $211.038 \pm 0.117$ & $1.3368 \pm 0.0714$ & $0.04891 \pm 0.00653$ \\
Source asymmetry & Final only & Window 8 & $211.038 \pm 0.117$ & $0.7773 \pm 0.0383$ & $0.15547 \pm 0.00696$ \\
Source asymmetry & Paired & Reset & $233.600 \pm 0.000$ & $1.4478 \pm 0.0025$ & $0.00832 \pm 0.00158$ \\
Source asymmetry & Paired & Cumulative & $233.600 \pm 0.000$ & $1.4500 \pm 0.0000$ & $0.00000 \pm 0.00000$ \\
Source asymmetry & Paired & Window 8 & $233.600 \pm 0.000$ & $1.4500 \pm 0.0000$ & $0.00000 \pm 0.00000$ \\
Source asymmetry & Randomized & Reset & $232.500 \pm 0.000$ & $1.3683 \pm 0.0051$ & $0.01962 \pm 0.00163$ \\
Source asymmetry & Randomized & Cumulative & $232.500 \pm 0.000$ & $1.4017 \pm 0.0132$ & $0.00510 \pm 0.00139$ \\
Source asymmetry & Randomized & Window 8 & $232.500 \pm 0.000$ & $1.3735 \pm 0.0102$ & $0.00806 \pm 0.00107$ \\
Source asymmetry & Omit echo & Reset & $233.600 \pm 0.000$ & $1.2506 \pm 0.0009$ & $0.04477 \pm 0.00519$ \\
Source asymmetry & Omit echo & Cumulative & $233.600 \pm 0.000$ & $1.2500 \pm 0.0000$ & $0.02108 \pm 0.00000$ \\
Source asymmetry & Omit echo & Window 8 & $233.600 \pm 0.000$ & $1.2500 \pm 0.0000$ & $0.02108 \pm 0.00000$ \\
Source asymmetry & Fixed exposed & Fixed & $0.000 \pm 0.000$ & $1.0000 \pm 0.0000$ & $0.06349 \pm 0.00000$ \\
Source asymmetry & Fixed blind & Fixed & $0.000 \pm 0.000$ & $1.1500 \pm 0.0000$ & $0.04066 \pm 0.00000$ \\
Stronger humans & Final only & Reset & $211.641 \pm 0.115$ & $0.5113 \pm 0.0122$ & $0.40155 \pm 0.00386$ \\
Stronger humans & Final only & Cumulative & $211.641 \pm 0.115$ & $1.7568 \pm 0.0648$ & $0.08766 \pm 0.01067$ \\
Stronger humans & Final only & Window 8 & $211.641 \pm 0.115$ & $1.1179 \pm 0.0290$ & $0.23444 \pm 0.00987$ \\
Stronger humans & Paired & Reset & $233.600 \pm 0.000$ & $1.5895 \pm 0.0085$ & $0.02702 \pm 0.00243$ \\
Stronger humans & Paired & Cumulative & $233.600 \pm 0.000$ & $1.5600 \pm 0.0000$ & $0.00000 \pm 0.00000$ \\
Stronger humans & Paired & Window 8 & $233.600 \pm 0.000$ & $1.5610 \pm 0.0016$ & $0.00048 \pm 0.00052$ \\
Stronger humans & Randomized & Reset & $232.500 \pm 0.000$ & $1.5928 \pm 0.0076$ & $0.03037 \pm 0.00239$ \\
Stronger humans & Randomized & Cumulative & $232.500 \pm 0.000$ & $1.5600 \pm 0.0000$ & $0.00000 \pm 0.00000$ \\
Stronger humans & Randomized & Window 8 & $232.500 \pm 0.000$ & $1.5610 \pm 0.0014$ & $0.00042 \pm 0.00041$ \\
Stronger humans & Omit echo & Reset & $233.600 \pm 0.000$ & $1.3511 \pm 0.0098$ & $0.18371 \pm 0.00574$ \\
Stronger humans & Omit echo & Cumulative & $233.600 \pm 0.000$ & $1.3020 \pm 0.0000$ & $0.12444 \pm 0.00000$ \\
Stronger humans & Omit echo & Window 8 & $233.600 \pm 0.000$ & $1.3053 \pm 0.0030$ & $0.12545 \pm 0.00129$ \\
Stronger humans & Fixed exposed & Fixed & $0.000 \pm 0.000$ & $1.0000 \pm 0.0000$ & $0.43790 \pm 0.00000$ \\
Stronger humans & Fixed blind & Fixed & $0.000 \pm 0.000$ & $1.1500 \pm 0.0000$ & $0.22340 \pm 0.00000$ \\
\end{longtable}\endgroup
\clearpage
\subsubsection*{E. Paired policy and memory contrasts}
\begingroup\footnotesize\setlength{\tabcolsep}{4pt}

\begin{longtable}{@{}l>{\raggedright\arraybackslash}p{.175\linewidth}>{\raggedright\arraybackslash}p{.175\linewidth}rrr@{}}
\toprule Environment & Policy A / memory & Policy B / memory & Error diff. (pp) & Net difference & Regret diff. \\\midrule\endfirsthead
\toprule Environment & Policy A / memory & Policy B / memory & Error diff. (pp) & Net difference & Regret diff. \\\midrule\endhead
\midrule\multicolumn{6}{r}{Continued on next page}\\\endfoot
\bottomrule\endlastfoot
No echo & Paired / Cumulative & Omit echo / Cumulative & $0.000 \pm 0.000$ & $-0.00005 \pm 0.00001$ & $0.00005 \pm 0.00001$ \\
No echo & Paired / Cumulative & Randomized / Cumulative & $0.000 \pm 0.004$ & $0.00142 \pm 0.00027$ & $-0.00168 \pm 0.00027$ \\
No echo & Paired / Window 8 & Paired / Cumulative & $0.000 \pm 0.000$ & $0.00000 \pm 0.00000$ & $0.00000 \pm 0.00000$ \\
No echo & Randomized / Window 8 & Randomized / Cumulative & $0.000 \pm 0.000$ & $-0.00012 \pm 0.00017$ & $0.00011 \pm 0.00017$ \\
No echo & Paired / Cumulative & Fixed blind / Fixed & $-1.748 \pm 0.002$ & $-0.00665 \pm 0.00008$ & $-0.05038 \pm 0.00008$ \\
High echo & Paired / Cumulative & Omit echo / Cumulative & $-8.052 \pm 0.001$ & $0.23755 \pm 0.00004$ & $-0.23755 \pm 0.00004$ \\
High echo & Paired / Cumulative & Randomized / Cumulative & $-0.009 \pm 0.008$ & $-0.00000 \pm 0.00025$ & $-0.00027 \pm 0.00025$ \\
High echo & Paired / Window 8 & Paired / Cumulative & $0.000 \pm 0.000$ & $0.00000 \pm 0.00000$ & $0.00000 \pm 0.00000$ \\
High echo & Randomized / Window 8 & Randomized / Cumulative & $0.000 \pm 0.000$ & $0.00000 \pm 0.00000$ & $0.00000 \pm 0.00000$ \\
High echo & Paired / Cumulative & Fixed blind / Fixed & $-1.747 \pm 0.003$ & $-0.01062 \pm 0.00009$ & $-0.04641 \pm 0.00009$ \\
Both dependences & Paired / Cumulative & Omit echo / Cumulative & $-8.516 \pm 0.365$ & $0.25134 \pm 0.01073$ & $-0.25134 \pm 0.01073$ \\
Both dependences & Paired / Cumulative & Randomized / Cumulative & $-0.075 \pm 0.061$ & $0.00154 \pm 0.00131$ & $-0.00181 \pm 0.00131$ \\
Both dependences & Paired / Window 8 & Paired / Cumulative & $0.140 \pm 0.034$ & $-0.00310 \pm 0.00073$ & $0.00309 \pm 0.00073$ \\
Both dependences & Randomized / Window 8 & Randomized / Cumulative & $0.132 \pm 0.040$ & $-0.00295 \pm 0.00084$ & $0.00293 \pm 0.00084$ \\
Both dependences & Paired / Cumulative & Fixed blind / Fixed & $-1.137 \pm 0.049$ & $-0.02921 \pm 0.00105$ & $-0.02783 \pm 0.00105$ \\
Echo shift & Paired / Cumulative & Omit echo / Cumulative & $-4.019 \pm 0.006$ & $0.11851 \pm 0.00017$ & $-0.11851 \pm 0.00017$ \\
Echo shift & Paired / Cumulative & Randomized / Cumulative & $-0.334 \pm 0.084$ & $0.01049 \pm 0.00239$ & $-0.01075 \pm 0.00239$ \\
Echo shift & Paired / Window 8 & Paired / Cumulative & $-0.007 \pm 0.006$ & $0.00019 \pm 0.00017$ & $-0.00019 \pm 0.00017$ \\
Echo shift & Randomized / Window 8 & Randomized / Cumulative & $-0.153 \pm 0.076$ & $0.00453 \pm 0.00224$ & $-0.00453 \pm 0.00224$ \\
Echo shift & Paired / Cumulative & Fixed blind / Fixed & $-1.741 \pm 0.006$ & $-0.00885 \pm 0.00018$ & $-0.04819 \pm 0.00018$ \\
Source asymmetry & Paired / Cumulative & Omit echo / Cumulative & $-0.853 \pm 0.010$ & $0.02158 \pm 0.00031$ & $-0.02158 \pm 0.00031$ \\
Source asymmetry & Paired / Cumulative & Randomized / Cumulative & $-0.244 \pm 0.046$ & $0.00590 \pm 0.00116$ & $-0.00616 \pm 0.00116$ \\
Source asymmetry & Paired / Window 8 & Paired / Cumulative & $0.000 \pm 0.000$ & $0.00000 \pm 0.00000$ & $0.00000 \pm 0.00000$ \\
Source asymmetry & Randomized / Window 8 & Randomized / Cumulative & $0.072 \pm 0.030$ & $-0.00183 \pm 0.00075$ & $0.00182 \pm 0.00074$ \\
Source asymmetry & Paired / Cumulative & Fixed blind / Fixed & $-1.547 \pm 0.004$ & $-0.01662 \pm 0.00013$ & $-0.04041 \pm 0.00013$ \\
Stronger humans & Paired / Cumulative & Omit echo / Cumulative & $-4.347 \pm 0.016$ & $0.12525 \pm 0.00048$ & $-0.12525 \pm 0.00048$ \\
Stronger humans & Paired / Cumulative & Randomized / Cumulative & $-0.016 \pm 0.013$ & $0.00021 \pm 0.00039$ & $-0.00048 \pm 0.00039$ \\
Stronger humans & Paired / Window 8 & Paired / Cumulative & $0.006 \pm 0.006$ & $-0.00018 \pm 0.00018$ & $0.00018 \pm 0.00018$ \\
Stronger humans & Randomized / Window 8 & Randomized / Cumulative & $0.007 \pm 0.006$ & $-0.00024 \pm 0.00019$ & $0.00024 \pm 0.00019$ \\
Stronger humans & Paired / Cumulative & Fixed blind / Fixed & $-7.697 \pm 0.009$ & $0.16567 \pm 0.00028$ & $-0.22270 \pm 0.00028$ \\
\end{longtable}\endgroup
\endgroup

\end{document}